\documentclass[12pt]{article}

\usepackage[pdftex,margin=1in]{geometry}
\usepackage[T1]{fontenc}
\usepackage[english]{babel}
\usepackage[authoryear]{natbib}
\usepackage[small]{titlesec}
\usepackage[justification=centering]{caption}

\usepackage[final]{pdfpages}

\usepackage{amsmath,amsthm,amssymb,graphicx,enumerate,booktabs,bigstrut,rotating,multirow,float,etoolbox,lmodern,comment,listings,qtree,a4wide,moresize,bbm,subcaption,tikz,pdfescape,mathtools,flafter,enumitem,setspace,ragged2e,colortbl,lineno,lscape,pdflscape,stackengine}

\usetikzlibrary{decorations.pathreplacing, positioning, arrows.meta, shapes, calc, arrows, shadows, decorations.pathmorphing}

\newcolumntype{L}[1]{>{\raggedright\let\newline\\\arraybackslash\hspace{0pt}}m{#1}}
\newcolumntype{C}[1]{>{\centering\let\newline\\\arraybackslash\hspace{0pt}}m{#1}}
\newcolumntype{R}[1]{>{\raggedleft\let\newline\\\arraybackslash\hspace{0pt}}m{#1}}

\makeatletter
\renewcommand\subsubsection{\@startsection{subsubsection}{3}{\z@}%
	{-3.25ex\@plus -1ex \@minus -.2ex}%
	{-1.5ex \@plus -.2ex}%
	{\normalfont\normalsize\bfseries}
}
\def\@biblabel#1{\hspace*{-\labelsep}}

\makeatother

\def\sym#1{\ifmmode^{#1}\else\(^{#1}\)\fi}

\pdfpageheight\paperheight
\pdfpagewidth\paperwidth

\usepackage{datatool}
\DTLsetseparator{,}
\DTLloaddb[keys={key}]{est}{results/kv_store.csv}
\newcommand{\getval}[1]{\DTLfetch{est}{key}{#1}{value}}

\usepackage{hyperref}
\hypersetup{colorlinks,linkcolor={red},citecolor={blue},urlcolor={blue}}

\definecolor{redcomment}{RGB}{213,94,0}
\definecolor{yellowcomment}{RGB}{240,228,66}
\definecolor{greencomment}{RGB}{0,158,115}
\definecolor{bluecomment}{RGB}{0,114,178}

\begin{document}
\begin{filecontents*}{results/kv_store.csv}
key,value,timestamp,label
n_intake,256,,N completed the sign-up survey and scheduled a Session One slot
n_students,211,,N attended S1
n_both_sessions,205,,N attended both sessions
s1_attend_pct,82.4,,S1 attendance rate (%)
both_attend_pct,80.1,,Both-sessions attendance rate (%)
s2_retention_pct,97.2,,S2 retention rate conditional on S1 (%)
mean_age,20.1,,Mean age
pct_male,35.1,,% male
pct_white,50.0,,% white
pct_international,26.9,,% international
pct_public_hs,54.3,,% public high school
pct_firstyear,37.4,,% first-year
pct_nat_sci,30.8,,% natural sciences
pct_soc_sci,36.5,,% social sciences
mean_gpa,3.68,,Mean GPA
mean_sat,1386,,Mean SAT score
mean_baseline_knowledge,1.5,,Mean baseline self-assessment
baseline_frac_correct,30.3,,Baseline fraction correct (%)
mean_comp_check,4.76,,Mean comprehension check score (0-5)
mean_gpa_control,3.69,,Control mean GPA
mean_gpa_treat,3.67,,Treatment mean GPA
mean_sat_control,1392,,Control mean SAT
mean_sat_treat,1383,,Treatment mean SAT
s1_attend_pct_control,82.5,,"S1 attendance rate, control (%)"
s1_attend_pct_treat,82.3,,"S1 attendance rate, treatment (%)"
bal_joint_F,1.19,,"Joint balance F-statistic (non-redundant baseline, strata FE)"
bal_joint_pval,0.278,,Joint balance F-test p-value
pct_no_admissions_test,22,,% who did not take admission test
avg_session_gap,6.98,,"Avg days between sessions, S2 attendees"
gap_7days_pct,61.0,,% of S2 attendees with exactly 7-day gap
gap_6to8_pct,80.0,,% of S2 attendees with 6-8 day gap
fs_revealed_pp,65.9,,First-stage: revealed preference (pp)
ai_takeup_pct,70,,% AI-allowed who used AI (raw share)
fs_selfreport_pp,60.9,,First-stage: self-reported any AI (pp)
fs_chatgpt_pp,60.2,,First-stage: ChatGPT usage (pp)
fs_other_ai_pp,2.9,,First-stage: other AI models (pp)
fs_other_ai_pval,0.182,,First-stage: other AI models p-value
ai_use_explain_pct,61.6,,% AI users: explain concepts
ai_use_summarize_pct,28.8,,% AI users: summarize readings
ai_use_draft_pct,20.5,,% AI users: draft responses
pct_sr_explain,46,,% treated: self-report explain concepts
pct_sr_summarize,23,,% treated: self-report summarize readings
pct_sr_draft,17,,% treated: self-report draft responses
pct_sr_edit,14,,% treated: self-report edit responses
pct_sr_proofread,14,,% treated: self-report proofreading
pct_sr_other,3,,% treated: self-report other uses
ai_helpful_users_pct,88,,% AI users reporting helpful
compliance_perceived_te_median,0,,"Median perceived AI effect (questions), treated S1 attendees (Table A1 split point)"
pct_revealed_explain,56,,% treated: revealed explain (ChatGPT logs)
pct_revealed_summarize,16,,% treated: revealed summarize (ChatGPT logs)
pct_revealed_draft,29,,% treated: revealed draft (ChatGPT logs)
pct_revealed_edit,23,,% treated: revealed edit (ChatGPT logs)
pct_revealed_proofread,2,,% treated: revealed proofread (ChatGPT logs)
pct_revealed_other,39,,% treated: revealed other uses (ChatGPT logs)
s1_test_itt_sd,0.28,,S1 test score ITT (SD)
s1_test_itt_sd3,0.285,,"S1 test score ITT (SD, 3 dp)"
s1_test_itt_pval,0.021,,S1 test score ITT p-value
s1_test_tot_sd,0.41,,S1 test score TOT (SD)
s1_test_tot_pval,0.022,,S1 test score TOT p-value
s1_test_itt_pp,7.2,,S1 test score ITT (pp)
s1_test_control_mean,56.3,,S1 control mean fraction correct (%)
s1_test_pct_improvement,13,,S1 test % improvement over control mean
s1_test_tot_pp,10.4,,S1 test score TOT (pp)
s1_selfknow_itt,-0.10,,S1 self-assessed knowledge ITT (points)
s1_selfknow_control_mean,4.79,,S1 control mean self-assessed knowledge (0-10)
s2_test_itt_sd,0.21,,S2 test score ITT (SD)
s2_test_itt_pval,0.068,,S2 test score ITT p-value
s2_test_itt_pp,3.9,,S2 test score ITT (pp)
s2_test_control_mean,50.0,,S2 control mean fraction correct (%)
s2_test_pct_improvement,7.8,,S2 test % improvement over control mean
s2_selfknow_itt,0.22,,S2 self-assessed knowledge ITT (points)
s2_selfknow_control_mean,3.83,,S2 control mean self-assessed knowledge (0-10)
s2_selfknow_itt_pval,0.250,,S2 self-assessed knowledge ITT p-value
s2_selfknow_pct_decline,20,,S2 self-assessed knowledge % decline from S1 (control)
s2_test_itt_persistence_pct,54,,S2 test ITT pp as % of S1 ITT pp
fade_s1_test_control_mean,56.9,,"Control mean % correct in S1, students in both sessions"
fade_s2_test_control_mean,50.0,,"Control mean % correct in S2, students in both sessions"
fade_s1_selfknow_control_mean,4.77,,"Control mean self-assessed knowledge S1, students in both sessions"
fade_s2_selfknow_control_mean,3.83,,"Control mean self-assessed knowledge S2, students in both sessions"
s1_to_s2_test_control_fade_p,0.012,,S1->S2 control test fade-out: paired t p-value
s1_to_s2_selfknow_control_fade_p,0.000,,S1->S2 control selfknow fade-out: paired t p-value
baseline_test_forbidden_mean,30.6,,"Baseline test %, AI-forbidden"
baseline_test_allowed_mean,30.1,,"Baseline test %, AI-allowed"
post_s1_test_forbidden_mean,56.3,,"Post-S1 test %, AI-forbidden"
post_s1_test_allowed_mean,61.9,,"Post-S1 test %, AI-allowed"
baseline_to_s1_test_forbidden_p,0.000,,Baseline->S1 within AI-forbidden test gain: paired t p-value
baseline_to_s1_test_allowed_p,0.000,,Baseline->S1 within AI-allowed test gain: paired t p-value
baseline_selfknow_forbidden_mean,1.56,,"Baseline self-knowledge, AI-forbidden"
baseline_selfknow_allowed_mean,1.48,,"Baseline self-knowledge, AI-allowed"
post_s1_selfknow_forbidden_mean,4.79,,"Post-S1 self-knowledge, AI-forbidden"
post_s1_selfknow_allowed_mean,4.86,,"Post-S1 self-knowledge, AI-allowed"
baseline_to_s1_selfknow_forbidden_p,0.000,,Baseline->S1 within AI-forbidden selfknow gain: paired t p-value
baseline_to_s1_selfknow_allowed_p,0.000,,Baseline->S1 within AI-allowed selfknow gain: paired t p-value
lit_grand_mean_sd,0.17,,"Lit grand mean (DL random-effects, SD, incl. ours)"
lit_grand_mean_lo_sd,0.09,,"Lit grand mean: 95% CI lower (SD, incl. ours)"
lit_grand_mean_hi_sd,0.25,,"Lit grand mean: 95% CI upper (SD, incl. ours)"
lit_n_estimates,28,,Lit grand mean: total number of estimates (incl. ours)
s1_above40_pp,8.3,,S1 effect on Pr(score>=40%) (pp)
s1_above60_pp,10.6,,S1 effect on Pr(score>=60%) (pp)
s2_above20_pp,2.0,,S2 effect on Pr(score>=20%) (pp)
s2_above60_pp,9.8,,S2 effect on Pr(score>=60%) (pp)
s1_above40_p,0.071,,S1 effect on Pr(score>=40%): p-value
s1_above60_p,0.094,,S1 effect on Pr(score>=60%): p-value
s2_above20_p,0.412,,S2 effect on Pr(score>=20%): p-value
s2_above60_p,0.122,,S2 effect on Pr(score>=60%): p-value
between_lookup_pct,5,,% looked up extra info between sessions
between_study_pct,0,,% studied topic between sessions
s1_length_index_sd,0.13,,S1 length index ITT (SD)
s1_length_index_pval,0.272,,S1 length index ITT p-value
s1_readability_index_sd,0.06,,S1 readability index ITT (SD)
s1_readability_index_pval,0.492,,S1 readability index ITT p-value
s1_lexdiv_index_sd,0.11,,S1 lex diversity index ITT (SD)
s1_lexdiv_index_pval,0.354,,S1 lex diversity index ITT p-value
s1_cosine_within_itt,0.003,,S1 within-group cosine ITT (signed)
s1_cosine_within_control_mean,0.788,,S1 within-group cosine control mean
s1_cosine_within_pct_change,0.4,,S1 within-group cosine % change vs control (signed)
s1_cosine_source_itt,0.000,,S1 source-reading cosine ITT (signed)
s1_cosine_source_control_mean,0.814,,S1 source-reading cosine control mean
s1_cosine_source_pct_change,0.0,,S1 source-reading cosine % change vs control (signed)
s1_ai_detect_pp,22.2,,S1 AI-detected ITT (pp)
s1_ai_detect_p,0.000,,S1 AI-detected ITT p-value
s1_ai_detect_control_mean,6.6,,S1 control mean AI-detected (%)
s1_ai_detect_pct_increase,337,,S1 AI-detected % increase over control
s1_ai_detect_tot_pp,31.8,,S1 AI-detected TOT (pp)
s1_ai_detect_tot_p,0.000,,S1 AI-detected TOT p-value
s2_ai_detect_pp,0.7,,S2 AI-detected fraction increase (pp)
s2_ai_detect_control_mean,0.3,,S2 control mean AI-detected (%)
s2_ai_detect_pct_change,251,,S2 AI-detected % change over control
s2_ai_detect_pval,0.494,,S2 AI-detected ITT p-value
s2_length_index_sd,0.04,,S2 length index ITT (SD)
s2_readability_index_sd,0.00,,S2 readability index ITT (SD)
s2_lexdiv_index_sd,-0.11,,S2 lex diversity index ITT (SD)
s2_lexdiv_index_sd_abs,0.11,,S2 lex diversity index ITT (abs SD value)
s1_length_control_mean,348,,S1 control mean essay length (tokens)
s1_extra_tokens,23,,S1 extra tokens from AI
s1_length_pct_increase,7,,S1 essay length % increase
s1_flesch_increase,4.2,,S1 Flesch Reading Ease increase
s1_sentence_length_decrease,4.7,,S1 mean sentence length decrease (words)
s2_extra_tokens,8.3,,S2 extra tokens from AI
s2_flesch_increase,0.6,,S2 Flesch Reading Ease increase
s2_sentence_length_decrease,1.8,,S2 mean sentence length decrease (words)
s2_ttr_decrease,0.013,,S2 TTR decrease
s2_hapax_decrease,0.013,,S2 hapax proportion decrease
s2_quality_human_increase,0.39,,S2 human overall quality increase (points)
s2_quality_ai_increase,0.23,,S2 AI overall quality increase (points)
s1_qual_overall_human_itt,0.21,,"S1 overall quality, human graders, ITT (points)"
s1_qual_overall_human_itt_sd,0.15,,"S1 overall quality, human graders, ITT (SD)"
s1_qual_overall_ai_itt,0.66,,"S1 overall quality, AI graders, ITT (points)"
s1_qual_overall_ai_itt_sd,0.42,,"S1 overall quality, AI graders, ITT (SD)"
s1_qual_overall_avg_itt,0.45,,"S1 overall quality, human+AI avg, ITT (points)"
s1_qual_overall_avg_itt_sd,0.34,,"S1 overall quality, human+AI avg, ITT (SD)"
s1_qual_overall_avg_pval,0.024,,"S1 overall quality, human+AI avg, ITT p-value"
s1_qual_overall_avg_control_mean,5.50,,"S1 overall quality, human+AI avg, control mean (points)"
s1_qual_overall_avg_pct_increase,8,,"S1 overall quality, human+AI avg, ITT % increase over control mean"
s1_qual_index_avg_itt,0.34,,"S1 quality index, human+AI avg, ITT (points)"
s1_qual_index_avg_itt_sd,0.28,,"S1 quality index, human+AI avg, ITT (SD)"
s1_qual_index_avg_pval,0.058,,"S1 quality index, human+AI avg, ITT p-value"
s1_qual_clarity_itt,0.40,,S1 writing style and clarity ITT (points)
s1_qual_clarity_itt_sd,0.29,,S1 writing style and clarity ITT (SD)
s1_qual_clarity_pval,0.042,,S1 writing style and clarity ITT p-value
s1_qual_relevance_itt,0.17,,S1 relevance to prompt ITT (points)
s1_qual_relevance_itt_sd,0.12,,S1 relevance to prompt ITT (SD)
s1_qual_relevance_pval,0.448,,S1 relevance to prompt ITT p-value
s1_qual_accuracy_itt,0.34,,S1 factual accuracy ITT (points)
s1_qual_accuracy_itt_sd,0.29,,S1 factual accuracy ITT (SD)
s1_qual_accuracy_pval,0.052,,S1 factual accuracy ITT p-value
s1_qual_structure_itt,0.45,,S1 structure and organization ITT (points)
s1_qual_structure_itt_sd,0.31,,S1 structure and organization ITT (SD)
s1_qual_structure_pval,0.036,,S1 structure and organization ITT p-value
s1_qual_evidence_itt,0.23,,S1 evidence and examples ITT (points)
s1_qual_evidence_itt_sd,0.16,,S1 evidence and examples ITT (SD)
s1_qual_evidence_pval,0.271,,S1 evidence and examples ITT p-value
s2_qual_overall_human_itt,0.39,,"S2 overall quality, human graders, ITT (points)"
s2_qual_overall_human_itt_sd,0.26,,"S2 overall quality, human graders, ITT (SD)"
s2_qual_overall_ai_itt,0.23,,"S2 overall quality, AI graders, ITT (points)"
s2_qual_overall_ai_itt_sd,0.15,,"S2 overall quality, AI graders, ITT (SD)"
s2_qual_overall_ai_pval,0.225,,"S2 overall quality, AI graders, ITT p-value"
s2_qual_overall_avg_itt,0.33,,"S2 overall quality, human+AI avg, ITT (points)"
s2_qual_overall_avg_itt_sd,0.24,,"S2 overall quality, human+AI avg, ITT (SD)"
s2_qual_overall_avg_pval,0.083,,"S2 overall quality, human+AI avg, ITT p-value"
s2_qual_overall_avg_control_mean,4.47,,"S2 overall quality, human+AI avg, control mean (points)"
s2_qual_overall_avg_pct_increase,7,,"S2 overall quality, human+AI avg, ITT % increase over control mean"
s2_qual_index_avg_itt,0.30,,"S2 quality index, human+AI avg, ITT (points)"
s2_qual_index_avg_itt_sd,0.24,,"S2 quality index, human+AI avg, ITT (SD)"
s2_qual_index_avg_pval,0.084,,"S2 quality index, human+AI avg, ITT p-value"
s2_qual_accuracy_itt,0.26,,S2 factual accuracy ITT (points)
s2_qual_accuracy_itt_sd,0.22,,S2 factual accuracy ITT (SD)
s2_qual_accuracy_pval,0.124,,S2 factual accuracy ITT p-value
s2_qual_evidence_itt,0.43,,S2 evidence and examples ITT (points)
s2_qual_evidence_itt_sd,0.35,,S2 evidence and examples ITT (SD)
s2_qual_evidence_pval,0.018,,S2 evidence and examples ITT p-value
s2_qual_relevance_itt,0.29,,S2 relevance to prompt ITT (points)
s2_qual_relevance_itt_sd,0.20,,S2 relevance to prompt ITT (SD)
s2_qual_relevance_pval,0.133,,S2 relevance to prompt ITT p-value
s2_qual_structure_itt,0.24,,S2 structure and organization ITT (points)
s2_qual_structure_itt_sd,0.15,,S2 structure and organization ITT (SD)
s2_qual_structure_pval,0.261,,S2 structure and organization ITT p-value
s2_qual_clarity_itt,0.38,,S2 writing style and clarity ITT (points)
s2_qual_clarity_itt_sd,0.27,,S2 writing style and clarity ITT (SD)
s2_qual_clarity_pval,0.029,,S2 writing style and clarity ITT p-value
n_graders,309,,N unique Prolific graders with at least one rating
n_graders_comp,286,,N graders with at least one rating who answered all three comprehension questions correctly
n_graders_comp_pct,93,,% of graders who answered all three comprehension questions correctly
n_graders_comp_fail,23,,N graders with at least one rating who failed the comprehension check (excluded)
n_essays_no_comp_grader,0,,Rated essays with no grader who passed the comprehension check
graders_comp_per_essay_min,2,,Min graders per essay who passed the comprehension check
graders_comp_per_essay_max,7,,Max graders per essay who passed the comprehension check
graders_comp_per_essay_mean,3.5,,Mean graders per essay who passed the comprehension check
s2_quality_evidence_increase,0.43,,S2 evidence dimension increase (points)
s2_quality_style_increase,0.43,,S2 style/clarity dimension increase (points)
s2_quality_org_increase,0.25,,S2 organization dimension increase (points)
mech_time_obj_decrease,0.2,,Objective time decrease (min)
mech_time_obj_decrease_sd,0.04,,Objective time decrease (SD)
mech_time_obj_control_mean,32.6,,Qualtrics learning-phase control mean (min)
mech_time_sr_decrease,0.5,,Self-reported time decrease (min)
mech_time_sr_decrease_sd,0.11,,Self-reported time decrease (SD)
mech_time_sr_control_mean,34.3,,Self-reported learning-phase control mean (min)
mech_writing_log,-0.122,,Writing time ITT (log points)
mech_writing_pct_decrease,12,,Writing time pct decrease (from log)
mech_writing_pval,0.046,,Writing time ITT p-value
mech_research_log,0.079,,Research time ITT (log points)
mech_research_pct_increase,8,,Research time pct increase (from log)
mech_research_pval,0.158,,Research time ITT p-value
mech_research_share_increase,4.4,,Research share ITT (pp increase)
mech_research_share_pval,0.050,,Research share ITT p-value
mech_research_share_control,44.5,,Control group research share (%)
mech_writing_share_decrease,5.3,,Writing share ITT (pp decrease)
mech_writing_share_pval,0.028,,Writing share ITT p-value
mech_writing_share_control,53.4,,Control group writing share (%)
mech_research_share_tot_increase,6.4,,Research share TOT (pp increase)
mech_research_share_tot_pval,0.050,,Research share TOT p-value
mech_writing_share_tot_decrease,7.7,,Writing share TOT (pp decrease)
mech_writing_share_tot_pval,0.026,,Writing share TOT p-value
mech_topic_recall_control,96,,Control group topic recall rate (%)
mech_enjoy_increase,0.58,,Enjoyment increase (0-10 scale)
mech_enjoy_control_mean,5.22,,Control mean enjoyment
mech_enjoy_pct_increase,11,,Enjoyment % increase
mech_enjoy_sd,0.25,,Enjoyment effect in SD
mech_enjoy_pval,0.063,,Enjoyment ITT p-value
mech_enjoy_above_med_pp,12.1,,Above-median enjoyment increase (pp)
mech_enjoy_above_med_control_pct,48,,Control share at or above the sample median of enjoyment (%)
mech_enjoy_above_med_pval,0.082,,Above-median enjoyment ITT p-value
mech_effective_increase,0.06,,Perceived effectiveness increase (points)
mech_effective_pval,0.834,,Perceived effectiveness ITT p-value
mech_efficacy_sd,0.03,,Self-efficacy effect in SD
mech_effective_above_med_pp,12.1,,Above-median effectiveness increase (pp)
mech_effective_above_med_pval,0.078,,Above-median effectiveness ITT p-value
mech_cheat_proctor_pp,5.6,,Proctor caught cheating increase (pp)
mech_cheat_proctor_control,2.9,,Control rate proctor caught (%)
mech_cheat_proctor_pval,0.076,,Proctor caught cheating ITT p-value
mech_cheat_sr_pp,9.6,,Self-reported cheating increase (pp)
mech_cheat_sr_control,1.0,,Control rate self-reported cheating (%)
mech_cheat_sr_pval,0.003,,Self-reported cheating ITT p-value
mech_cheat_combined_pp,13.6,,Combined cheating increase (pp)
mech_cheat_combined_control,3.9,,Control rate any cheating (%)
mech_cheat_combined_pval,0.002,,Combined cheating ITT p-value
mech_draft_time_decrease,1.23,,Drafting time decrease (min)
mech_draft_control_mean,11.07,,Control mean drafting time (min)
mech_draft_pct_decrease,11,,Drafting time % decrease
mech_reading_time_increase,0.71,,Reading time increase (min)
mech_search_time_increase,0.38,,Searching time increase (min)
cosine_within_topic,0.79,,Within-topic cosine similarity
corr_human_ai_s1,0.67,,"Human-AI correlation, overall quality, S1"
corr_human_ai_s1_p,0.000,,"Human-AI correlation p-value, S1"
corr_human_ai_s2,0.69,,"Human-AI correlation, overall quality, S2"
corr_human_ai_s2_p,0.000,,"Human-AI correlation p-value, S2"
belief_own_control_mean,3.78,,Control mean of own-performance belief (raw questions)
belief_own_control_pct,37.8,,Control mean of own-performance belief (% correct)
belief_own_treat_mean,3.72,,Treated mean of own-performance belief (raw questions)
belief_own_treat_pct,37.2,,Treated mean of own-performance belief (% correct)
s2_actual_q_correct_mean,5.12,,"S2 actual questions correct, S1 attendees who attended S2"
s2_actual_pct_correct,51.2,,"S2 actual % correct, S1 attendees who attended S2"
belief_oth_control_mean,5.08,,Control mean of belief about others' performance (raw questions)
belief_oth_control_pct,50.8,,Control mean of belief about others' performance (% correct)
belief_oth_treat_mean,5.09,,Treated mean of belief about others' performance (raw questions)
belief_oth_treat_pct,50.9,,Treated mean of belief about others' performance (% correct)
belief_own_minus_oth_control_pp,-12.7,,Control: mean within-respondent self-minus-others belief gap (pp)
belief_own_minus_oth_treat_pp,-13.6,,Treated: mean within-respondent self-minus-others belief gap (pp)
belief_forbidden_self_pred,2.55,,"AI-forbidden: predicted self gain (questions), belief-table sample"
belief_forbidden_self_pred_pp,25.5,,"AI-forbidden: predicted self gain (pp), control mean"
belief_allowed_self_pred,0.30,,"AI-allowed: predicted self gain (questions), treated raw mean"
belief_allowed_self_pred_pp,3.0,,"AI-allowed: predicted self gain (pp), treated raw mean"
belief_self_itt_pp,-16.8,,"Perceived self gain: double-lasso ITT of AI access (pp), belief table"
belief_forbidden_self_ratio,6.5,,Control-group predicted own AI gain as a multiple of the S2 test ITT (pp over pp)
belief_forbidden_other_pred,1.42,,"AI-forbidden: predicted other gain (questions), belief-table sample"
belief_forbidden_other_pred_pp,14.2,,"AI-forbidden: predicted other gain (pp), control mean"
belief_allowed_other_pred,1.12,,"AI-allowed: predicted other gain (questions), treated raw mean"
belief_allowed_other_pred_pp,11.2,,"AI-allowed: predicted other gain (pp), treated raw mean"
belief_other_itt_pp,-3.8,,"Perceived other gain: double-lasso ITT of AI access (pp), belief table"
s2_test_itt_q,0.39,,"S2 test score ITT (questions, 0-10)"
belief_subgroup_corr,0.35,,Cross-subgroup Pearson corr: perceived vs actual own TE (former Fig 7B)
belief_subgroup_corr_p,0.099,,p-value of the cross-subgroup Pearson correlation (former Fig 7B)
belief_subgroup_spearman,0.06,,Cross-subgroup Spearman corr: perceived vs actual own TE (former Fig 7B)
s1_auto_test_itt,0.29,,"S1 test ITT, automation users (SD)"
s1_auto_qual_overall_itt,1.50,,"S1 overall quality ITT, automation users (SD)"
s1_auto_qual_overall_pval,0.006,,"S1 overall quality ITT p-value, automation users"
s1_auto_qual_index_itt,1.28,,"S1 quality index ITT, automation users (SD)"
s1_aug_test_itt,0.34,,"S1 test ITT, augmentation users (SD)"
s1_aug_test_pval,0.031,,"S1 test ITT p-value, augmentation users"
s1_aug_qual_overall_itt,0.09,,"S1 overall quality ITT, augmentation users (SD)"
s1_aug_qual_index_itt,0.06,,"S1 quality index ITT, augmentation users (SD)"
s2_auto_test_itt,-0.24,,"S2 test ITT, automation users (SD)"
s2_auto_test_pval,0.538,,"S2 test ITT p-value, automation users"
s2_auto_qual_overall_itt,-0.14,,"S2 overall quality ITT, automation users (SD)"
s2_auto_qual_index_itt,-0.09,,"S2 quality index ITT, automation users (SD)"
s2_aug_test_itt,0.24,,"S2 test ITT, augmentation users (SD)"
s2_aug_test_pval,0.095,,"S2 test ITT p-value, augmentation users"
s2_aug_qual_overall_itt,0.23,,"S2 overall quality ITT, augmentation users (SD)"
s2_aug_qual_overall_pval,0.217,,"S2 overall quality ITT p-value, augmentation users"
s2_aug_qual_index_itt,0.14,,"S2 quality index ITT, augmentation users (SD)"
s1_mixed_test_itt,0.58,,"S1 test ITT, mixed users (SD)"
s1_mixed_test_pval,0.009,,"S1 test ITT p-value, mixed users"
s1_mixed_qual_overall_itt,0.89,,"S1 overall quality ITT, mixed users (SD)"
s1_mixed_qual_overall_pval,0.005,,"S1 overall quality ITT p-value, mixed users"
s2_mixed_test_itt,0.28,,"S2 test ITT, mixed users (SD)"
s2_mixed_test_pval,0.220,,"S2 test ITT p-value, mixed users"
s2_mixed_qual_overall_itt,0.25,,"S2 overall quality ITT, mixed users (SD)"
s2_mixed_qual_overall_pval,0.320,,"S2 overall quality ITT p-value, mixed users"
s1_auto_aug_test_diff_pval,0.854,,"S1 test ITT, p-value for automation = augmentation (suest)"
s2_auto_aug_test_diff_pval,0.189,,"S2 test ITT, p-value for automation = augmentation (suest)"
fade_auto_test_p,0.058,,"p-value, S2 minus S1 test-score ITT within automation users (suest)"
fade_aug_test_p,0.584,,"p-value, S2 minus S1 test-score ITT within augmentation users (suest)"
fade_auto_overall_fall,1.64,,"Fall in the overall-quality ITT from S1 to S2 within automation users (SD, suest)"
fade_auto_overall_p,0.023,,p-value of the S2 minus S1 overall-quality change within automation users (suest)
fade_aug_overall_p,0.458,,p-value of the S2 minus S1 overall-quality change within augmentation users (suest)
fade_dd_overall_gap,1.78,,"Absolute double difference, (automation minus augmentation) x (S2 minus S1), overall quality (SD, suest)"
fade_dd_overall_p,0.014,,"p-value of the double difference, overall quality (suest)"
fade_dd_index_gap,1.45,,"Absolute double difference, (automation minus augmentation) x (S2 minus S1), index quality (SD, suest)"
fade_dd_index_p,0.056,,"p-value of the double difference, index quality (suest)"
s1_s2_test_diff_pval,0.321,,"p-value: S2 minus S1 test-score ITT in pp (suest, student clusters)"
s1_ai_detect_auto_mean,85.7,,"S1 mean AI-detected fraction, automation users (%)"
s1_ai_detect_aug_mean,21.3,,"S1 mean AI-detected fraction, augmentation users (%)"
s1_ai_detect_mixed_mean,57.3,,"S1 mean AI-detected fraction, mixed users (%)"
pct_aug_users,61,,% of treated AI users classified as pure augmentation
pct_auto_users,9,,% of treated AI users classified as pure automation
pct_mixed_users,27,,% of treated AI users classified as mixed
pct_other_users,3,,% of treated AI users classified as other
s1_ai_detect_auto_itt_pp,80.3,,"S1 AI-detected ITT, automation subsample (pp)"
s1_ai_detect_aug_itt_pp,14.4,,"S1 AI-detected ITT, augmentation subsample (pp)"
s1_ai_detect_auto_tot_pp,80.3,,"S1 AI-detected TOT, automation subsample (pp)"
s1_ai_detect_aug_tot_pp,14.4,,"S1 AI-detected TOT, augmentation subsample (pp)"
s1_ai_detect_diff_pval,0.000,,"S1 AI-detected auto vs aug difference p-value (treated only, strata FE, cluster id)"
s1_lowgpa_test_itt,0.38,,"S1 test ITT, low-GPA students (SD)"
s1_freq_ai_test_itt,0.53,,"S1 test ITT, frequent AI users (SD)"
s1_freq_ai_test_diff,0.52,,"S1 test interaction (frequent vs infrequent AI users, SD)"
s1_freq_ai_test_diff_pval,0.046,,S1 test interaction p-value (frequent vs infrequent AI users)
s2_freq_ai_test_itt,0.33,,"S2 test ITT, frequent AI users (SD)"
s2_freq_ai_test_diff,0.23,,"S2 test interaction (frequent vs infrequent AI users, SD)"
s2_freq_ai_test_diff_pval,0.328,,S2 test interaction p-value (frequent vs infrequent AI users)
s1_high_prof_qual_overall_itt,0.43,,"S1 overall quality ITT, high-proficiency students (SD)"
s1_high_prof_qual_overall_diff,0.26,,"S1 overall quality interaction (high vs low AI proficiency, SD)"
s1_high_prof_qual_overall_diff_pval,0.412,,S1 overall quality interaction p-value (high vs low AI proficiency)
s1_high_prof_qual_index_itt,0.39,,"S1 quality index ITT, high-proficiency students (SD)"
s1_high_prof_qual_index_diff,0.28,,"S1 quality index interaction (high vs low AI proficiency, SD)"
s1_high_prof_qual_index_diff_pval,0.375,,S1 quality index interaction p-value (high vs low AI proficiency)
s1_female_test_itt,0.45,,"S1 test ITT, female students (SD)"
s1_male_test_itt,0.01,,"S1 test ITT, male students (absolute SD; sign rendered in TeX)"
s1_female_test_diff,0.44,,"S1 test interaction (female vs male, SD)"
s1_female_test_diff_pval,0.087,,S1 test interaction p-value (female vs male)
s1_test_itt_gpa_q1_sd,0.05,,S1 test diff-in-means by GPA quartile 1 (SD)
s1_test_itt_sat_q1_sd,0.06,,S1 test diff-in-means by SAT quartile 1 (SD)
s1_test_itt_gpa_q2_sd,0.29,,S1 test diff-in-means by GPA quartile 2 (SD)
s1_test_itt_sat_q2_sd,0.15,,S1 test diff-in-means by SAT quartile 2 (SD)
s1_test_itt_gpa_q3_sd,0.19,,S1 test diff-in-means by GPA quartile 3 (SD)
s1_test_itt_sat_q3_sd,0.40,,S1 test diff-in-means by SAT quartile 3 (SD)
s1_test_itt_gpa_q4_sd,0.35,,S1 test diff-in-means by GPA quartile 4 (SD)
s1_test_itt_sat_q4_sd,0.07,,S1 test diff-in-means by SAT quartile 4 (SD)
n_experiment_base,256,,N in experiment_base (intake completers who signed up for S1)
n_sat_observed,189,,N with observed SAT score
n_sat_imputation,189,,N in the SAT imputation regression (2-merge-sessions.do)
pct_sat_missing,26,,% with missing SAT score (full sample)
between_lookup_pval,0.192,,p-value: treatment effect on looked-up info between sessions
between_study_n,0,,N students reporting they studied the topic between sessions
n_prompts_classified,435,,Total ChatGPT prompts sent to the LLM classifier
n_prompts_refused,2,,ChatGPT prompts the LLM declined to classify
prompt_pct_explaining_concepts,49.0,,% of prompts: Explaining concepts
prompt_pct_summarizing_readings,5.5,,% of prompts: Summarizing readings
prompt_pct_drafting_responses,10.3,,% of prompts: Drafting responses
prompt_pct_editing_responses,12.6,,% of prompts: Editing responses
prompt_pct_proofreading,0.5,,% of prompts: Proofreading
prompt_pct_other,21.6,,% of prompts: Other
n_convs_classified,85,,Total ChatGPT conversations (all classified by LLM)
conv_pct_augmentation,49,,% of conversations: Augmentation
conv_pct_automation,26,,% of conversations: Automation
conv_pct_mixed,18,,% of conversations: Mixed
conv_pct_other,7,,% of conversations: Other
val_agree_pct,78,,"Human coder vs LLM: % agreement, four categories"
val_kappa,0.65,,"Human coder vs LLM: Cohen kappa, four categories"
val_auto_agree_pct,87,,"Human coder vs LLM: % agreement, automation incl. Mixed"
val_auto_kappa,0.73,,"Human coder vs LLM: Cohen kappa, automation incl. Mixed"
val_auto_share_llm,44,,"% of conversations with automation (incl. Mixed), LLM"
val_auto_share_human,33,,"% of conversations with automation (incl. Mixed), human coder"
val_aug_agree_pct,82,,"Human coder vs LLM: % agreement, augmentation incl. Mixed"
val_aug_kappa,0.55,,"Human coder vs LLM: Cohen kappa, augmentation incl. Mixed"
val_aug_share_llm,67,,"% of conversations with augmentation (incl. Mixed), LLM"
val_aug_share_human,82,,"% of conversations with augmentation (incl. Mixed), human coder"
val_llm_auto_n,22,,Conversations the LLM labels Automation
val_llm_auto_h_auto,8,,LLM Automation conversations the human coder labels Automation
val_llm_auto_h_mixed,7,,LLM Automation conversations the human coder labels Mixed
val_llm_auto_h_aug,7,,LLM Automation conversations the human coder labels Augmentation
val_other_agree_pct,100,,"Human coder vs LLM: % agreement, Other (off-topic)"
val_other_kappa,1.00,,"Human coder vs LLM: Cohen kappa, Other (off-topic)"
val_other_share_llm,7,,"% of conversations labeled Other, LLM"
val_other_share_human,7,,"% of conversations labeled Other, human coder"
val_auto_llm_n,37,,Conversations the LLM flags as auto (three-way split)
val_auto_both_n,27,,Conversations both coders flag as auto (three-way split)
val_aug_llm_n,57,,Conversations the LLM flags as aug (three-way split)
val_aug_both_n,56,,Conversations both coders flag as aug (three-way split)
val_other_llm_n,6,,Conversations the LLM flags as other (three-way split)
val_other_both_n,6,,Conversations both coders flag as other (three-way split)
ri_within_sim_s1_p,0.730,,"Randomization-inference p-value, within-group similarity, S1 (strata FE, 2000 permutations)"
ri_within_sim_s2_p,0.995,,"Randomization-inference p-value, within-group similarity, S2 (strata FE, 2000 permutations)"
ri_n_perm,2000,,Permutations in the randomization inference for within-group similarity
cosine_within_topic_all,0.76,,"Mean pairwise cosine sim., same topic (any prompt)"
cosine_across_topic,0.16,,"Mean pairwise cosine sim., different topic"
cosine_diff_prompt,0.72,,"Mean pairwise cosine sim., same topic + different prompt"
cosine_same_prompt,0.79,,"Mean pairwise cosine sim., same topic + same prompt"
vader_takeup_b,0.041,,"VADER: sentiment to experimental take-up, coef"
vader_takeup_se,0.090,,"VADER: sentiment to experimental take-up, SE"
vader_takeup_p,0.647,,"VADER: sentiment to experimental take-up, p-value"
vader_semester_b,0.107,,"VADER: sentiment to semester AI use, coef"
vader_semester_se,0.064,,"VADER: sentiment to semester AI use, SE"
vader_semester_p,0.096,,"VADER: sentiment to semester AI use, p-value"
sat_q1_mean,1185,,"Mean SAT, quartile 1 (S1 attendees, reported SAT)"
sat_q2_mean,1380,,"Mean SAT, quartile 2 (S1 attendees, reported SAT)"
sat_q3_mean,1464,,"Mean SAT, quartile 3 (S1 attendees, reported SAT)"
sat_q4_mean,1530,,"Mean SAT, quartile 4 (S1 attendees, reported SAT)"
auto_aug_research_pct,32,,Auto vs aug-only: % difference in self-reported time_research relative to the augmentation-only mean (treated S1)
auto_aug_research_p,0.010,,Auto vs aug: p-value for time_research difference
auto_aug_writenonedit_pct,40,,Auto vs aug-only: % difference in self-reported time_writenonedit relative to the augmentation-only mean (treated S1)
auto_aug_writenonedit_p,0.143,,Auto vs aug: p-value for time_writenonedit difference
auto_aug_edit_pct,208,,Auto vs aug-only: % difference in self-reported time_editing relative to the augmentation-only mean (treated S1)
auto_aug_edit_p,0.053,,Auto vs aug: p-value for time_editing difference
auto_aug_total_pct,15,,Auto vs aug-only: % difference in self-reported time_learning_self relative to the augmentation-only mean (treated S1)
auto_aug_total_p,0.080,,Auto vs aug: p-value for time_learning_self difference
mixed_aug_research_pct,5,,Mixed vs aug-only: signed % difference in self-reported time_research relative to the augmentation-only mean (treated S1)
mixed_aug_research_p,0.578,,Mixed vs aug: p-value for time_research difference
mixed_aug_total_pct,-5,,Mixed vs aug-only: signed % difference in self-reported time_learning_self relative to the augmentation-only mean (treated S1)
mixed_aug_total_p,0.190,,Mixed vs aug: p-value for time_learning_self difference
belief_aug_auto_pp_diff,31.3,,"Belief gap: augmentation minus automation, perceived own AI gain (pp)"
belief_aug_auto_p,0.008,,Belief gap aug vs auto: p-value
cheat_score_gap_pp,17.8,,"Test-score gap, cheaters vs non-cheaters (pp)"
cheat_implied_pp,2.4,,Implied cheating-driven test-score ITT (pp)
cheat_share_of_effect,33,,Cheating share of S1 test ITT (percent)
cheat_score_gap_proctor_pp,22.1,,"Test-score gap, proctor-observed violators vs others (pp)"
cheat_implied_proctor_pp,1.2,,"Implied violation-driven test-score ITT, proctor measure (pp)"
cheat_share_proctor,17,,"Violation share of S1 test ITT, proctor measure (percent)"
cheat_score_gap_sr_pp,16.4,,"Test-score gap, self-reported violators vs others (pp)"
cheat_implied_sr_pp,1.6,,"Implied violation-driven test-score ITT, self-report measure (pp)"
cheat_share_sr,22,,"Violation share of S1 test ITT, self-report measure (percent)"
time_at_cap_pct,64,,Share of S1 attendees at the 35-minute cap (percent)
time_at_cap_pct_control,70,,"Share at the 35-minute cap, AI-Forbidden (percent)"
time_at_cap_pct_treated,58,,"Share at the 35-minute cap, AI-Allowed (percent)"
narr_nonuser_n,30,,Treated non-users with codable narratives (N)
narr_user_aug_pct,24,,Augmentation framing among treated users (%)
narr_ctrl_aug_pct,17,,Augmentation framing among control students (%)
narr_both_pct,69,,% naming both helping and harming channel
narr_explain_pct,40,,% of coded narratives invoking explanation/tutoring channel
narr_shortcut_pct,43,,% of coded narratives invoking shortcutting/cheating channel
narr_helps_concede_pct,20,,% of net-positive narratives naming a harming channel
narr_harms_concede_pct,27,,% of net-negative narratives naming a promoting channel
narr_n_coded,201,,Open-ended responses codable as causal graphs
narr_usemode_pct,55,,% of coded narratives with the mode-of-use node
narr_mean_factors,3.1,,Mean mechanisms per coded narrative (excl. mode of use)
narr_mean_links,6.5,,Mean links per coded narrative
narr_cond_pct,69,,% coded narratives framing AI effect as conditional
narr_multi_pct,94,,% coded narratives naming more than one mechanism
narr_cl_top4_pct,94,,% of coded narratives in the four displayed clusters
narr_cl_double_pct,47,,% of coded narratives in double-edged cluster
narr_cl_aug_pct,28,,% of coded narratives in pure-augmentation cluster
narr_cl_auto_pct,10,,% of coded narratives in pure-automation cluster
narr_cl_deskill_pct,8,,% of coded narratives in deskilling cluster
narr_irr_factor,0.88,,"Two-coder agreement on mechanisms (pooled Dice, coder1 vs GPT-5.5)"
narr_irr_group,0.97,,"Two-coder agreement on any promoting channel (coder1 vs GPT-5.5, n=202 shared responses)"
narr_irr_group_harm,0.97,,"Two-coder agreement on any harming channel (coder1 vs GPT-5.5, n=202 shared responses)"
narr_irr_link,0.73,,Two-coder agreement on signed links
narr_irr_factor_rand,0.33,,Random-coding benchmark for mechanism agreement
narr_irr_link_rand,0.22,,Random-coding benchmark for link agreement
narr_irr_valence_pct,94,,% two-coder agreement on overall valence
narr_irr_framing_pct,85,,% two-coder agreement on aug-vs-auto framing
loo_s1_n_min,173,,"Smallest leave-one-slot-out regression sample, Session One"
loo_s1_n_max,190,,"Largest leave-one-slot-out regression sample, Session One"
loo_s2_n_min,171,,"Smallest leave-one-slot-out regression sample, Session Two"
loo_s2_n_max,185,,"Largest leave-one-slot-out regression sample, Session Two"
narr_mv_aug_prof_pp,16.3,,"High AI proficiency on augmentation framing (pp, conditional)"
narr_mv_aug_prof_pval,0.024,,p-value: high proficiency on augmentation
narr_mv_auto_prof_pp,12.4,,"High AI proficiency on automation framing (pp magnitude, conditional)"
narr_mv_auto_prof_pval,0.054,,p-value: high proficiency on automation
narr_mv_aug_freq_pp,1.8,,"Frequent AI user on augmentation framing (pp, conditional)"
narr_mv_aug_freq_pval,0.807,,p-value: frequent user on augmentation
narr_mv_auto_treat_pp,9.8,,"Randomized AI access on automation framing (pp magnitude, conditional)"
narr_mv_auto_treat_pval,0.083,,p-value: treatment on automation
narr_mv_multi_treat_pp,7.3,,"Randomized AI access on the multi-causal indicator (pp magnitude, conditional)"
narr_mv_multi_treat_pval,0.092,,p-value: treatment on the multi-causal indicator
narr_mv_male_factors,0.41,,"Male on number of mechanisms (magnitude, conditional)"
narr_mv_male_factors_pval,0.042,,p-value: male on mechanisms
narr_mv_gpa_factors,0.45,,High GPA on number of mechanisms (conditional)
narr_mv_gpa_factors_pval,0.025,,p-value: high GPA on mechanisms
narr_mv_male_links,0.78,,"Male on number of links (magnitude, conditional)"
narr_mv_male_links_pval,0.055,,p-value: male on links
narr_mv_gpa_multi_pp,6.9,,"High GPA on multi-causal indicator (pp, conditional)"
narr_mv_gpa_multi_pval,0.141,,p-value: high GPA on multi-causal
narr_mv_natsci_cond_pp,30.6,,"Natural-science major on conditional framing (pp, conditional)"
narr_mv_natsci_cond_pval,0.001,,p-value: natsci on conditional
narr_mv_natsci_auto_pp,21.4,,"Natural-science major on automation framing (pp magnitude, conditional)"
narr_mv_natsci_auto_pval,0.002,,p-value: natsci on automation
narr_mv_perc_auto_pp,16.4,,"High perceived AI effect on automation framing (pp magnitude, conditional)"
narr_mv_perc_auto_pval,0.006,,p-value: high perceived AI effect on automation
narr_mv_perc_cond_pp,17.2,,"High perceived AI effect on conditional framing (pp, conditional)"
narr_mv_perc_cond_pval,0.044,,p-value: high perceived AI effect on conditional
bench_session_gain_sd,1.02,,Control S1 test gain over baseline (control S1 SD)
bench_gpa_s1_sd,0.22,,"S1 test SD per 1 SD of GPA, controls (bivariate)"
bench_gpa_s2_sd,0.25,,"S2 test SD per 1 SD of GPA, controls (bivariate)"
s1_test_chance_sd,0.05,,S1 test ITT magnitude (SD) with violators at chance - sign is negative and the tex adds the minus
s1_test_chance_pval,0.688,,p-value: S1 test ITT with violators at chance
s1_test_chance_both_sd,0.22,,"S1 test ITT (SD, signed) with students flagged by both a proctor and a self-report at chance"
s1_test_chance_both_pval,0.070,,p-value: S1 test ITT with students flagged by both at chance
s1_test_violdum_proctor_sd,0.25,,"S1 test ITT (SD) with a proctor-observed violation indicator added, paper controls"
s1_test_violdum_proctor_pval,0.046,,p-value: S1 test ITT with a proctor-observed violation indicator added
s1_test_violdum_sr_sd,0.21,,"S1 test ITT (SD) with a self-reported violation indicator added, paper controls"
s1_test_violdum_sr_pval,0.094,,p-value: S1 test ITT with a self-reported violation indicator added
s1_test_violdum_both_sd,0.25,,"S1 test ITT (SD) with an indicator for students flagged by both added, paper controls"
s1_test_violdum_both_pval,0.042,,p-value: S1 test ITT with an indicator for students flagged by both added
s1_test_violdum_any_sd,0.18,,"S1 test ITT (SD) with an indicator for students flagged by either source added, paper controls"
s1_test_violdum_any_pval,0.145,,p-value: S1 test ITT with an indicator for students flagged by either source added
s1_test_noviol_proctor_sd,0.24,,S1 test ITT (SD) excluding proctor-observed violators
s1_test_noviol_proctor_pval,0.056,,p-value: S1 test ITT excluding proctor-observed violators
s1_noviol_proctor_n,198,,N in the S1 sample excluding proctor-observed violators
s1_test_noviol_proctor_eq_pval,0.229,,p-value: full-sample S1 test ITT = ITT excluding proctor-observed violators (suest)
s1_test_noviol_sr_sd,0.15,,S1 test ITT (SD) excluding self-reported violators
s1_test_noviol_sr_pval,0.245,,p-value: S1 test ITT excluding self-reported violators
s1_noviol_sr_n,198,,N in the S1 sample excluding self-reported violators
s1_test_noviol_sr_eq_pval,0.021,,p-value: full-sample S1 test ITT = ITT excluding self-reported violators (suest)
s1_test_noviol_both_sd,0.27,,S1 test ITT (SD) excluding students flagged by both a proctor and a self-report
s1_test_noviol_both_pval,0.029,,p-value: S1 test ITT excluding students flagged by both
s1_noviol_both_n,208,,N in the S1 sample excluding students flagged by both
s1_test_noviol_both_eq_pval,0.342,,p-value: full-sample S1 test ITT = ITT excluding students flagged by both (suest)
s1_test_noviol_sd,0.16,,S1 test ITT (SD) excluding violators flagged by either source (union; kept for the record)
s1_test_noviol_pval,0.217,,p-value: S1 test ITT excluding violators flagged by either source (union)
s1_noviol_n,188,,N in the S1 sample excluding violators flagged by either source (union)
s1_test_noviol_eq_pval,0.015,,"p-value: full-sample S1 test ITT = ITT excluding violators flagged by either source, union (suest)"
s2_test_noviol_sd,0.13,,S2 test ITT (SD) excluding S1 violators
s2_test_noviol_pval,0.369,,p-value: S2 test ITT excluding S1 violators
lit_grand_mean_cl_sd,0.19,,Lit grand mean (DL across studies - arms aggregated within study - ours excluded)
lit_grand_mean_cl_lo_sd,0.08,,Lit grand mean clustered: 95% CI lower
lit_grand_mean_cl_hi_sd,0.29,,Lit grand mean clustered: 95% CI upper
lit_n_studies_cl,14,,Lit grand mean clustered: number of studies
lit_ws_bastani_diff,0.18,,Within-study contrast bastani: difference in unaided learning (SD)
lit_ws_bastani_lo,0.01,,Within-study contrast bastani: 95% CI lower (SD)
lit_ws_bastani_hi,0.35,,Within-study contrast bastani: 95% CI upper (SD)
lit_ws_gallegos_diff,0.22,,Within-study contrast gallegos: difference in unaided learning (SD)
lit_ws_gallegos_lo,0.02,,Within-study contrast gallegos: 95% CI lower (SD)
lit_ws_gallegos_hi,0.41,,Within-study contrast gallegos: 95% CI upper (SD)
lit_ws_hou_diff,0.78,,Within-study contrast hou: difference in unaided learning (SD)
lit_ws_hou_lo,0.25,,Within-study contrast hou: 95% CI lower (SD)
lit_ws_hou_hi,1.29,,Within-study contrast hou: 95% CI upper (SD)
lit_ws_bassner_diff,-0.06,,Within-study contrast bassner: difference in unaided learning (SD)
lit_ws_bassner_lo,-0.26,,Within-study contrast bassner: 95% CI lower (SD)
lit_ws_bassner_hi,0.14,,Within-study contrast bassner: 95% CI upper (SD)
lit_ws_kreijkes_ret_diff,0.13,,Within-study contrast kreijkes_ret: difference in unaided learning (SD)
lit_ws_kreijkes_ret_lo,0.01,,Within-study contrast kreijkes_ret: 95% CI lower (SD)
lit_ws_kreijkes_ret_hi,0.26,,Within-study contrast kreijkes_ret: 95% CI upper (SD)
lit_ws_kreijkes_comp_diff,0.14,,Within-study contrast kreijkes_comp: difference in unaided learning (SD)
lit_ws_kreijkes_comp_lo,0.00,,Within-study contrast kreijkes_comp: 95% CI lower (SD)
lit_ws_kreijkes_comp_hi,0.28,,Within-study contrast kreijkes_comp: 95% CI upper (SD)
lit_ws_liu_diff,0.29,,Within-study contrast liu: difference in unaided learning (SD)
lit_ws_liu_lo,0.03,,Within-study contrast liu: 95% CI lower (SD)
lit_ws_liu_hi,0.55,,Within-study contrast liu: 95% CI upper (SD)
\end{filecontents*}

	
	\title{Experimental Evidence on the Learning Impact of Generative AI\thanks{\noindent Contractor: Middlebury College (\href{mailto:zcontractor@middlebury.edu}{zcontractor@middlebury.edu}); Reyes (corresponding author): Middlebury College and IZA (\href{mailto:greyes@middlebury.edu}{greyes@middlebury.edu}). Authors are listed in alphabetical order. For helpful discussions and comments, we thank Chris Campos, Jeff Carpenter, Amy Collier, Sam Hirshman, David Huffman, Guy Ishai, Cory Koedel, Seunghoon Lee, Peter Matthews, David Munro, Caitlin Myers, Ted O'Donoghue, Evan Riehl, Andrea Robbett, Nick Swanson, Christiane Szerman, and participants at the Advances with Field Experiments Conference, the Liberal Arts Colleges Behavioral and Experimental Economics Workshop (LACBEE), the NBER Economics of Education Spring Meeting, the NBER Digital Economics and Artificial Intelligence Summer Institute, the New England Experimental Economics Workshop, the Northeastern Economics of Education Workshop, the Royal Economic Society Conference, the Society of Labor Economists Annual Meeting, the Valle Nevado Economics of Education Workshop, and seminars at Cornell University, Middlebury College (Carol Rifelj Faculty Lecture), Ohio University, Universidad Nacional de La Plata, the University of Missouri, and the University of Texas Rio Grande Valley. Financial support from Middlebury College's Office of the Provost is gratefully acknowledged. We thank Kimberly Barros, Rebecca Deranian, Brooke Dolan, Alice Gindin, Sarah Hayward, Ember Pikramenos, Kevin Ramirez, Ephraim Shinko, and Juan Marcelo Verdugo for their assistance with experiment proctoring. The study was reviewed and approved by the Institutional Review Board of Middlebury College (protocol 447). This RCT was registered as \href{https://www.socialscienceregistry.org/trials/15806}{AEARCTR-0015806} \citep{contractor.reyes2026}. Survey instruments and additional materials are available in the \href{https://www.dropbox.com/scl/fi/pzzmqwikxweqzrofrnzt9/ai-learning-supplementary.pdf?rlkey=zjg1lf7cmzno304na3bgq58bn&st=gtj3nlqa&dl=0}{Supplementary Materials}.}}
	
	\author{Zara Contractor \and Germ\'an Reyes} 
	
	\renewcommand{\today}{\ifcase \month \or January\or February\or March\or %
		April\or May \or June\or July\or August\or September\or October\or November\or %
		December\fi \ \number \year} 
	\date{\today \vspace{-1cm}}
	
	\maketitle	
	
	\begin{abstract}
		\begin{singlespace}
			\noindent We study whether access to off-the-shelf generative AI changes how much students learn in a timed assignment. In a proctored, in-person randomized experiment, undergraduates study an unfamiliar topic and write an essay with or without AI, then complete unaided assessments immediately and one week later. AI access raises test scores, and a smaller gain persists a week later. Gains remain for students who use AI as a tutor (``augmentation'') and fade for students who have AI write for them (``automation''). We document two mechanisms: students shift time from writing toward reading and searching for information, and report greater learning enjoyment.
		\end{singlespace}
		
	\end{abstract}
	
	\clearpage
	\section{Introduction}
	
	Generative artificial intelligence (AI) is now widely used in higher education. ChatGPT has over 900 million weekly users \citep{openai2026scaling}, college-age users send nearly half of its messages, and more than a quarter of those messages relate to learning \citep{chatterji.etal2025, openai2025}. Despite this widespread use, evidence on whether the off-the-shelf chatbots students use help them learn is scarce.
	
	Generative AI can change how much students learn through two margins. The first is time allocation: students may spend less time on academic tasks when AI can automate some of them \citep{kazemitabaar.etal2023, kestin.etal2025, kreijkes.etal2026}, and whether that helps or harms learning depends on what students do with the time they save. The second margin is how much students learn from a fixed amount of study time (henceforth, ``learning productivity''). Holding time fixed, AI may raise learning by, for example, acting as an on-demand tutor that explains difficult concepts and provides personalized feedback \citep{nickow.etal2024}, or lower it by offloading the cognitive effort that builds durable knowledge \citep{chi.wylie2014, risko.gilbert2016}. 
	
	In this paper, we focus on the second margin and study how AI affects learning productivity in a common class of academic tasks: time-bounded assignments in which students must learn unfamiliar material, identify relevant information, and demonstrate understanding through writing and tests. Our randomized experiment with \getval{n_students} undergraduates unfolds over two in-person sessions approximately one week apart. In the first session, all students have the same amount of time to learn about a technical topic and write an analytical essay, and then take a knowledge test. We randomly assign students to an \emph{AI-allowed} or an \emph{AI-forbidden} condition. The AI-allowed group may use any generative AI tool along with traditional resources, while the AI-forbidden group may use only traditional resources. We monitor AI usage and compliance through direct proctor observation, ChatGPT conversation logs, and self-reports. In the second session, all students complete another test and write an essay on the same topic, this time without AI or any other resources.
	
	We measure learning in two ways. Knowledge tests target factual and conceptual understanding: we ask students to recall key information and identify core mechanisms. The open-ended essays measure understanding of the material beyond memorization: we ask students to analyze relationships between concepts and to support their arguments with specific evidence. Graders give each essay an overall quality rating and scores on five dimensions: accuracy of content, use of evidence and examples, relevance to the prompt, organization and structure, and writing style and clarity. Students complete all tests and the Session Two essay without AI or other resources, so these assessments measure what they learned. The Session One essay, which AI can help write, measures the work they submit. We also measure characteristics of the essay text: the fraction flagged as AI-generated by a state-of-the-art detector, length, readability, lexical diversity, and textual similarity among student essays.
	
	We first document that our random assignment generated substantial variation in AI usage. About \getval{ai_takeup_pct} percent of students in the AI-allowed condition used generative AI during the first session, compared with near-zero usage in the control group. Users value this access: \getval{ai_helpful_users_pct} percent report that AI is helpful for the learning task. ChatGPT conversation logs reveal that students use AI most often to explain concepts (\getval{pct_revealed_explain} percent of treated students). Drafting responses (\getval{pct_revealed_draft} percent), editing responses (\getval{pct_revealed_edit} percent), and summarizing readings (\getval{pct_revealed_summarize} percent) are the next most common substantive uses.
	
	We turn next to our main result: AI access improves learning outcomes, and a smaller gain remains after the tool is removed. Students with AI access score \getval{s1_test_itt_pp} percentage points (pp) higher on the immediate knowledge test (relative to a control mean of \getval{s1_test_control_mean} percent). Standardized by the control-group standard deviation (SD), this is equivalent to a \getval{s1_test_itt_sd} SD gain in test scores. One week later, AI-allowed students score \getval{s2_test_itt_pp} pp higher on the Session Two test (relative to a control mean of \getval{s2_test_control_mean} percent), or \getval{s2_test_itt_sd} SD. 
	
	The essays provide further evidence that AI access improves learning. In Session One, the share of treated students' text flagged as AI-generated rises by \getval{s1_ai_detect_pp} pp, consistent with some treated students delegating the writing to AI. Graders assess these essays as \getval{s1_qual_overall_avg_itt_sd} SD higher in overall quality, yet the essays are no more homogeneous than control students' essays. This gain could reflect AI-generated text, learning, or both. We separate the two with the Session Two essays, which students write one week later without AI. In those essays, the AI-detection effect disappears. Consistent with the test-score results, overall quality is \getval{s2_qual_overall_avg_itt_sd} SD higher, and the point estimates for all five dimensions are positive, with the largest effects on use of evidence and examples and on writing style and clarity.
	
	To understand why AI access improves learning, we examine time use and the learning experience. Although both groups spend the same amount of time learning on average, AI access shifts the composition of learning time away from producing text: treated students report spending \getval{mech_writing_share_decrease} pp less of their learning time on writing activities and \getval{mech_research_share_increase} pp more on reading and searching for information, relative to control-group shares of \getval{mech_writing_share_control} and \getval{mech_research_share_control} percent. AI access also makes learning more enjoyable: treated students report \getval{mech_enjoy_pct_increase} percent higher enjoyment ratings.
	
	To interpret these results, we write down a model in which a student divides effort between research (reading and searching) and writing. Both activities produce knowledge, and writing also produces the essay, whose quality depends on the student's writing and on what the student knows. We model AI as giving the student two capabilities. Used as a tutor, it raises the return to research effort; used to write the essay, it lets the student delegate, so the student's own work no longer raises the essay's quality. With that incentive gone, the student exerts less effort on both activities and, consequently, learns less. The model rationalizes the test-score gain, the shift in learning time, and the higher enjoyment. The test-score gain implies that, averaged across treated students, the tutoring gain exceeded the losses from delegation. Both capabilities lower the share of effort spent on writing, which produces the shift from writing toward research. Both also raise the utility the student obtains from the task, consistent with the higher enjoyment.
	
	The model also predicts that outcomes should differ with how students use AI. To test this implication, we use treated students' ChatGPT conversation logs and classify \getval{pct_aug_users} percent of AI users as ``augmentation'' users (who work \textit{with} AI), \getval{pct_auto_users} percent as ``automation'' users (who use AI to do the work \textit{for} them), and \getval{pct_mixed_users} percent as ``mixed'' users (who do both). This distinction shows up in both usage and output: augmentation users prompt AI mainly to explain concepts, automation users prompt it to draft their essays for them, mixed users do both, and AI-detection rates rise from augmentation to mixed to automation users. Automation and mixed users submit higher-rated essays in Session One, but automation users' advantage is absent from their unaided essays one week later. Augmentation and mixed users show positive test-score differences relative to the control group in both sessions, and both groups' point estimates exceed automation users' on every Session Two outcome. These patterns are consistent with the model's prediction that what students learn from AI depends on how they use it.
	
	Students choose whether and how to use AI, and those choices may depend on what they believe AI does to their learning. Are those beliefs accurate? At the end of the experiment, we ask students to estimate how much AI access changed (or would have changed) their test scores. Both treated and control students expect positive effects, but only the treated group's prediction comes close to our estimate; the control group overestimates it. To understand why students expect these effects, we elicit an open-ended account of how AI affects student learning in general and code each response as a causal graph from AI use to learning. Most students name both a channel through which AI helps learning and one through which it harms it, say the effect depends on how AI is used, and organize the channels around the augmentation-versus-automation distinction along which our use-type results differ. 
	
	We contribute to an emerging literature on the learning impacts of AI \citep[e.g.,][]{bastani.etal2025, franco.etal2026, stromberg.etal2026}. We benchmark our estimates against the experimental literature in Section~\ref{sub:tests}, but note three challenges for interpreting existing evidence. First, many studies evaluate AI systems that are scaffolded, restricted, or otherwise customized beyond a standard off-the-shelf chatbot.\footnote{For example, \citet{bastani.etal2025} test GPT-4 with guardrails that provide teacher-designed hints rather than direct solutions in Turkish high school math classes; \citet{xu.etal2025} add metacognitive prompts to a generative AI learning environment for Chinese college students; \citet{kim.etal2025} restrict AI to tutoring after students attempt each problem on a math practice platform; \citet{desimone.etal2025} study teacher-guided sessions with Microsoft Copilot in after-school English classes in Nigeria; \citet{poulidis.etal2025} compare a system-regulated AI that auto-delivers tips at key moments with a version that students could also query on demand in a 12-week chess training program; and \citet{kumar.etal2025} expose online adults practicing math problems to explanations pre-generated by GPT-4 under a hidden tutoring prompt, rather than to the chatbot itself.} These designs inform the particular system studied but speak less directly to the chatbots students use in practice. Second, some studies compare AI with active alternatives rather than a business-as-usual counterfactual, so the estimates reflect AI's value relative to those alternatives rather than its absolute effect on learning.\footnote{For example, \citet{kestin.etal2025} compare AI with in-class active learning in a college physics course, \citet{learnlm.team2025} with static pre-written hints and human tutoring on a secondary school mathematics platform, \citet{kreijkes.etal2026} with note-taking on reading-comprehension tasks in secondary schools, and \citet{chung.etal2026} compare an AI tutor that adapts the order and difficulty of practice problems to each student with a fixed-sequence version of the same tutor in a high school Python course.} Third, in many designs students choose how much time to spend on the task \citep[e.g.,][]{kazemitabaar.etal2023, kestin.etal2025, kreijkes.etal2026}, so the estimated effects combine the two margins: time allocation and learning productivity. Our design addresses all three. Beyond these design features, we contribute by measuring what students retain one week later, after AI is removed; distinguishing augmentation from automation uses; tracing how AI shifts time use and the learning experience; assessing learning with essays as well as test scores; and comparing students' beliefs about AI's effects with those we estimate.
	
	We also contribute to the literature on digital technologies and student learning. Prior research has examined laptops and personal computers \citep{malamud2011home, carter2017laptops, cristia2017technology}; online courses \citep{figlio2013online, bettinger2017virtual}; internet access \citep{vigdor2014scaling, dettling2018internet}; and interactive classroom technologies such as clickers and whiteboards \citep{caldwell2007clickers, lewin2008interactive}. Simply expanding access to computers and internet connectivity has not consistently improved academic achievement \citep{bulman.fairlie2016}. More encouraging evidence comes from computer-assisted learning programs that adapt instruction to students' levels or provide targeted practice and immediate feedback, particularly in mathematics \citep{escueta2020upgrading}. Generative AI combines features of both kinds of technology: it is a general-purpose tool that can give personalized explanations and respond to students' questions in real time, but it can also do the work for students. We study what students learn when both capabilities are available and students choose how to use them.
	
	Finally, we contribute to the literature on the productivity effects of AI. Prior studies show that AI assistance can generate substantial productivity gains in professional writing, software development, consulting, customer support, legal work, sales, team-based problem solving, and taxi driving.\footnote{See, among others, \cite{noy.zhang2023, peng.etal2023, choi2024lawyering, jia2024and, brynjolfsson.etal2025, dell2025cybernetic, cruces.etal2026, cui.etal2026, dellacqua.etal2026, kanazawa.etal2025}.} Much of this research measures task performance while AI is available. But AI may also raise productivity by facilitating skill accumulation. A few studies examine whether AI-assisted work improves subsequent unassisted performance, with mixed results. \citet{brynjolfsson.etal2025} provide evidence consistent with learning among customer-support agents, and \citet{cruces.etal2026} find that adults with lower levels of education retain some performance gains on an unassisted follow-up task. \citet{autor.etal2026} also find that an AI drafting assistant improves subsequent unassisted performance among senior patent lawyers, though not among junior patent lawyers. \citet{shen.tamkin2026} find that developers who used AI to learn an unfamiliar programming library scored substantially worse on a subsequent unassisted test. We provide evidence that AI access can raise learning productivity in an academic context: given the same amount of time, students acquire more knowledge, with gains evident on unassisted assessments immediately and one week later. These findings suggest a human-capital channel through which AI may raise longer-run productivity beyond its immediate effects on task performance.
	
	\section{Experimental Design} \label{sec:design}
	
	The experiment consists of two in-person sessions conducted approximately one week apart. Survey instruments (including tests) are available in the \href{https://www.dropbox.com/scl/fi/pzzmqwikxweqzrofrnzt9/ai-learning-supplementary.pdf?rlkey=zjg1lf7cmzno304na3bgq58bn&st=gtj3nlqa&dl=0}{Supplementary Materials}.
	
	\subsection{Session One}\label{sub:session1}
	
	\subsubsection{Overview.}
	
	The first session employs a between-subjects design with random assignment to either an AI-allowed or AI-forbidden condition. We schedule eight time slots with two parallel labs per slot. We randomly assign each student who signed up for a time slot to a specific computer in one of two labs. For each time slot, we then randomly assign one lab to the AI-allowed condition and the other to the AI-forbidden condition. Each workstation has a privacy divider to minimize distractions and prevent students from viewing other screens (see Appendix Figure~\ref{fig:lab}). We administer all components of the session through Qualtrics, a survey platform that tracks each student's progress, enforces section time limits, and records how long students spend on each component.
	
	\subsubsection{Session Structure.}
	
	The first session follows a fixed 60-minute structure with the following components (Figure~\ref{fig:timelines}, Panel A).
	
	\textit{Welcome and Instructions.} Each session begins with a staff member reading a condition-specific welcome script and asking students to put their phones away for the duration of the session. We tell students that their primary task is to learn about a prespecified topic and that they will demonstrate their understanding through a series of assessments. To ensure comprehension of the experimental procedures, students complete a series of comprehension checks, with laboratory staff providing clarification if needed. Students correctly answer an average of \getval{mean_comp_check} out of 5 comprehension questions, and the median student answers all questions correctly.
	
	\textit{Topic Assignment and Baseline Assessment.} We select three topics rich in factual content, about which most students have limited prior knowledge: blockchain technology, carbon capture systems, and CRISPR gene editing. We randomly assign each student to one of these three topics, and students cannot change their assignment. Students then complete a five-question multiple-choice test to assess their baseline knowledge of the assigned topic. In all tests, we randomize the order of questions and of the answer options within each question. Students may not use notes, the internet, or AI tools during tests. 
	
	\textit{Learning Phase.} Following the baseline test, students enter the learning phase. During this phase, they have up to 35 minutes to learn about their assigned topic. We instruct them to ``use the same learning approach you would typically follow for a college assignment'' and emphasize that their performance affects their payoffs, as described below.
	
	During this phase, students also write an approximately 500-word analytical essay demonstrating their understanding of the topic. Each essay prompt asks students to apply critical thinking skills by analyzing relationships between concepts, comparing elements, and supporting their analysis with specific evidence (see Appendix~\ref{app:prompts}).\footnote{For example, a student assigned to carbon capture may be asked to ``analyze the three main barriers to scaling carbon capture technologies (technical limitations, economic costs, and policy challenges),'' identify which barrier is most critical, and support their analysis with specific examples.} For each topic, we develop two prompts and randomly assign each student one for Session One and the other for Session Two. We provide all students with a link to an introductory reading about their assigned topic, created to be of comparable length and reading difficulty across the three topics.\footnote{Texts range from 1,248 to 1,382 words (Appendix Table~\ref{tab:readability}). At an average silent reading speed of approximately 250 words per minute for college students \citep{brysbaert2019many}, the estimated reading time is 5.0 to 5.5 minutes per text. The full text of the learning materials is available in the \href{https://www.dropbox.com/scl/fi/pzzmqwikxweqzrofrnzt9/ai-learning-supplementary.pdf?rlkey=zjg1lf7cmzno304na3bgq58bn&st=gtj3nlqa&dl=0}{Supplementary Materials}.} 
	
	Students can allocate their time freely across reading, searching, drafting, editing, or other activities such as surfing the web. Students can finish before the 35-minute maximum but cannot exceed it. The interface automatically advances at 35 minutes.
	
	\textit{Post-Learning Survey.} After the learning phase, students complete a brief survey that measures self-assessed current knowledge of the topic (0--10 scale); time allocation during the learning phase; subjective task assessment (enjoyment, feeling skilled or effective); and whether they had previously done similar tasks in their current academic year. The survey also includes an attention check.
	
	\textit{Post-Learning Test.} Students next complete five multiple-choice questions to test factual knowledge and conceptual understanding of the assigned topic. These questions differ from those on the baseline test. Students complete this assessment without access to external resources, the internet, or AI tools regardless of treatment condition.
	
	\textit{Waiting Room.} Students who complete all tasks before the 60-minute session ends are directed to a virtual waiting room. The ``Finish Session 1'' button is enabled only after 50 minutes have elapsed from the session start time. We tell students that we record their responses only after they click this button. This rule prevents early departures from disrupting others or pressuring them to finish early. During this waiting period, students may browse the internet freely.
	
	\subsubsection{Treatment.}
	
	We randomly vary students' access to generative AI tools during the learning phase. The treatment has two components: explicit instructions about whether AI use is permitted, and a logged-in ChatGPT account for students in the AI-allowed condition.
	
	Students in the \emph{AI-allowed} condition receive explicit instructions that permit the use of any generative AI tools during the learning phase. The instructions are delivered orally through the proctor-read welcome script and in writing through the survey interface (Appendix Figure~\ref{fig:treatment_instructions}, Panel A). The script states: ``[I]f you typically use generative AI tools such as ChatGPT, feel free to use them here as well. Using ChatGPT or other generative AI is completely allowed. We provide you with a ChatGPT account, already logged in on one of the tabs, so you don't need to use your personal account.'' To ease access, each computer has three browser tabs already open: a dedicated ChatGPT (GPT-4o) account, Wikipedia, and the Middlebury College Library website.
	
	Students in the \emph{AI-forbidden} condition receive explicit instructions prohibiting use of generative AI tools. The instructions are delivered orally through the proctor-read welcome script and in writing through the survey interface (Appendix Figure~\ref{fig:treatment_instructions}, Panel B). The script states: ``You are not allowed to use generative AI tools such as ChatGPT, Claude, or other AI assistants.'' Each computer has three browser tabs already open: Google, Wikipedia, and the Middlebury College Library website.
	
	\subsubsection{Compliance Monitoring.}
	
	We monitor treatment compliance through three mechanisms. First, two proctors per lab directly observe students throughout the session using classroom seating charts to record any unauthorized resource use (Appendix Figure~\ref{fig:map_labs}). Proctors note instances of students accessing AI tools in the AI-forbidden condition or violating other experimental protocols such as accessing external resources during the tests. Second, the Session Two exit survey elicits self-reported AI usage during the learning phase, including which specific AI tools were used (ChatGPT, Claude, Gemini, etc.). Third, the platform periodically captures students' writing interfaces to detect sudden appearances of large text blocks that may indicate copying from external sources.
	
	\subsubsection{Compensation and Incentives.}
	
	Students receive both attendance compensation and performance-based incentives. For attendance, students receive \$5 for completing Session One alone but \$50 for completing both sessions. For performance, students earn lottery tickets: one ticket per correct response on each test and one per point on their rounded essay-quality score. At the end of the experiment, we randomly draw 30 lottery tickets, each worth \$100. To further support attendance, we send email and text reminders on the day of each scheduled session and let students schedule the second session flexibly, rather than requiring exactly seven days between sessions.
	
	\subsection{Session Two}\label{sub:session2}
	
	Session Two takes place approximately one week after Session One.\footnote{When scheduling, we encouraged students to sign up for a Session Two slot approximately one week after their Session One slot, but allowed flexibility to accommodate scheduling constraints. Among students who attended Session Two, the average gap between sessions was \getval{avg_session_gap} days. Most returned exactly seven days after Session One (\getval{gap_7days_pct} percent), and \getval{gap_6to8_pct} percent returned within six to eight days.} This session measures knowledge retention when all students, regardless of Session One treatment assignment, work without access to AI or external resources. We did not tell students that they would complete additional assessments, and we find no evidence that students prepared between sessions: only \getval{between_lookup_pct} percent of students reported looking up extra information about the topic, with no significant difference between treatment and control groups, and no student reported studying it.
	
	\subsubsection{Session Structure.} The session includes two assessments presented in randomized order (Figure~\ref{fig:timelines}, Panel B): a ten-item multiple-choice knowledge test and a 20-minute analytical essay that uses the complementary prompt developed for each topic.
	
	Session Two concludes with an exit survey on four topics. First, students' beliefs about AI's impact on their own and others' test scores. Second, self-reported AI skills (Likert scale), frequency of use across academic tasks, and when students began using AI. Third, retrospective Session One behavior, including which tools they used and for what purposes (explaining concepts, summarizing, drafting, proofreading, editing) and whether they studied the topic between sessions. Fourth, an open-ended question about generative AI's impact on learning.\footnote{We note four main deviations from the experimental protocol that occurred during early time slots. First, the initial Session One verbal welcome script did not mention AI permissions or restrictions; students learned their treatment condition only through written platform instructions. We introduced verbal acknowledgment of treatment conditions starting with the second Session One time slot. Second, the initial Session One included a timer that malfunctioned, displaying available time after the 35-minute limit had elapsed. We removed the timer for subsequent sessions. Third, the first Session Two time slot lacked a waiting room, requiring students to manually track time before leaving. We introduced a virtual waiting room for all following sessions. Fourth, a technical glitch during one Session Two time slot prevented students from completing the essay on the platform. We added extra time, asked affected students to write their essays in Microsoft Word, and entered those essays into the dataset manually. All analyses include session fixed effects to account for these protocol variations.}
	
	\section{Data and Research Design} \label{sec:data}
	
	\subsection{Sample and Recruitment}
	
	The experiment took place at Middlebury College during Spring 2025. At the time of the experiment, AI adoption at Middlebury exceeded 80 percent \citep{contractor2026generative}, so our treatment varies AI access among students already familiar with the tool. We recruited students through campus-wide email announcements, posted flyers, student group chats, and a dedicated recruitment website. To minimize selection concerns, we framed the study as ``a study on learning'' without mentioning AI in any recruitment materials.
	
	\subsection{Summary Statistics and Balance}
	
	Table~\ref{tab:summ} reports summary statistics for three groups: the \getval{n_intake} students who completed the sign-up survey and scheduled a Session One slot (column~1), the \getval{n_students} (\getval{s1_attend_pct} percent) who attended Session One (column~2), and the \getval{n_both_sessions} (\getval{both_attend_pct} percent) who attended both sessions (column~3).\footnote{Our sample is relatively large by the standards of in-person laboratory experiments in economics. For example, \citet{camerer.etal2016} replicate 18 between-subjects laboratory experiments published in the \emph{American Economic Review} and the \emph{Quarterly Journal of Economics} between 2011 and 2014. Among these, the median sample size is 120, and only two exceed ours.} We focus on column~2. The average age is \getval{mean_age}, and \getval{pct_male} percent of students are male, \getval{pct_white} percent white, \getval{pct_international} percent international, and just over half (\getval{pct_public_hs} percent) attended a public high school (Panel~A).\footnote{ Appendix Table~\ref{tab:sample_vs_body} compares the sample with the Middlebury student body. The sample overrepresents women and first-year students and matches the student body on college GPA.} The mean GPA is \getval{mean_gpa} and the mean SAT score is \getval{mean_sat} (Panel~B).\footnote{We asked students for both SAT and ACT scores. Most students took the SAT. When students provided ACT but not SAT scores, we use concordance tables to convert ACT scores to SAT equivalents.} Most students (\getval{s2_retention_pct} percent) returned for Session Two (Panel~C). Students self-reported little baseline knowledge of their assigned topic (\getval{mean_baseline_knowledge} on a 0--10 scale) and, consistent with this self-assessment, answered only \getval{baseline_frac_correct} percent of baseline multiple-choice questions correctly (Panel D).
	
	The sample is balanced on most observable characteristics across treatment groups. Table~\ref{tab:bal} compares mean characteristics of students in the AI-forbidden versus AI-allowed conditions (columns~1~and~2). Only the coefficients on age and on having declared a major are statistically significant at the 10 percent level. Measures of academic ability, strong predictors of learning outcomes, are well balanced: mean GPA (\getval{mean_gpa_control} versus \getval{mean_gpa_treat}) and SAT scores (\getval{mean_sat_control} versus \getval{mean_sat_treat}) are nearly identical across groups. To address any residual imbalance, we use the double-lasso procedure \citep{belloni.etal2014}, as described below. We find no differential attendance in Session One (\getval{s1_attend_pct_control} versus \getval{s1_attend_pct_treat} percent) and no differential attrition between sessions.
	
	\subsection{Regression Model} \label{sec:model}
	
	We estimate linear models of the form
	\begin{align}\label{eq:main}
		Y_{i} = \alpha + \beta \text{AI-Allowed}_{i} + \mathbf{X}_i'\gamma + \varepsilon_{i},
	\end{align}
	where $Y_{i}$ is an outcome for student $i$, $\text{AI-Allowed}_{i}$ is an indicator that equals one if the student was randomly assigned to the AI-allowed condition, $\mathbf{X}_i$ is a vector of baseline control variables to improve precision, and $\varepsilon_{i}$ is an error term. The vector $\mathbf{X}_i$ contains fixed effects for the randomization strata (the eight Session One time slots) and additional controls selected through the double-lasso procedure \citep{belloni.etal2014}, which selects controls that predict either the outcome or the treatment assignment from a pool of potential covariates.\footnote{The pool of potential controls includes: (1) demographic characteristics---age, gender, race/ethnicity indicators (Black, Latino, Asian, white), international student status, and public/private high school attendance; (2) academic background---cohort fixed effects, declared major status, field-of-study indicators (natural sciences, social sciences, humanities/arts), self-reported study hours per week, college GPA and SAT score (each entered continuously and as decile bins), high school GPA (as four letter-grade bands), and an indicator for taking a standardized admissions test; (3) baseline measures---pre-learning self-assessed knowledge, pre-learning test score, prior experience with similar tasks, and comprehension check score; (4) experimental design features---topic fixed effects, writing prompt version, Session Two assessment order, and lab location; and (5) missing-data indicators. For students who did not take a standardized admissions test, we impute SAT scores from baseline covariates to assign them to bins, as described in Appendix~\ref{app:sat_imputation}.} We report heteroskedasticity-robust standard errors.
	
	The coefficient of interest, $\beta$, measures the causal effect of AI access on a given outcome. This intent-to-treat (ITT) estimate captures the effect of being \textit{allowed} to use AI during the learning phase, regardless of whether students actually used AI. The ITT may be the relevant parameter for policy evaluation: institutions can choose whether to permit AI use, but they cannot force students to use it. To estimate the effect of actual AI use rather than assignment, we complement our ITT estimates with a treatment-on-the-treated (TOT) analysis. We instrument actual AI use with the randomly assigned treatment status (AI-allowed) and estimate by two-stage least squares (2SLS). These estimates identify the local average treatment effect (LATE) for compliers, students whose treatment assignment influenced their AI use.
	
	\subsection{Main Outcomes}  \label{sec:outcomes}
	
	We estimate equation \eqref{eq:main} separately for several categories of outcomes:
	
	\subsubsection{AI Usage.}
	
	As a first-stage measure, we examine whether students used generative AI during the learning phase. Our primary measure is an indicator that equals one if the ChatGPT account provided to the student during Session One contained at least one prompt. We measure intensity through the number of prompts sent and the number of distinct conversations (i.e., separate chat threads) initiated. We complement these with self-reported measures of whether students used any generative AI tool, ChatGPT specifically, or alternative AI models (Claude, Gemini, Copilot, DeepSeek, or Llama).
	
	\subsubsection{Test-Based Learning Outcomes.}
	
	We measure knowledge acquisition and retention using self-reported and objective measures. Self-assessed knowledge captures students' subjective evaluation of their understanding on a 0--10 scale, with 0 indicating ``I know nothing about this topic'' and 10 indicating ``I am an expert.'' The fraction correct is the proportion of multiple-choice questions answered correctly (out of five questions in Session One, and ten in Session Two). We construct standardized test scores by normalizing raw scores to have mean zero and standard deviation (SD) one in the control group (separately for each session). We also define binary indicators for scoring at or above multiple thresholds to trace treatment effects across the score distribution.
	
	\subsubsection{Essay-Based Learning Outcomes.}
	
	Essays measure students' ability to reason with, synthesize, and argue from the information they learned. To evaluate the essays, we recruited \getval{n_graders} graders through Prolific, restricting eligibility to people with a master's or PhD degree. We exclude \getval{n_graders_comp_fail} graders who failed a comprehension check. Each essay was independently evaluated on a 0--10 scale by three or four graders blind to treatment condition and student identity; each grader scored five essays. Graders provided an overall quality rating and scores on five dimensions: accuracy of content, use of evidence and examples, relevance to the prompt, organization and structure, and writing style and clarity (see Appendix~\ref{app:prolific_graders} for details). We also construct a quality index as the average of the five dimension scores, each on the 0--10 scale; we standardize it by the control-group standard deviation when we report it in SD units. As specified in our pre-analysis plan, the overall quality rating and the quality index are our primary essay-based outcomes, and the individual dimensions are secondary outcomes. We complement human grading with AI grading: a large language model (LLM) evaluates each essay on the same dimensions and rubric (see Appendix~\ref{app:ai_grading} for the procedure and prompts).\footnote{Two considerations motivate our use of AI grading: a growing literature finds that LLM ratings of text quality track human ratings \citep{chiang2023can, liu2023geval, zheng2023judging}, and our learning topics (blockchain, carbon capture, CRISPR) are technical subjects on which human graders recruited from a general population may have limited expertise.} For our main estimates, we use the average of human and AI grader scores for each dimension and show robustness to using only the human or only the AI ratings. %
	
	\subsubsection{Essay Writing Characteristics.}
	
	Beyond quality, we measure three characteristics of each essay to examine how AI access changes the writing students submit. First, we use Pangram 4, a commercial AI content detector \citep{emi2024pangram, masrour.etal2025, thai.etal2026, glickenhaus.etal2026}, to measure the fraction of text classified as AI-generated.\footnote{\citet{jabarian.imas2025} evaluate four AI text detectors and find that Pangram achieves near-zero false positive and false negative rates, even when AI-generated text is modified using ``humanizer'' tools.} We also use Pangram to measure the fraction flagged as plagiarized.\footnote{Pangram's plagiarism checker matches sentences against the open web: webpages, books, news articles, and other publicly indexable sources \citep{pangram2025plagiarism}.} Second, we measure writing style through three indices: a length index (number of tokens, words, and sentences), a readability index (sentence length, syllables per word, the Flesch--Kincaid grade level, and the Flesch Reading Ease score, with difficulty measures reversed so that higher values indicate easier-to-read text), and a lexical diversity index (the ratio of unique to total words and the share of words used only once). Each index averages its components after standardizing them to mean zero and SD one in the control group. Third, to examine whether AI access homogenizes student writing, we follow \citet{brynjolfsson.etal2025} in computing sentence embeddings. We measure within-group textual similarity as the average pairwise cosine similarity between a student's essay and all other essays in the same treatment group, topic, and prompt cell, and reading material similarity as the cosine similarity between the essay and the provided reading material.\footnote{As a validation exercise, we confirm that the embeddings capture content variation: average within-topic cosine similarity is \getval{cosine_within_topic_all}, compared with \getval{cosine_across_topic} across topics, and essays responding to the same prompt are more similar to each other (\getval{cosine_same_prompt}) than essays on the same topic but a different prompt (\getval{cosine_diff_prompt}).}
	
	\section{Effects of AI Access on Learning} \label{sec:results_s1}
	
	\subsection{First Stage: AI Adoption and Usage Patterns}
	
\begin{figure}[H]
		\caption{The Impact of AI Access on Generative AI Usage During the Learning Phase}\label{fig:ai_impact_usage}
		\centering
		\includegraphics[width=.75\linewidth]{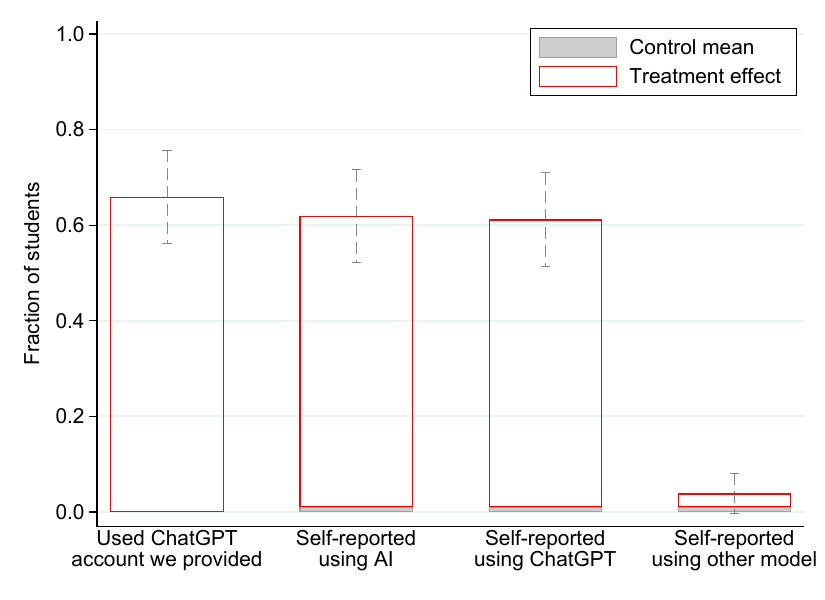}
		\hfill						
		{\footnotesize
			\singlespacing \justify
			
			\textit{Notes:} This figure shows the impact of AI access on generative AI usage during the learning phase in Session One. Gray bars represent control-group means, while red outlines show treatment effects. The first measure is constructed from activity in the ChatGPT account provided to each student. The remaining three measures are based on self-reported usage collected in the exit survey. Vertical bars represent 95 percent confidence intervals. \par
			
		}
	\end{figure}

	We begin with the first-stage estimates of how random assignment to AI access translated into AI use. Table~\ref{tab:first_stage} reports estimates from equation \eqref{eq:main} using several measures of AI usage as outcomes. Panel A presents measures recorded in treated students' ChatGPT accounts. Panel B presents self-reported measures. Figure~\ref{fig:ai_impact_usage} illustrates these first-stage effects.
	
	The experimental manipulation generated substantial variation in AI usage. Assignment to the AI-allowed group increased AI usage by $\hat{\beta} = \getval{fs_revealed_pp}$ pp as measured by activity in the provided ChatGPT account ($p < 0.001$). Self-reported measures yield similar though slightly attenuated effects: any generative AI use increased by $\hat{\beta} = \getval{fs_selfreport_pp}$ pp ($p < 0.001$). This effect represents a large increase from the near-zero baseline in the control group. AI adoption was almost entirely ChatGPT: a $\getval{fs_chatgpt_pp}$ pp increase in ChatGPT use ($p < 0.001$), versus just a $\getval{fs_other_ai_pp}$ pp increase in the use of other AI models ($p = \getval{fs_other_ai_pval}$).\footnote{The near-exclusive use of ChatGPT is consistent with surveys of adoption among college students \citep{hirabayashi2024, stohr.etal2024, contractor2026generative}.} Consistent with this strong first stage, \getval{ai_helpful_users_pct} percent of AI users reported that AI was somewhat or very helpful for the learning task.\footnote{Appendix Table~\ref{tab:compliance} reports correlates of AI take-up among treated students. Frequent AI use for academic purposes is the strongest and most robust predictor. Self-assessed proficiency also predicts take-up in bivariate specifications, while early adoption does not. White students are less likely to use the provided account in the bivariate specification. We find no significant association between take-up and either AI's perceived benefits or GPA.}
	
	Treated students use AI for a mix of purposes: most often to explain concepts, but also to draft and edit text. Figure~\ref{fig:ai_usage_type} reports the fraction of treated students who used AI for each of six purposes, measured through self-reports (Panel~A) and ChatGPT conversation logs (Panel~B), which we classify using an LLM (see Appendix~\ref{app:chatgpt_classification}). In the logs, explaining concepts is the most common use (\getval{pct_revealed_explain} percent of treated students), followed among the substantive categories by drafting responses (\getval{pct_revealed_draft} percent), editing responses (\getval{pct_revealed_edit} percent), and summarizing readings (\getval{pct_revealed_summarize} percent). The self-reports give a similar picture except for writing: explaining concepts is again the most common use (\getval{pct_sr_explain} percent), but students report drafting (\getval{pct_sr_draft} percent) and editing (\getval{pct_sr_edit} percent) less often than the logs show.\footnote{The high rate of ``other'' uses in the logs largely reflects conversational turns (e.g., ``yes,'' ``sure''), synonym lookups, and off-topic questions: prompts that students would likely not consider a distinct use of AI when responding to our survey.} 
	
	\subsection{Effects on Test Scores} \label{sub:tests}
	
	AI access improves students' test performance but not their self-assessed knowledge (Table~\ref{tab:test_score_long}, columns~1--3). Both treated and control students perform better after the learning phase, consistent with skill accumulation. Self-assessed knowledge in the control group rises sharply, from \getval{baseline_selfknow_forbidden_mean} to \getval{post_s1_selfknow_forbidden_mean} on the 0--10 scale ($p < 0.001$). The rise is nearly identical among treated students, leaving an ITT of $\hat{\beta} = \getval{s1_selfknow_itt}$ points, which is statistically indistinguishable from zero. Test performance, by contrast, responds to AI access. Control students' fraction of questions answered correctly rises from \getval{baseline_test_forbidden_mean} percent to \getval{post_s1_test_forbidden_mean} percent ($p < 0.001$); AI access adds a further $\hat{\beta} = \getval{s1_test_itt_pp}$ pp, a \getval{s1_test_pct_improvement} percent improvement over the control mean. In standardized units, this is a $\hat{\beta} = \getval{s1_test_itt_sd}$~SD effect (column~2, $p = \getval{s1_test_itt_pval}$). The corresponding TOT estimate is $\getval{s1_test_tot_pp}$ pp, or $\getval{s1_test_tot_sd}$~SD ($p = \getval{s1_test_tot_pval}$, column~3). The test-score estimates are robust to excluding any one time slot (Appendix Figure~\ref{fig:loo_slots}) or any one test item (Appendix Figure~\ref{fig:item_level}).\footnote{One concern in interpreting these results as pure learning is that AI access increases the likelihood of rule violations on the Session One test (Appendix Table~\ref{tab:integrity}). Proctor-observed violations rise by $\hat{\beta} = \getval{mech_cheat_proctor_pp}$ pp ($p = \getval{mech_cheat_proctor_pval}$), while self-reported violations rise by $\hat{\beta} = \getval{mech_cheat_sr_pp}$ pp ($p = \getval{mech_cheat_sr_pval}$). As a back-of-the-envelope calculation, we multiply each measure's treatment effect by the test-score gap between violators and non-violators (\getval{cheat_score_gap_proctor_pp} pp for proctor-observed violations and \getval{cheat_score_gap_sr_pp} pp for self-reported ones). The implied contribution of violations to the ITT is \getval{cheat_implied_proctor_pp} pp under the proctor measure and \getval{cheat_implied_sr_pp} pp under the self-report measure, or \getval{cheat_share_proctor} and \getval{cheat_share_sr} percent of the estimate. Excluding students who violated the rules as captured by proctors, by self-reports, or by both a proctor and a self-report reduces the learning effect to \getval{s1_test_noviol_proctor_sd}~SD, \getval{s1_test_noviol_sr_sd}~SD, and \getval{s1_test_noviol_both_sd}~SD, respectively. Adding a violation indicator to the regression gives similar results: the Session One ITT falls to \getval{s1_test_violdum_proctor_sd}~SD under the proctor measure and \getval{s1_test_violdum_sr_sd}~SD under self-reports. We note that no student violated the rules in Session Two, by either proctor observation or self-report.}
	
	A smaller learning gain remains one week later (Table~\ref{tab:test_score_long}, columns~4--6). Both treated and control students perform worse in Session Two than in Session One, consistent with skill depreciation. Self-assessed knowledge among control students who attended both sessions falls from \getval{fade_s1_selfknow_control_mean} to \getval{fade_s2_selfknow_control_mean} on the 0--10 scale ($p < 0.001$). The decline is of similar size among treated students, leaving an ITT of $\hat{\beta} = \getval{s2_selfknow_itt}$ points ($p = \getval{s2_selfknow_itt_pval}$). Test performance shows a similar fade-out pattern. Among these control students, the fraction of questions answered correctly falls from \getval{fade_s1_test_control_mean} percent in Session One to \getval{fade_s2_test_control_mean} percent in Session Two ($p = \getval{s1_to_s2_test_control_fade_p}$). Treated students decline as well but remain ahead of the control group. The resulting ITT of $\hat{\beta} = \getval{s2_test_itt_pp}$ pp ($p = \getval{s2_test_itt_pval}$) is \getval{s2_test_itt_persistence_pct} percent of the \getval{s1_test_itt_pp} pp Session One effect (though given the standard errors, the two estimates are not statistically distinguishable, $p = \getval{s1_s2_test_diff_pval}$). Standardized, this corresponds to a $\hat{\beta} = \getval{s2_test_itt_sd}$ SD effect. AI-assisted learning therefore produces knowledge gains that are still present one week later.
	
\begin{figure}[H]
		\caption{Distributional Effects of AI Access on Test Performance}\label{fig:distributional}
		
		\begin{subfigure}[t]{\textwidth}
			\caption*{Panel A. Treatment effects by score threshold}
			\centering
			\includegraphics[width=.75\linewidth]{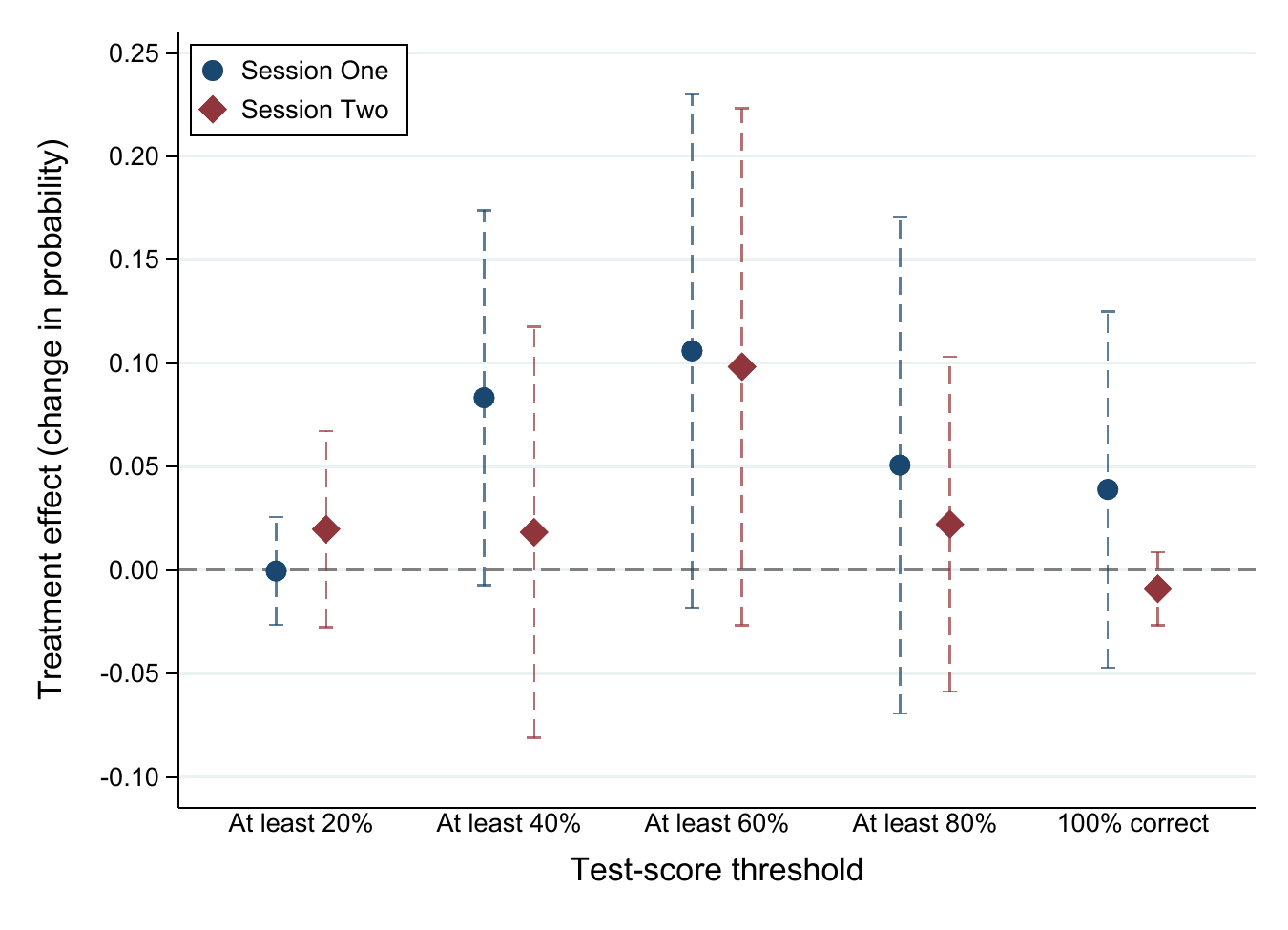}
		\end{subfigure}
		
		\vspace{0.3cm}
		
		\begin{subfigure}[t]{.48\textwidth}
			\caption*{Panel B. Session One (knowledge acquisition)}
			\centering
			\includegraphics[width=\linewidth]{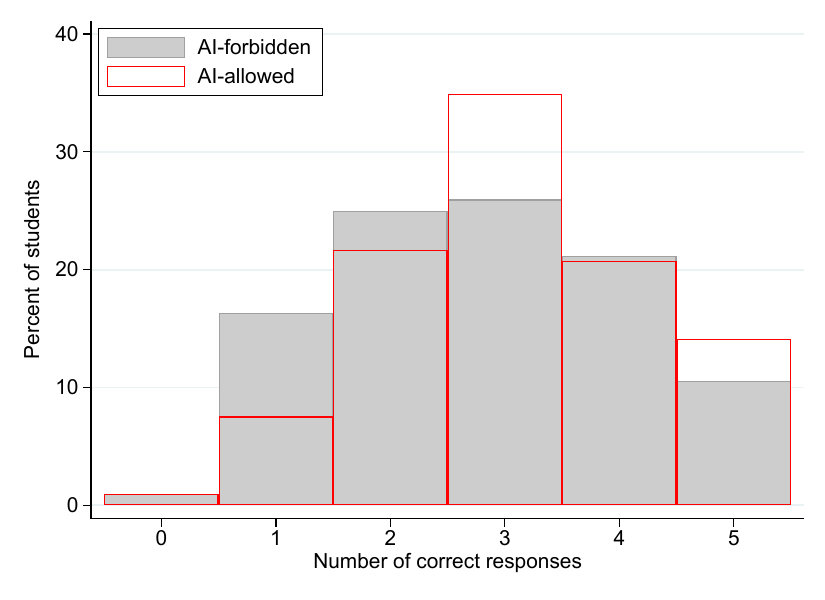}
		\end{subfigure}
		\hfill
		\begin{subfigure}[t]{0.48\textwidth}
			\caption*{Panel C. Session Two (knowledge retention)}
			\centering
			\includegraphics[width=\linewidth]{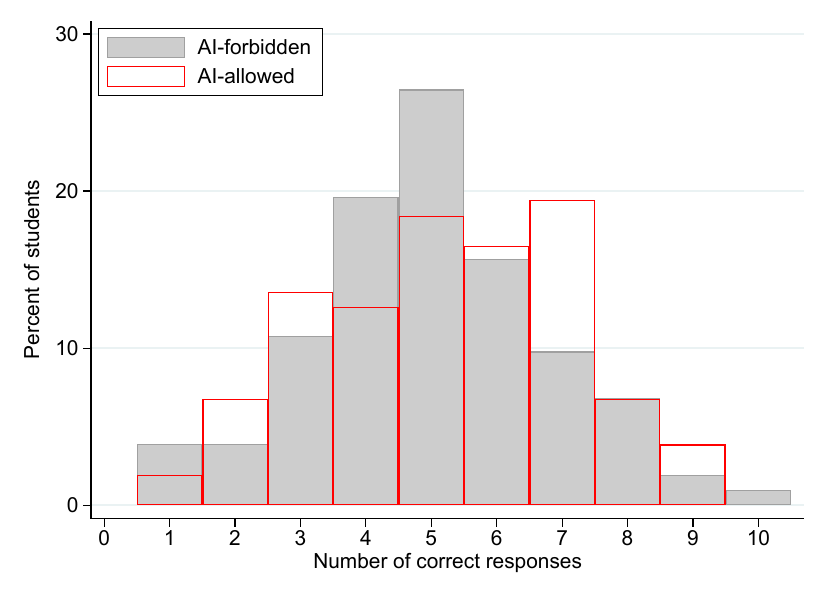}
		\end{subfigure}
		
		\hfill
		{\footnotesize
			\singlespacing \justify
			
			\textit{Notes:} Panel~A plots ITT estimates of the effect of AI access on indicators for scoring at or above each threshold, separately for Session One (navy circles) and Session Two (maroon diamonds). Each estimate comes from a regression that includes strata fixed effects and controls selected through a double-lasso procedure \citep{belloni.etal2014}. Vertical bars show 95 percent confidence intervals based on heteroskedasticity-robust standard errors. Panels~B and~C show the distribution of test scores by treatment assignment. The $x$-axis shows the number of correct responses; the $y$-axis shows the percentage of students achieving each score. The Session One test had five questions; the Session Two test had ten questions. \par
			
		}
	\end{figure}

	The point estimates place the learning gains from AI access in the middle of the performance distribution in both sessions (Figure~\ref{fig:distributional}, Panel~A). In Session One, treated students are \getval{s1_above40_pp} pp more likely to score at least 40 percent correct ($p = \getval{s1_above40_p}$) and \getval{s1_above60_pp} pp more likely to score at least 60 percent correct ($p = \getval{s1_above60_p}$). The estimates at higher thresholds are also positive, but smaller and imprecise. In Session Two, the largest threshold estimate is an increase of $\hat{\beta} = \getval{s2_above60_pp}$ pp at the 60 percent threshold; the other threshold estimates are smaller. Panels~B and~C of Figure~\ref{fig:distributional} show the underlying score distributions: AI access shifts the middle of the performance distribution rightward while leaving the tails largely unchanged. 
	
	\subsubsection{Benchmarking the Test-Score Effect.}
	
	The effect of AI access on test scores is large by the standards of the educational-intervention literature. \citet{kraft2020} reports a median effect of 0.10~SD across 1,942 effect sizes from 747 RCTs with standardized test outcomes. Our ITT of $\hat{\beta} = \getval{s1_test_itt_sd}$~SD is substantially larger. To make this magnitude concrete, we compare it with two in-sample benchmarks. A single learning session raises AI-forbidden students' test scores by \getval{bench_session_gain_sd}~SD above baseline, and AI access adds roughly another quarter of that gain. Among control students, a one-SD increase in college GPA predicts about \getval{bench_gpa_s1_sd}~SD higher test performance, so the effect of AI access exceeds the gain associated with a one-SD increase in GPA. Our estimate is similar to the 0.29~SD pooled gain from structured human tutoring programs \citep{nickow.etal2024}.
	
	Figure~\ref{fig:literature_comparison} places our estimates alongside 24 others from 14 studies (see Appendix~\ref{app:literature_comparison} for the studies and inclusion criteria). Aggregating estimates within each study and excluding our own, the random-effects grand mean across the \getval{lit_n_studies_cl} studies is $\getval{lit_grand_mean_cl_sd}$~SD (95 percent confidence interval $\getval{lit_grand_mean_cl_lo_sd}$ to $\getval{lit_grand_mean_cl_hi_sd}$). Thus, our estimates are in line with the AI-and-learning experiments.
	
	\subsection{Effects on Essay-Based Outcomes}
	
	We examine how AI access affects students' writing in three steps. First, we check whether AI-generated text enters treated students' essays. Second, we ask whether AI homogenizes the writing students submit. Third, we ask whether AI access improves essay quality. Because Session One essays are written during the learning phase, when treated students have AI access, the Session One estimates reflect both AI-generated text and any learning that occurred during the session. With the Session Two essays, which students write one week later without AI, we isolate the learning channel.
	
	\subsubsection{AI Detection.}
	
	AI use is detectable in treated students' essays (Figure~\ref{fig:essay_quality} and Table~\ref{tab:writ_index_long}, Panel A). The fraction of text classified as AI-generated rises by $\hat{\beta} = \getval{s1_ai_detect_pp}$ pp in the AI-allowed group ($p < 0.001$), a \getval{s1_ai_detect_pct_increase} percent increase over the control mean of \getval{s1_ai_detect_control_mean} percent.\footnote{That \getval{s1_ai_detect_control_mean} percent of control-group text is flagged as AI-generated likely reflects a combination of control students using AI, false positives, and students whose natural writing style resembles AI-generated text, a pattern consistent with recent evidence that exposure to LLMs shifts human spoken language toward words that LLMs favor \citep{yakura.etal2024}.} The corresponding TOT estimate is $\getval{s1_ai_detect_tot_pp}$ pp ($p < 0.001$). Despite this large increase in the fraction of AI-generated text, we find little evidence of stylistic changes in student essays: the effects on the length, readability, and lexical diversity indices are small and statistically insignificant (Appendix Table~\ref{tab:writ_style_long} and Appendix Figure~\ref{fig:writing_char_detailed}). By Session Two, AI traces disappear, consistent with students writing without AI access (Table~\ref{tab:writ_index_long}, Panel A). The treatment effect collapses from $\hat{\beta} = \getval{s1_ai_detect_pp}$ pp to $\hat{\beta} = \getval{s2_ai_detect_pp}$ pp ($p = \getval{s2_ai_detect_pval}$). We find no evidence of copying from online sources: the fraction of text flagged by Pangram's plagiarism checker is near zero in both groups, with no significant treatment effect in either session.
	
	\subsubsection{Homogenization.} 
	
	Despite concerns that AI homogenizes writing \citep{doshi.hauser2024,meincke.etal2025,moon.etal2025,moon.etal2025creative}, we see no such effect in our data (Table~\ref{tab:writ_index_long}, Panel B). Treatment effects on cosine similarity, our measure of homogeneity, are near zero and statistically insignificant: within-group textual similarity changes by $\hat{\beta} = \getval{s1_cosine_within_itt}$, a \getval{s1_cosine_within_pct_change} percent change relative to the control mean of \getval{s1_cosine_within_control_mean}. Similarity to the provided reading material changes by less than 0.001, relative to a control mean of \getval{s1_cosine_source_control_mean}. Both similarity effects remain near zero in Session Two (columns~4--6). These null results contrast with the convergence patterns documented by \citet{brynjolfsson.etal2025} among customer-support agents and with the reductions in collective creative diversity observed in experiments on AI-assisted writing and brainstorming \citep{doshi.hauser2024,meincke.etal2025}. One possible explanation is task structure: customer-support and ideation tasks have well-defined correct answers or converge on common AI-suggested ideas, whereas essay prompts are open-ended and admit multiple valid responses.
	
	\subsubsection{Essay Quality and Dimension-Level Ratings.}
	
\begin{figure}[H]
		\caption{Effects of AI Access on Essay Outcomes}\label{fig:essay_quality}
		\centering
		\includegraphics[width=.85\linewidth]{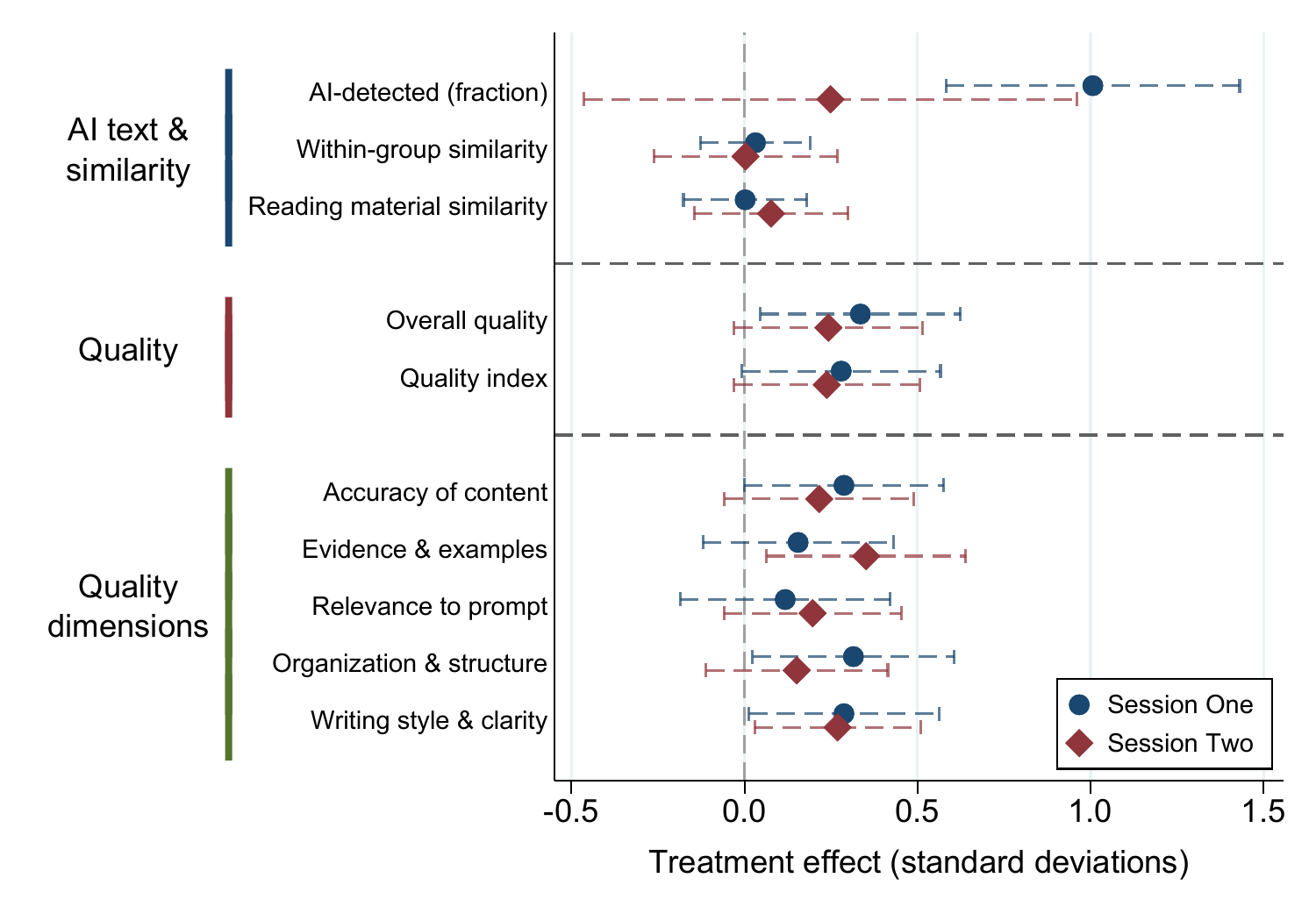}
		
		\hfill
		{\footnotesize
			\singlespacing \justify
			
			\textit{Notes:} This figure presents treatment effects of AI access on AI detection and textual similarity (top panel), overall essay quality (middle panel), and individual quality dimensions (bottom panel), measured in standard deviations. The top-panel measures are the AI-detected fraction (the share of text classified as AI-generated by Pangram 4, standardized in the control group), within-group similarity (the average pairwise cosine similarity within treatment$\times$topic$\times$prompt cells), and reading material similarity (between the student essay and the provided reading material). The overall quality measures are overall quality (the grader's holistic assessment) and the quality index (averaging the five dimensions below), each averaged across the human and AI graders. The quality dimensions are the five sub-components of a standardized rubric (accuracy of content, use of evidence and examples, relevance to the prompt, organization and structure, and writing style and clarity), each averaged across the human and AI graders. Circles represent Session One effects (essays written with or without AI access); diamonds represent Session Two effects (essays written one week later without AI access). Horizontal lines represent 95 percent confidence intervals. Point estimates underlying each coefficient in raw units are reported in Table~\ref{tab:writ_index_long} and Appendix Table~\ref{tab:essay_quality_dims}. The five quality dimensions split by grader appear in Appendix Figure~\ref{fig:essay_quality_detailed}. \par
			
		}
	\end{figure}

	Effects on overall essay quality are positive in both sessions (Figure~\ref{fig:essay_quality} and Table~\ref{tab:writ_index_long}, Panel C). In Session One, treated essays score $\hat{\beta} = \getval{s1_qual_overall_avg_itt}$~points higher in overall quality (or $\getval{s1_qual_overall_avg_itt_sd}$~SD, $p = \getval{s1_qual_overall_avg_pval}$), an \getval{s1_qual_overall_avg_pct_increase} percent increase over the control mean of \getval{s1_qual_overall_avg_control_mean} points, with a similar effect on the quality index ($\hat{\beta} = \getval{s1_qual_index_avg_itt}$ points, or $\getval{s1_qual_index_avg_itt_sd}$~SD, $p = \getval{s1_qual_index_avg_pval}$). All five dimension-level point estimates are positive (Appendix Table~\ref{tab:essay_quality_dims} and Appendix Figure~\ref{fig:essay_quality_detailed}). Organization and structure shows the largest effect ($\hat{\beta} = \getval{s1_qual_structure_itt}$ points, or $\getval{s1_qual_structure_itt_sd}$~SD, $p = \getval{s1_qual_structure_pval}$), followed by writing style and clarity ($\hat{\beta} = \getval{s1_qual_clarity_itt}$ points, or $\getval{s1_qual_clarity_itt_sd}$~SD, $p = \getval{s1_qual_clarity_pval}$) and accuracy of content ($\hat{\beta} = \getval{s1_qual_accuracy_itt}$ points, or $\getval{s1_qual_accuracy_itt_sd}$~SD, $p = \getval{s1_qual_accuracy_pval}$). The estimates for use of evidence and examples ($\hat{\beta} = \getval{s1_qual_evidence_itt}$ points, or $\getval{s1_qual_evidence_itt_sd}$~SD, $p = \getval{s1_qual_evidence_pval}$) and relevance to the prompt ($\hat{\beta} = \getval{s1_qual_relevance_itt}$ points, or $\getval{s1_qual_relevance_itt_sd}$~SD, $p = \getval{s1_qual_relevance_pval}$) are smaller and imprecise. In Session Two, overall quality rises by $\hat{\beta} = \getval{s2_qual_overall_avg_itt}$ points (or $\getval{s2_qual_overall_avg_itt_sd}$~SD, $p = \getval{s2_qual_overall_avg_pval}$), with a similar effect on the quality index ($\hat{\beta} = \getval{s2_qual_index_avg_itt}$ points, or $\getval{s2_qual_index_avg_itt_sd}$~SD, $p = \getval{s2_qual_index_avg_pval}$). Use of evidence and examples shows the largest effect ($\hat{\beta} = \getval{s2_qual_evidence_itt}$ points, or $\getval{s2_qual_evidence_itt_sd}$~SD, $p = \getval{s2_qual_evidence_pval}$), followed by writing style and clarity ($\hat{\beta} = \getval{s2_qual_clarity_itt}$ points, or $\getval{s2_qual_clarity_itt_sd}$~SD, $p = \getval{s2_qual_clarity_pval}$). The estimates for relevance to the prompt ($\hat{\beta} = \getval{s2_qual_relevance_itt}$ points, or $\getval{s2_qual_relevance_itt_sd}$~SD, $p = \getval{s2_qual_relevance_pval}$) and accuracy of content ($\hat{\beta} = \getval{s2_qual_accuracy_itt}$ points, or $\getval{s2_qual_accuracy_itt_sd}$~SD, $p = \getval{s2_qual_accuracy_pval}$) are smaller and imprecise, and organization and structure shows the smallest effect ($\hat{\beta} = \getval{s2_qual_structure_itt}$ points, or $\getval{s2_qual_structure_itt_sd}$~SD, $p = \getval{s2_qual_structure_pval}$). 
	
	Together with the test-score gains of Section~\ref{sub:tests}, these estimates indicate that treated students learn more about the topic and express that learning in two ways: in the facts they recall and in how clearly they explain the topic and argue about it in their own writing.
	
	\section{How AI Changes the Learning Process} \label{sec:mechanisms}
	
	To understand why AI access improves learning, we first examine changes in the way students work during the learning phase: the time they allocate across learning activities and how they experience the task. We then present a model that organizes these findings and yields one further implication: outcomes should differ with how students use AI. We examine that implication with descriptive evidence from our experiment and with evidence from other experiments.
	
	\subsection{Effects on How Students Learn} \label{sub:production}
	
	\subsubsection{Time Allocation Across Learning Activities.} \label{sub:time_use}
	
	AI access does not change time spent on the learning phase (Table~\ref{tab:mech_long}, Panel A). We measure time in two ways: Qualtrics records the duration of the learning phase, and the post-learning survey provides a self-reported measure. Control students spend an average of \getval{mech_time_obj_control_mean} minutes by the Qualtrics measure and \getval{mech_time_sr_control_mean} minutes by self-report. Most students use the full 35 minutes (Appendix Figure~\ref{fig:time_learning_dist}). Treatment effects are small and not statistically significant: a \getval{mech_time_obj_decrease}-minute decrease by Qualtrics and a \getval{mech_time_sr_decrease}-minute decrease by self-report. 
	
	Although total learning time is unchanged, its composition shifts away from producing text (Table~\ref{tab:mech_long}, Panel B). Treated students spend about \getval{mech_writing_share_decrease} pp less of their learning time on writing activities (drafting, editing, note-taking, and organizing), down from a control-group share of \getval{mech_writing_share_control} percent ($p = \getval{mech_writing_share_pval}$), and about \getval{mech_research_share_increase} pp more on research activities (reading and searching), up from \getval{mech_research_share_control} percent ($p = \getval{mech_research_share_pval}$). The corresponding TOT effects are a \getval{mech_writing_share_tot_decrease} pp decrease in the writing share ($p = \getval{mech_writing_share_tot_pval}$) and a \getval{mech_research_share_tot_increase} pp increase in the research share ($p = \getval{mech_research_share_tot_pval}$). Activity by activity, students move from writing text to finding information and reading (Appendix Figure~\ref{fig:mech_inputs}). Editing is the one writing activity that rises, which plausibly reflects students revising AI-generated text rather than drafting their own. This pattern is consistent with a broader finding in professional settings: AI shifts effort from task execution to task stewardship \citep{lee.etal2025}. For example, \citet{mozannar.etal2024} find that GitHub Copilot users spend about a third of their coding time double-checking and editing AI-generated suggestions, and more than half on Copilot-related activities.
	
	\subsubsection{The Learning Experience.} \label{sub:experience}
	
	AI access makes learning more enjoyable but does not change how effective students feel (Table~\ref{tab:mech_long}, Panel C). Enjoyment of the learning task rises by $\hat{\beta} = \getval{mech_enjoy_increase}$ points on the 0--10 scale ($p = \getval{mech_enjoy_pval}$), an \getval{mech_enjoy_pct_increase} percent improvement over the control mean of \getval{mech_enjoy_control_mean}. Consistent with this, the share of students rating their enjoyment at or above the sample median rises by \getval{mech_enjoy_above_med_pp} pp ($p = \getval{mech_enjoy_above_med_pval}$). The effect on how skilled or effective students feel is also positive, but small and statistically insignificant  ($\hat{\beta} = \getval{mech_effective_increase}$ points, $p = \getval{mech_effective_pval}$). These findings align with \citet{noy.zhang2023}, who find that ChatGPT access increases enjoyment of the task by $0.47$~SD and self-efficacy by $0.19$~SD among knowledge workers (though the latter is not statistically significant). 
	
	\subsection{A Model of Learning with AI} \label{sub:framework}
	
	The evidence so far shows that AI access raises test scores, shifts time from writing toward research, and makes the task more enjoyable. To organize this evidence, we build a model in which a student produces two outputs during the learning phase: the essay the student submits and the knowledge the student acquires. The goal of the model is to rationalize these results and to generate additional testable implications. Here, we provide an abbreviated version of the model and its implications and present the complete model in Appendix~\ref{app:framework}.
	
	\subsubsection{Summary of the Model.}
	
	In the model, a student chooses how much cognitive effort to exert on two activities: ``research'' (which includes reading and searching) and ``writing'' (which includes drafting, editing, and note-taking). Effort in these activities produces two outputs. Both activities produce knowledge, and writing also produces the essay the student submits, whose quality reflects both the student's writing and the knowledge the student has acquired. The student chooses the two effort levels to maximize a weighted sum of knowledge and essay quality, net of the cost of effort. In our experiment, the performance incentives contribute to these weights, alongside the student's own interest in learning about the topic and any intrinsic motivation to write a high-quality essay.
	
	We model AI as a technology that gives the student two non-mutually exclusive capabilities \citep{brynjolfsson2017machine, acemoglu2019automation}. First, we assume that AI used as a personalized tutor raises the return to research effort: with AI, each unit of reading and searching produces more knowledge, because the student can ask for explanations and pose follow-up questions. Second, we assume that the student can delegate the essay to AI, which replaces the student's own writing with an AI-written essay of a given quality. We model delegation as a choice variable and show that a student delegates when the AI-written essay is good enough relative to the essay the student would write, and the delegation threshold rises with the value the student places on knowledge. %
	
	The model shows that, under these assumptions, the effect of AI on learning combines a gain from tutoring with two losses from delegation. Tutoring raises knowledge because each unit of research effort produces more of it and because the student endogenously supplies more research effort in response. Delegation removes the essay-quality reward from the student's own work, so the student supplies less effort in both activities and forgoes the knowledge that this effort would have produced. Delegation also lowers the gain from tutoring: the knowledge that tutoring produces no longer improves the submitted essay, so research effort responds less to the higher return. %
	
	\subsubsection{Model Implications.}
	
	The model can rationalize the three main facts documented so far. The test-score gain (Section~\ref{sub:tests}) implies that, averaged across treated students, the tutoring gain exceeded the losses from delegation. The shift of learning time from writing toward research (Table~\ref{tab:mech_long}, Panel~B) follows from both capabilities: tutoring raises research effort, and delegation lowers writing effort by a larger proportion than research effort, so each lowers the share of effort spent on writing. The enjoyment gain (Table~\ref{tab:mech_long}, Panel~C) is consistent with each capability raising the utility the student obtains from the task and with a lower cost of effort.
	
	The model also predicts that outcomes should differ with how students use AI. In the model, an AI-written essay raises the quality of the submitted essay without adding knowledge. Because students write the Session Two essays without AI, the essay advantage of students who delegated in Session One should be absent from their unaided work one week later, whereas the knowledge that tutoring produces should persist and show up on the tests and in the unaided essay. We next examine this implication.
	
	\subsection{Outcomes by Type of AI Use} \label{sub:aug_auto}
	
	We provide two types of evidence on whether the learning effect of AI depends on how students use it: how outcomes in our experiment vary with the way students use AI, and experiments in the AI-and-learning literature that randomize or observe how learners use AI.
	
	\subsubsection{Evidence from Our Experiment.}
	
	To classify how students use AI, we had a research assistant read each ChatGPT conversation log and label it \textit{Automation} if the AI does the work \textit{for} the student, \textit{Augmentation} if the AI works \textit{with} the student, \textit{Mixed} if AI does both, or \textit{Other} if the conversation is off-topic. Of treated AI users, \getval{pct_aug_users} percent are pure augmentation users, \getval{pct_auto_users} percent are pure automation users, \getval{pct_mixed_users} percent are mixed, and the remaining \getval{pct_other_users} percent are classified as \textit{Other}. In the results below we compare each of these groups with the control group.\footnote{To check the hand coding, we also ask an LLM to label all conversations with the same four categories (see Appendix~\ref{app:chatgpt_classification} for the full prompt). We compare the two coders on three indicators: whether a conversation includes automation (labeled Automation or Mixed), whether it includes augmentation (labeled Augmentation or Mixed), and whether it is off-topic (labeled Other). Agreement is \getval{val_auto_agree_pct} percent on the automation indicator, \getval{val_aug_agree_pct} percent on the augmentation indicator, and \getval{val_other_agree_pct} percent on the off-topic indicator.}
	
	Three pieces of evidence support this classification. First, the three groups prompt AI differently: automation users use AI mainly for drafting and editing, augmentation users mainly for explaining concepts, and mixed users for both (Appendix Figure~\ref{fig:ai_usage_by_type}). Second, automation users spend about \getval{auto_aug_research_pct} percent less time on research activities than augmentation users ($p = \getval{auto_aug_research_p}$), and \getval{auto_aug_total_pct} percent less time overall ($p = \getval{auto_aug_total_p}$). Mixed users' time use is indistinguishable from augmentation users' ($p = \getval{mixed_aug_research_p}$ for research activities, $p = \getval{mixed_aug_total_p}$ overall). Third, these behavioral differences show up in essay outputs: Pangram flags \getval{s1_ai_detect_auto_mean} percent of automation users' text as AI-generated, \getval{s1_ai_detect_mixed_mean} percent of mixed users' text, and \getval{s1_ai_detect_aug_mean} percent of augmentation users' text.
	
	Students show different patterns of effects depending on how they use AI (Table~\ref{tab:heterogeneity_ai}). On the Session One test, the two groups that used AI for explanation score above the control group: mixed users by $\hat{\beta} = \getval{s1_mixed_test_itt}$~SD ($p = \getval{s1_mixed_test_pval}$) and augmentation users by $\hat{\beta} = \getval{s1_aug_test_itt}$~SD ($p = \getval{s1_aug_test_pval}$), compared with $\hat{\beta} = \getval{s1_auto_test_itt}$~SD for automation users. On the Session One essay, the pattern reverses: the two groups that had AI do some of the writing are rated above the control group, automation users by $\hat{\beta} = \getval{s1_auto_qual_overall_itt}$~SD ($p = \getval{s1_auto_qual_overall_pval}$) and mixed users by $\hat{\beta} = \getval{s1_mixed_qual_overall_itt}$~SD ($p = \getval{s1_mixed_qual_overall_pval}$), compared with $\hat{\beta} = \getval{s1_aug_qual_overall_itt}$~SD for augmentation users.
	
	A week later, when students work without AI, augmentation users' test-score difference remains positive ($\hat{\beta} = \getval{s2_aug_test_itt}$~SD, $p = \getval{s2_aug_test_pval}$) and does not change significantly between sessions, and their essay-quality difference is positive but imprecise in both sessions ($\hat{\beta} = \getval{s1_aug_qual_overall_itt}$~SD and $\getval{s2_aug_qual_overall_itt}$~SD). Mixed users' Session Two differences are positive but imprecise ($\hat{\beta} = \getval{s2_mixed_test_itt}$~SD, $p = \getval{s2_mixed_test_pval}$, on the test and $\hat{\beta} = \getval{s2_mixed_qual_overall_itt}$~SD, $p = \getval{s2_mixed_qual_overall_pval}$, on essay quality). In contrast, automation users' advantages disappear: their essay-quality difference falls by \getval{fade_auto_overall_fall}~SD ($p = \getval{fade_auto_overall_p}$) to $\hat{\beta} = \getval{s2_auto_qual_overall_itt}$~SD, and their test-score difference is negative but imprecise ($\hat{\beta} = \getval{s2_auto_test_itt}$~SD, $p = \getval{s2_auto_test_pval}$). The fall for automation users exceeds the change for augmentation users by \getval{fade_dd_overall_gap}~SD ($p = \getval{fade_dd_overall_p}$). While descriptive, these results suggest that the durability of learning gains depends on how students use AI, and that the gains are less durable for students who had AI do the work for them.
	
	\subsubsection{Evidence from Other Experiments.}
	
	To assess whether the use-type pattern extends beyond our experiment, and in particular to settings that randomize how students use AI, we turn to other experiments. We identified six experiments that randomize how learners are steered to use the same AI tool: a tutoring prompt that withholds answers \citep{bastani.etal2025}, a scaffolded tutor \citep{bassner.etal2026}, a structured sequence of prompts \citep{hou.etal2026}, written guidance \citep{gallegos2026, franco.etal2026}, and a note-taking requirement \citep{kreijkes.etal2026}. We also identified four studies that compare learners by their observed use of the same tool, as we do \citep{liu.etal2026, shen.tamkin2026, franco.etal2026, stromberg.etal2026}.
	
	The divergence between assisted output and unaided learning that we find for automation users recurs across settings. Steering learners toward augmentation raises subsequent test scores in four of the six experiments: by \getval{lit_ws_bastani_diff}~SD in \citet{bastani.etal2025}, \getval{lit_ws_gallegos_diff}~SD in \citet{gallegos2026}, \getval{lit_ws_hou_diff}~SD in \citet{hou.etal2026}, and \getval{lit_ws_kreijkes_ret_diff} to \getval{lit_ws_kreijkes_comp_diff}~SD in \citet{kreijkes.etal2026}. The exceptions are a scaffolded tutor whose knowledge gains are indistinguishable from ChatGPT's \citep{bassner.etal2026} and written guidance that reduces question pasting without raising exam scores \citep{franco.etal2026}. Comparisons by observed use of the same tool point the same way as ours. Participants who ask for hints score \getval{lit_ws_liu_diff}~SD higher on unaided problems than those who ask for answers \citep{liu.etal2026}, programmers who ask the assistant to explain generated code perform better than those who delegate or debug by repeated queries \citep{shen.tamkin2026}, heavy copy-pasters score below the control mean \citep{franco.etal2026}, and the exam-score loss from AI among Chinese secondary students is concentrated among the fastest homework finishers, while users who take as long as non-AI students score the same \citep{stromberg.etal2026}. Taken together, the evidence supports the model's distinction between AI that produces the output and AI that raises the return to the student's own effort.
	
	\section{Beliefs About AI's Impact on Learning} \label{sec:beliefs}
	
	Students' beliefs about AI's effect on learning may shape their adoption decisions. In surveys, about half of students say AI chatbots improve their learning \citep{stohr.etal2024, ravselj.etal2025}. Yet these beliefs may not track actual effects: students hold biased beliefs in other educational settings \citep{jensen2010perceived, wiswall2015determinants}, and even experienced professionals misjudge AI's productivity effects.\footnote{\citet{becker.etal2025} find that experienced open-source developers predicted AI tools would speed them up by 24 percent and, after the study, still believed AI had accelerated their work by roughly 20 percent. The actual measured effect was a 19 percent \textit{slowdown}.}
	
	We test the accuracy of students' beliefs---and how treatment shapes them---in two ways. First, after the Session Two test, students estimate (1) how many questions they answered correctly, (2) how many they would have answered correctly under the opposite treatment condition, (3) how many questions other students in their group answered correctly, and (4) the same for students in the opposite group. From these we compute perceived treatment effects on own and others' performance, in percentage points, and compare them with the actual estimates from Section~\ref{sub:tests}. Second, we analyze responses to an open-ended question about the effects of generative AI on student learning in college. We validate students' open-ended responses against their AI usage patterns (Appendix~\ref{app:open_ended_validation}) and code each response as a causal graph of the mechanisms it describes (Appendix~\ref{app:open_ended_method}).
	
	\subsection{Beliefs About AI's Effect on Test Scores}
	
\begin{figure}[H]
		\caption{Actual and Perceived Treatment Effects on Test Performance}\label{fig:beliefs}
		\centering
		\includegraphics[width=0.7\textwidth]{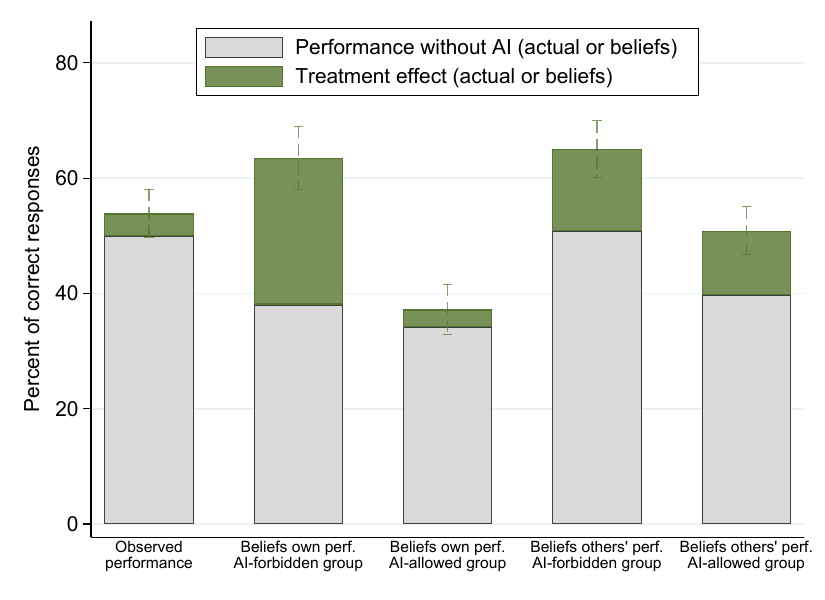}
		{\footnotesize
			\singlespacing \justify
			
			\textit{Notes:} This figure shows actual and perceived treatment effects of AI access on test performance. The first bar shows observed performance: gray is the control mean, and green is the estimated treatment effect. The remaining bars show students' beliefs about own and others' performance, separately by treatment group: gray is the believed score without AI and green the perceived AI gain. AI-forbidden and AI-allowed belief bars are group means. Vertical bars represent 95 percent confidence intervals. \par
			
		}
	\end{figure}

	Control students overestimate the effect of AI on their test performance, whereas treated students predict an effect close to the one we estimate (Figure~\ref{fig:beliefs}). Both groups give similar predictions of their own raw score (control \getval{belief_own_control_pct} percent correct, treated \getval{belief_own_treat_pct} percent, both below the actual mean of \getval{s2_actual_pct_correct} percent); what differs is what they attribute to AI. Control students predict an AI gain for themselves of \getval{belief_forbidden_self_pred_pp} pp, \getval{belief_forbidden_self_ratio} times the actual effect of \getval{s2_test_itt_pp} pp. Treated students, most of whom used AI in Session One, predict only \getval{belief_allowed_self_pred_pp} pp, close to the effect we estimate. The lower average perceived effect among treated students reflects two shifts in the distribution of beliefs relative to control students (Appendix Figure~\ref{fig:hist_perceived_te}): more mass at zero (no perceived effect) and less mass at large positive values (five or more questions). Predictions about others' performance show the same pattern in muted form: control students predict \getval{belief_forbidden_other_pred_pp} pp and treated \getval{belief_allowed_other_pred_pp} pp, both still inflated relative to the actual effect. %
	
	\subsection{Students' Mental Models of AI and Learning}
	
\begin{figure}[H]
		\caption{The Average Narrative About AI's Effect on Learning} \label{fig:avg_narrative}
		\centering
		\includegraphics[width=\linewidth]{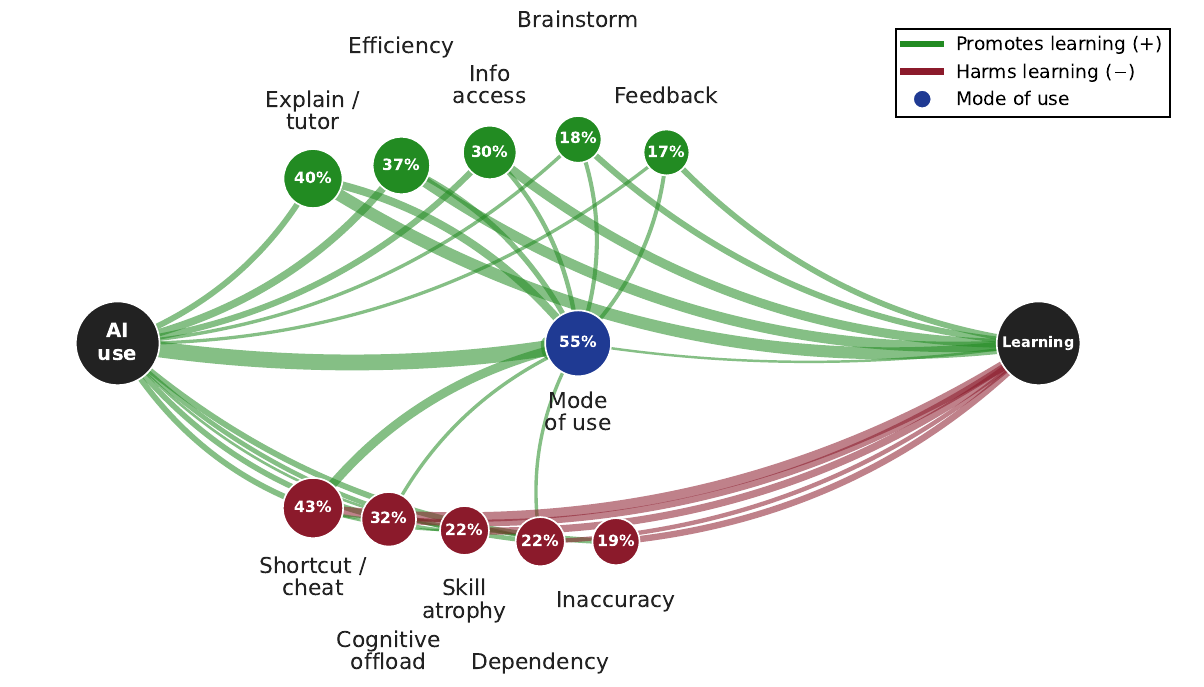}
		{\footnotesize
			\singlespacing \justify
			
			\textit{Notes:} This figure shows the aggregate signed causal graph across the \getval{narr_n_coded} open-ended responses codable as a narrative, constructed following \citet{andre.etal2026}. We code each student's answer to the question ``In your opinion, how does generative AI (e.g., ChatGPT) affect student learning in college? Please explain your reasoning'' as a causal graph from AI use (left) to learning (right) through the mechanisms the student names; the figure overlays these graphs. Node size is proportional to the share of students who name a mechanism, and link width to the share drawing the link; we omit mechanisms below 10 percent and links below 2 percent. Green links denote mechanisms described as promoting learning, red links mechanisms that harm it, and navy the mode-of-use node. Appendix~\ref{app:open_ended_method} describes the coding and its reliability. \par
		}
	\end{figure}

	From these predictions we learn how much students expect AI to help, but not the mechanisms through which they think it works. To recover these ``mental models,'' we asked an open-ended question on the exit survey: ``In your opinion, how does generative AI (e.g., ChatGPT) affect student learning in college? Please explain your reasoning.'' Following \citet{andre.etal2026}, we code each short narrative as a causal graph running from AI use to learning through the mechanisms the student names (see Appendix~\ref{app:open_ended_method} for the coding procedure and its reliability, and Appendix~\ref{app:narratives_results} for additional results). We aggregate students' mental models into a single causal graph (Figure~\ref{fig:avg_narrative}), in which each node is one such mechanism (sized by how often students name it), and each link is green where they describe the mechanism as promoting learning and red where they describe it as harming learning.
	
	Most students reason about the effects of AI on learning through the same augmentation-versus-automation distinction along which outcomes differ in Section~\ref{sub:aug_auto}. Among students naming any learning channel, \getval{narr_both_pct} percent name both one through which AI helps and one through which it harms. The most common single node, present in \getval{narr_usemode_pct} percent of coded responses, is mode of use, through which the helping and harming channels branch according to how AI is used; as one student put it, AI is ``a double-edged sword, with each individual student's use determining whether it is helpful or hurtful for their education.'' The two most common channels are the two sides of this distinction: AI explaining concepts and tutoring, as an augmentation channel (\getval{narr_explain_pct} percent of coded responses), and AI shortcutting the work, as an automation channel (\getval{narr_shortcut_pct} percent).\footnote{Appendix Table~\ref{tab:narr_covariates} reports how students' narratives about AI and learning vary with their characteristics. Students with GPA at or above the median describe more channels through which AI affects learning. Students at or above the median in self-assessed AI proficiency are more likely to frame AI as augmentation and less likely to frame it as automation.} Taken together, students' reasoning matches our estimates: they expect AI to raise their learning, their average estimate of the magnitude moves toward ours once they have access to AI, and they locate the gains in AI that augments their work rather than does it for them.
	
	\section{Conclusion}
	
	We provide causal evidence on how access to generative AI affects student learning. In a randomized experiment with undergraduate students, AI access produces learning gains, and a smaller gain remains one week later. AI access also changes how students approach learning a new topic: they spend less time writing and more time reading and searching for information, and report higher enjoyment. Taken together, these results provide proof of concept that off-the-shelf AI can improve learning, with outcomes that differ by how students choose to use it.
	
	We do not view our findings as implying that widespread AI adoption will raise learning overall. We find no effect on total learning time (likely because the lab setting offered few competing uses of time), so we estimate the learning gains holding time on task fixed. In ordinary academic work, students choose how long to spend on each task, and many likely use AI to save time. Whether AI raises or lowers total learning therefore depends on how students reallocate the time they save and on whether learning-productivity gains more than compensate for any reduction in time spent learning.
	
	Our findings have implications for incentive design in higher education. In our model, using AI as a tutor raises knowledge, while delegating the essay to AI displaces the learning that writing would have produced. Before AI, students who cared only about their grade still learned, because earning a good grade required reading, searching, and writing, the activities through which students build knowledge. With AI, students who place less value on learning are more likely to delegate. That value combines students' own interest in learning with external rewards for it, such as grades and the returns to skill, so how students use AI depends on the incentives they face. These incentives differ across colleges and across students. For example, grade inflation weakens the link between effort and grades \citep{babcock2010, bowden.etal2026}, which may push students to seek distinction outside their courses \citep{claybaugh2025} and to delegate coursework to AI to free up time. Similarly, students who view college primarily as a signal rather than as human-capital accumulation have more reason to delegate, since learning matters less for their perceived returns to education. How colleges can design incentives under which AI raises learning is a question for future research.
	
	\clearpage
	\section*{Figures and Tables}
	
	\begin{figure}[H]
		\caption{Experimental Session Timelines} \label{fig:timelines}
		\centering
		
		\textbf{Panel A. Session One timeline} \vspace{.2cm}
		
		\begin{tikzpicture}[scale=1.2, every node/.style={font=\small}]
			\draw[thick,->] (0,0) -- (12,0);
			\node[below] at (0,-0.3) {0 min};
			\node[below] at (12,-0.3) {60 min};
			
			\draw[thick] (1,0) -- (1,0.5);
			\draw[fill=gray!20] (0,0.5) rectangle (2,1.5);
			\node[align=center] at (1,1) {Welcome \&\\Instructions};
			
			\draw[thick] (3,0) -- (3,0.5);
			\draw[fill=bluecomment!20] (2.2,0.5) rectangle (3.7,1.5);
			\node[align=center] at (2.95,1) {Baseline\\Test};
			
			\draw[thick] (6.2,0) -- (6.2,0.5);
			\draw[fill=purple!20] (4.2,0.5) rectangle (8.2,1.5);
			\node[align=center] at (6.2,1) {Learning Phase\\(35 min max)};
			
			\draw[thick] (9.5,0) -- (9.5,0.5);
			\draw[fill=yellowcomment!20] (8.4,0.5) rectangle (10.4,1.5);
			\node[align=center] at (9.5,1) {Post-Learning\\Survey};
			
			\draw[thick] (11.5,0) -- (11.5,0.5);
			\draw[fill=redcomment!20] (10.6,0.5) rectangle (12.35,1.5);
			\node[align=center] at (11.5,1) {Post-Learning\\Test};
		\end{tikzpicture}
		
		\vspace{.5cm}
		
		\textbf{Panel B. Session Two timeline} \vspace{.2cm}
		
		\begin{tikzpicture}[scale=1.1, every node/.style={font=\small}]
			\draw[thick,->] (0,0) -- (12,0);
			\node[below] at (0,-0.3) {0 min};
			\node[below] at (12,-0.3) {45 min};
			
			\draw[thick] (1,0) -- (1,0.5);
			\draw[fill=gray!20] (0,0.5) rectangle (2,1.5);
			\node[align=center] at (1,1) {Welcome \& \\Instructions};
			
			\draw[thick] (4.5,0) -- (4.5,0.5);
			\draw[fill=bluecomment!20] (2.5,0.5) rectangle (6.5,1.5);
			\node[align=center] at (4.5,1) {Endline Test\\(10 Questions)};
			
			\draw[thick] (8.5,0) -- (8.5,0.5);
			\draw[fill=redcomment!20] (6.8,0.5) rectangle (10.2,1.5);
			\node[align=center] at (8.5,1) {Essay \\(20 min max)};
			
			\draw[red, dashed, thick] (2.3,0.3) rectangle (10.4,1.7);
			
			\node[red, above] at (6.35,1.7) {\textit{Randomized order}};
			
			\draw[thick] (11.2,0) -- (11.2,0.5);
			\draw[fill=yellowcomment!20] (10.5,0.5) rectangle (11.9,1.5);
			\node[align=center] at (11.2,1) {Exit\\Survey};
		\end{tikzpicture}
		
		\vspace{.5cm}
		
		\begin{singlespace}
			\footnotesize \justifying \noindent 
			\textit{Notes:} This figure shows the timeline of the two experimental sessions. Panel A shows the structure of Session One, during which students were randomly assigned to AI-allowed or AI-forbidden conditions. Panel B shows Session Two, conducted approximately one week later (mean of \getval{avg_session_gap} days), during which all students completed tasks without AI access. The red dashed box in Panel B indicates that the order of the Endline Test and the Essay was randomized across students. The Baseline Test and Post-Learning Test in Session One each contained five multiple-choice questions. The Endline Test in Session Two contained ten multiple-choice questions. \par
			
		\end{singlespace}
		
	\end{figure}
	
	\begin{figure}[H]
		\caption{Types of Generative AI Use During the Learning Phase}\label{fig:ai_usage_type}
		\centering
		\begin{subfigure}[t]{.48\textwidth}
			\caption*{Panel A. Self-reported usage}
			\centering
			\includegraphics[width=\linewidth]{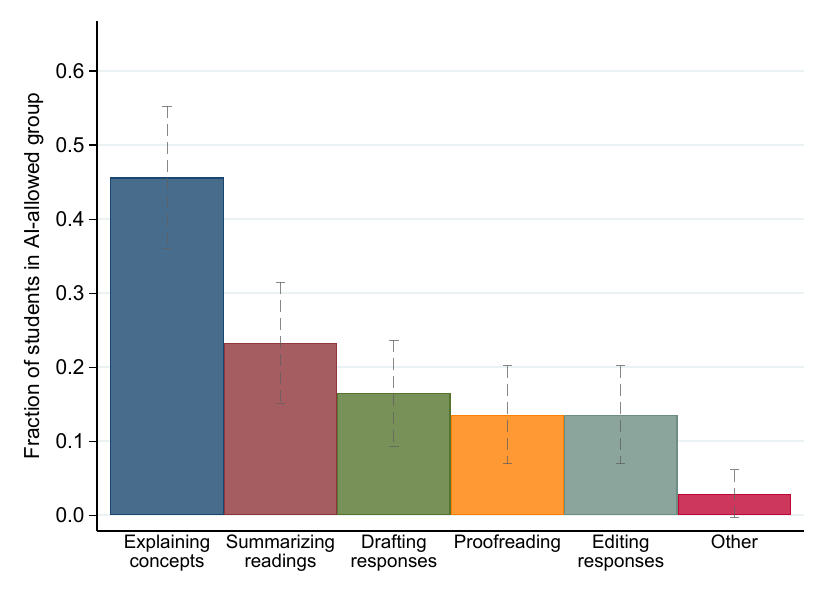}
		\end{subfigure}
		\hfill
		\begin{subfigure}[t]{0.48\textwidth}
			\caption*{Panel B. ChatGPT logs}
			\centering
			\includegraphics[width=\linewidth]{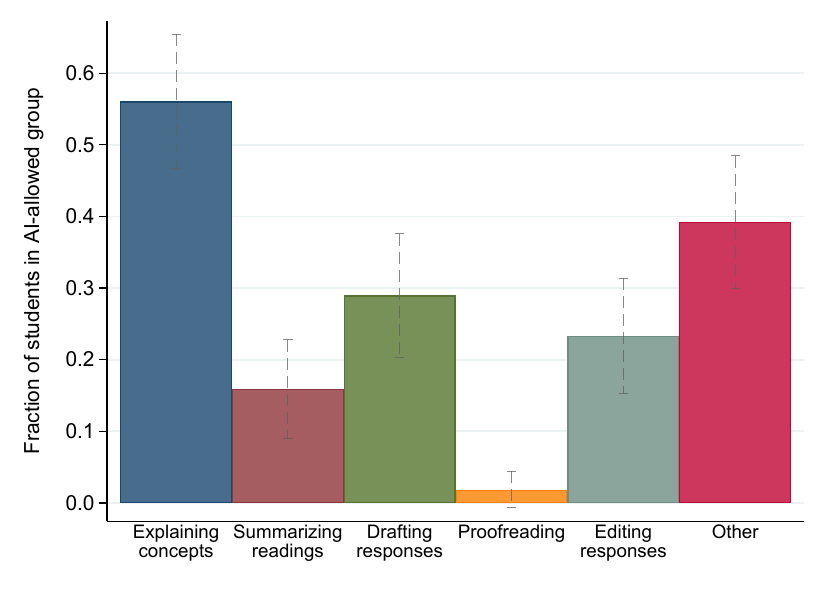}
		\end{subfigure}
		{\footnotesize
			\singlespacing \justify
			
			\textit{Notes:} This figure shows AI usage types among treated students. Panel~A presents self-reported usage types collected in the exit survey; students could select multiple categories. Panel~B is constructed from the actual ChatGPT conversation logs. We first use an LLM to classify each student prompt into one of six categories: explaining concepts, summarizing readings, drafting responses, proofreading, editing responses, and other uses (see Appendix~\ref{app:chatgpt_classification} for the classification procedure). We then construct student-level indicators: for each category, the indicator equals one if any of the student's prompts fell into that category. Students who were assigned to the AI-allowed condition but did not use the provided ChatGPT account are coded as zero for all categories. \par
			
		}
	\end{figure}
	
	\clearpage
	\begin{figure}[H]
		\caption{Effect Sizes Across AI-and-Learning Experiments}\label{fig:literature_comparison}
		\centering
		\includegraphics[width=\linewidth]{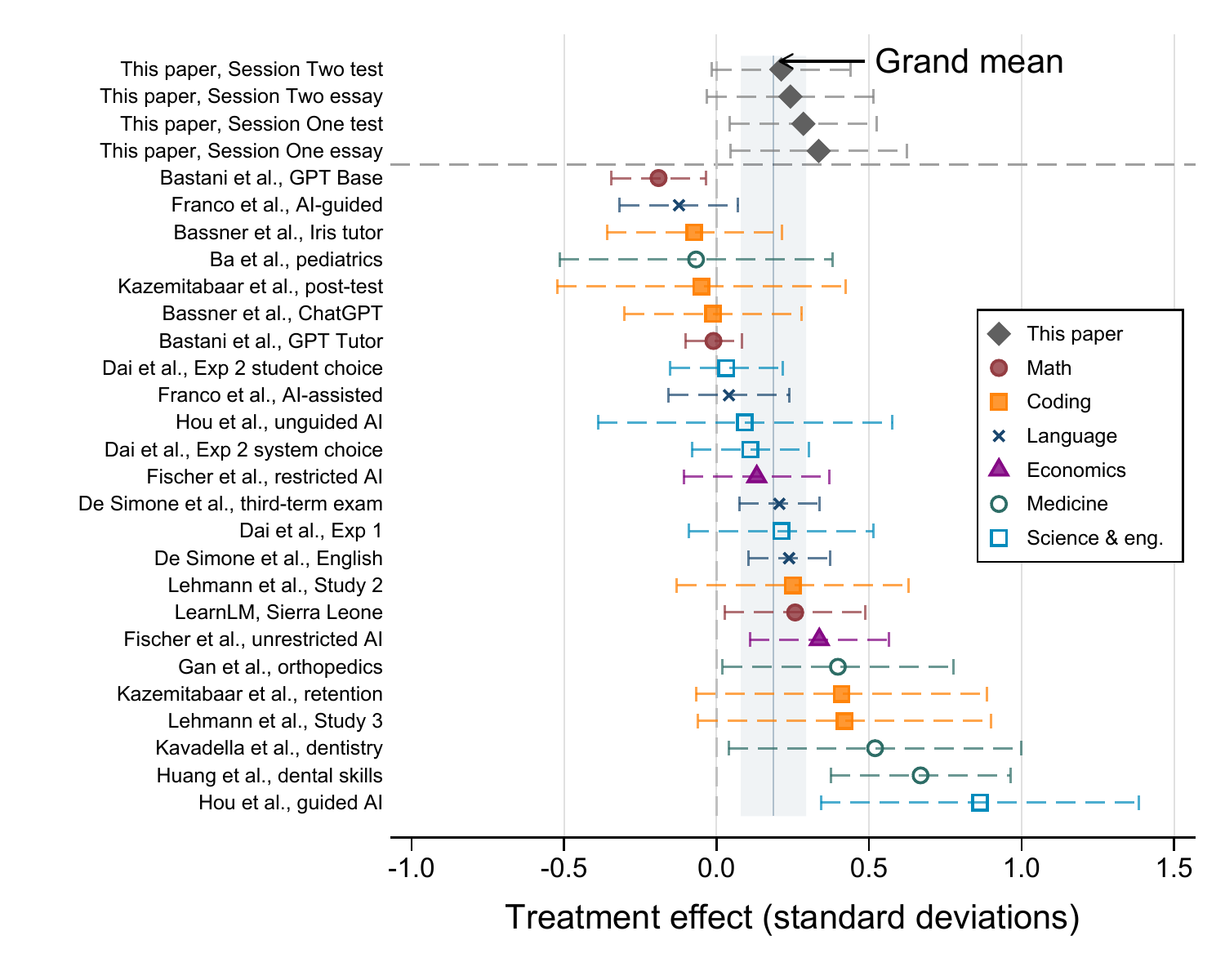}
		{\footnotesize
			\singlespacing \justify
			
			\textit{Notes:} This figure compares treatment effect sizes (in standard deviations) across randomized experiments that provide students with AI tools during a learning task and measure performance on unassisted assessments. Marker shapes indicate the outcome domain: diamonds for this paper, circles for math, squares for coding, crosses for language learning, triangles for economics, hollow circles for medicine, and hollow squares for science and engineering. Our estimates include both knowledge test scores and essay quality for Session One (immediate) and Session Two (retention), shown above the dashed separator line. Literature estimates come from the \getval{lit_n_studies_cl} studies described in Appendix~\ref{app:literature_comparison} and summarized in Appendix Table~\ref{tab:literature_summary}. Horizontal lines show 95 percent confidence intervals. The shaded band and vertical line show the random-effects grand mean across the other studies and its 95 percent confidence interval, estimated via the \citet{dersimonian.laird1986} method after aggregating each study's estimates by inverse-variance weighting (our own estimates are excluded from the pool). The dashed vertical line marks zero. \par
			
		}
	\end{figure}
	
	\clearpage
	
	\clearpage
	
	\clearpage
	\begin{table}[H]
		\caption{\centering Summary Statistics} \label{tab:summ}
		{\footnotesize
			\begin{centering}
				\begin{tabular}{lccc}
					\addlinespace\addlinespace \midrule
					& Signed up & Attended Session One & Attended Both Sessions \\ 
					& (1) & (2) & (3) \\ \midrule
					
					\multicolumn{4}{l}{\hspace{-1em}\textbf{Panel A. Demographic characteristics}} \\ 
					Age &20.211 &20.147 &20.127 \\
Male &0.359 &0.351 &0.351 \\
White &0.498 &0.500 &0.510 \\
International student &0.273 &0.269 &0.277 \\
Public high school &0.541 &0.543 &0.539 \\
 \midrule
					
					\multicolumn{4}{l}{\hspace{-1em}\textbf{Panel B. Academic background}} \\ 
					Took SAT/ACT &0.760 &0.778 &0.777 \\
SAT score (conditional) &1387.593 &1385.969 &1386.571 \\
SAT score (including predicted) &1379.218 &1375.322 &1375.656 \\
College GPA &3.680 &3.683 &3.686 \\
Freshman &0.359 &0.374 &0.385 \\
Already declared major &0.780 &0.780 &0.773 \\
Natural-science major &0.301 &0.308 &0.312 \\
Social-science major &0.367 &0.365 &0.356 \\
Humanities/Arts/Languages major &0.105 &0.100 &0.098 \\
Study hours per week &16.890 &16.621 &16.585 \\
 \midrule            
					
					\multicolumn{4}{l}{\hspace{-1em}\textbf{Panel C. Attendance and comprehension}} \\ 
					Attended Session One &0.824 &1.000 &1.000 \\
Attended Session Two &0.801 &0.972 &1.000 \\
Attended Session Two (conditional on S1) &0.972 &0.972 &1.000 \\
Attended both sessions &0.801 &0.972 &1.000 \\
Days between sessions &6.939 &6.971 &6.980 \\
Comprehension score (0--5) &4.763 &4.763 &4.761 \\
 \midrule
					
					\multicolumn{4}{l}{\hspace{-1em}\textbf{Panel D. Experimental conditions}} \\ 
					Assigned blockchain topic &0.365 &0.365 &0.371 \\
Assigned CRISPR topic &0.327 &0.327 &0.322 \\
Assigned carbon capture topic &0.308 &0.308 &0.307 \\
Baseline self-assessed knowledge (0--10) &1.517 &1.517 &1.512 \\
Fraction correct (baseline) &0.303 &0.303 &0.308 \\
Shared info between sessions &0.059 &0.059 &0.059 \\
Looked up info about topic &0.049 &0.049 &0.049 \\
Studied topic for Session Two &0.000 &0.000 &0.000 \\
 \midrule
					
					\(N\) (Students) &256 &211 &205 \\
 \midrule \addlinespace\addlinespace
				\end{tabular}
				\par\end{centering}
			
			\begin{singlespace}\vspace{-0.5cm}
				\noindent\justify \textit{Notes:} This table presents summary statistics for three groups of students: students who completed the sign-up survey and scheduled a Session One slot (column~1), those who attended Session One (column~2), and those who attended both sessions (column~3). GPA is on the $0$--$4$ scale. SAT score (conditional) is the mean among students with an SAT score or an ACT score converted to the SAT scale, and SAT score (including predicted) adds the imputed scores described in Appendix~\ref{app:sat_imputation}. \par
			\end{singlespace}
		}
	\end{table}
	
	\begin{table}[H]{\footnotesize
			\begin{center}
				\caption{Balance of Baseline Characteristics by Treatment Assignment} \label{tab:bal}
				\begin{tabular}{lcccrll}
					\midrule
					& \multicolumn{2}{c}{Treatment group:} & & \multicolumn{2}{c}{Difference:} \\ \cmidrule{2-3} \cmidrule{5-6}
					
					& AI-forbidden   & AI-allowed & & $\hat{\beta}$ & (SE)  & \\
					& (1)            & (2)        & & (3)     &  (4)  &  \\
					\midrule
					
					\multicolumn{5}{l}{\hspace{-1em} \textbf{Panel A. Demographic characteristics}}  \\
					Age & 20.016 & 20.400 & & 0.382\rlap{\sym{*}} & (0.227) \\
Male & 0.405 & 0.315 & & $-$0.091 & (0.060) \\
White & 0.528 & 0.469 & & $-$0.059 & (0.063) \\
International student & 0.286 & 0.260 & & $-$0.027 & (0.056) \\
Public high school & 0.540 & 0.543 & & 0.004 & (0.063) \\
 
					\midrule
					
					\multicolumn{5}{l}{\hspace{-1em} \textbf{Panel B. Academic background}}  \\
					Took SAT/ACT & 0.790 & 0.730 & & $-$0.059 & (0.054) \\
SAT score (conditional) & 1392.371 & 1382.554 & & $-$7.862 & (21.039) \\
SAT score (including predicted) & 1377.516 & 1380.868 & & 2.615 & (18.855) \\
College GPA & 3.694 & 3.666 & & $-$0.025 & (0.038) \\
Freshman & 0.357 & 0.362 & & 0.007 & (0.061) \\
Already declared major & 0.825 & 0.734 & & $-$0.092\rlap{\sym{*}} & (0.052) \\
Natural-science major & 0.302 & 0.300 & & $-$0.002 & (0.058) \\
Social-science major & 0.413 & 0.323 & & $-$0.093 & (0.061) \\
Humanities/Arts/Languages major & 0.111 & 0.100 & & $-$0.009 & (0.038) \\
Study hours per week & 16.864 & 16.915 & & 0.146 & (1.200) \\
 
					\midrule
					
					\multicolumn{5}{l}{\hspace{-1em} \textbf{Panel C. Attendance and comprehension}}  \\
					Attended Session One & 0.825 & 0.823 & & $-$0.006 & (0.048) \\
Attended Session Two & 0.810 & 0.792 & & $-$0.020 & (0.050) \\
Attended Session Two (conditional on S1) & 0.981 & 0.963 & & $-$0.017 & (0.022) \\
Attended both sessions & 0.810 & 0.792 & & $-$0.020 & (0.050) \\
Days between sessions & 6.966 & 6.913 & & $-$0.036 & (0.155) \\
Comprehension score (0--5) & 4.760 & 4.766 & & 0.001 & (0.065) \\
 
					\midrule
					
					\multicolumn{5}{l}{\hspace{-1em} \textbf{Panel D. Experimental conditions}}  \\
					Assigned blockchain topic & 0.375 & 0.355 & & $-$0.023 & (0.066) \\
Assigned CRISPR topic & 0.327 & 0.327 & & 0.002 & (0.064) \\
Assigned carbon capture topic & 0.298 & 0.318 & & 0.021 & (0.062) \\
Baseline self-assessed knowledge (0--10) & 1.558 & 1.477 & & $-$0.086 & (0.251) \\
Fraction correct (baseline) & 0.306 & 0.301 & & $-$0.007 & (0.038) \\
 
					\midrule
					
					\(N\) (Students) & 126 & 130 & & 256 &     \\
 
				\end{tabular}
			\end{center}
			\begin{singlespace}  \vspace{-.5cm}
				\noindent \justify \textit{Notes:} This table shows average student characteristics by treatment assignment. Students in the AI-allowed group could use generative AI tools during the learning phase; students in the AI-forbidden group could not. Panels A and B report characteristics measured before treatment assignment. Panel C reports session attendance and instruction comprehension, realized after assignment; Panel D reports the randomly assigned essay topics and baseline knowledge measures collected at the start of Session One, before the learning phase. Heteroskedasticity-robust standard errors in parentheses.  $^{***}$ $p < 0.01$; $^{**}$ $p < 0.05$; $^{*}$ $p < 0.10$. \par
				
			\end{singlespace}
		}
	\end{table}
	
	\clearpage
	\begin{table}[H]{\footnotesize
			\begin{center}
				\caption{The First-Stage Impact of Generative AI Access on AI Usage}
				\label{tab:first_stage}
				\newcommand\w{1.8}
				\begin{tabular}{l@{}lR{\w cm}@{}L{0cm}R{\w cm}@{}L{0.45cm}}
					\midrule
					&& Control && \\
					Outcome && mean && Effect ($\hat{\beta}$) \\
					&& (1) && (2) \\\midrule
					\multicolumn{5}{l}{\hspace{-1em}\textbf{Panel A. Revealed measures}} \\ \addlinespace
					Used ChatGPT account&&0.000&&0.659&\sym{***}\\
&&&&(0.049&)\\

					Number of prompts&&0.000&&4.073&\sym{***}\\
&&&&(0.454&)\\

					Number of conversations&&0.000&&0.759&\sym{***}\\
&&&&(0.059&)\\

					\midrule
					\multicolumn{5}{l}{\hspace{-1em}\textbf{Panel B. Self-reported measures}} \\ \addlinespace
					Used any AI&&0.010&&0.609&\sym{***}\\
&&&&(0.050&)\\

					Used ChatGPT&&0.010&&0.602&\sym{***}\\
&&&&(0.050&)\\

					Used other AI model&&0.010&&0.029&\\
&&&&(0.022&)\\

					\(N\) (Students)&&104&&211&\\
 \midrule
				\end{tabular}
			\end{center}
			
			\begin{singlespace}\vspace{-0.5cm}
				\footnotesize\noindent\justify
				\textit{Notes:} This table reports first-stage effects of AI-access assignment on measures of AI usage. Each row reports the effect of being assigned to the AI-allowed group on the indicated measure. Panel A presents revealed measures: Used ChatGPT account equals one if at least one prompt was recorded in the ChatGPT account provided to the student; Number of prompts is the total number of prompts sent to the provided ChatGPT account during the session; Number of conversations counts the number of distinct ChatGPT conversations (i.e., separate chat threads) initiated during the session. Panel B presents self-reported measures collected in Session Two. All specifications include controls selected through a double-lasso procedure \citep{belloni.etal2014}. Heteroskedasticity-robust standard errors in parentheses. $^{***}$ $p < 0.01$; $^{**}$ $p < 0.05$; $^{*}$ $p < 0.10$. \par
				
		\end{singlespace}}
	\end{table}
	
	\clearpage
	\begin{table}[H]{\footnotesize
			\begin{center}
				\caption{The Impact of Generative AI Access on Test Scores}
				\label{tab:test_score_long}
				\newcommand\w{1.5}
				\begin{tabular}{l@{}lR{\w cm}@{}L{0cm}R{\w cm}@{}L{0.45cm}R{\w cm}@{}L{0.45cm}R{\w cm}@{}L{0cm}R{\w cm}@{}L{0.45cm}R{\w cm}@{}L{0.45cm}}
					\midrule
					&& \multicolumn{5}{c}{Session One} && \multicolumn{5}{c}{Session Two} \\ \cmidrule{3-7} \cmidrule{9-13}
					&& Control &&  &&  && Control &&  &&  \\
					Outcome && mean && ITT && TOT && mean && ITT && TOT \\
					&& (1) && (2) && (3) && (4) && (5) && (6) \\\midrule
					Self-assessed&&4.788&&$-$0.098&&$-$0.142&&3.833&&0.223&&0.310&\\
&&&&(0.203&)&(0.296&)&&&(0.193&)&(0.268&)\\

					Fraction correct&&0.563&&0.072&\sym{**}&0.104&\sym{**}&0.500&&0.039&\sym{*}&0.055&\sym{*}\\
&&&&(0.031&)&(0.045&)&&&(0.021&)&(0.031&)\\

					Test score (SD)&&0.000&&0.285&\sym{**}&0.408&\sym{**}&0.000&&0.213&\sym{*}&0.300&\sym{*}\\
&&&&(0.123&)&(0.176&)&&&(0.116&)&(0.166&)\\

					\midrule
					\(N\) (Students)&&104&&210&&210&&102&&205&&205&\\

					\midrule
				\end{tabular}
			\end{center}
			
			\begin{singlespace}\vspace{-0.5cm}
				\footnotesize\noindent\justify
				\textit{Notes:} This table reports treatment effects of AI access on learning outcomes in Session One, immediately after the learning phase, and in Session Two, one week later. Students took both tests without access to any resources. The AI-allowed group had AI access only during the Session One learning phase. Each row reports the effect of being assigned to the AI-allowed group on the indicated outcome. Self-assessed knowledge is measured on a 0--10 scale, with 0 indicating ``I know nothing about this topic'' and 10 indicating ``I am an expert.'' Fraction correct is measured on a 0--1 scale. Test score (SD) is the fraction correct standardized to have mean zero and standard deviation one in the control group. Columns 1 and 4 report the control-group mean. Columns 2 and 5 report intent-to-treat (ITT) estimates. Columns 3 and 6 report treatment-on-the-treated (TOT) estimates from two-stage least squares, instrumenting actual ChatGPT use with random treatment assignment. All specifications include controls selected through a double-lasso procedure \citep{belloni.etal2014}. Heteroskedasticity-robust standard errors in parentheses. $^{***}$ $p < 0.01$; $^{**}$ $p < 0.05$; $^{*}$ $p < 0.10$. \par
				
		\end{singlespace}}
	\end{table}
	
	\clearpage
	\begin{table}[H]{\footnotesize
			\begin{center}
				\caption{Effects of Generative AI Access on Essay Outcomes}
				\label{tab:writ_index_long}
				\newcommand\w{1.25}
				\begin{tabular}{l@{}lR{\w cm}@{}L{0cm}R{\w cm}@{}L{0.45cm}R{\w cm}@{}L{0.45cm}R{\w cm}@{}L{0cm}R{\w cm}@{}L{0.45cm}R{\w cm}@{}L{0.45cm}}
					\midrule
					&& \multicolumn{5}{c}{Session One (with AI)} && \multicolumn{5}{c}{Session Two (without AI)} \\ \cmidrule{3-7} \cmidrule{9-13}
					&& Control &&  &&  && Control &&  &&  \\
					Outcome && mean && ITT && TOT && mean && ITT && TOT \\
					&& (1) && (2) && (3) && (4) && (5) && (6) \\\midrule
					\multicolumn{13}{l}{\hspace{-1em}\textbf{Panel A. AI detection and plagiarism}} \\ \addlinespace
					AI-detected (fraction)&&0.066&&0.222&\sym{***}&0.318&\sym{***}&0.003&&0.007&&0.010&\\
&&&&(0.048&)&(0.066&)&&&(0.010&)&(0.014&)\\

					Plagiarism (fraction)&&0.000&&0.000&&0.000&&0.000&&0.000&&0.000&\\
&&&&(0.000&)&(0.000&)&&&(0.000&)&(0.001&)\\

					\midrule
					\(N\) (Students)&&104&&210&&210&&102&&204&&204&\\

					\midrule
					\multicolumn{13}{l}{\hspace{-1em}\textbf{Panel B. Homogeneity and similarity}} \\ \addlinespace
					Within-group sim.&&0.788&&0.003&&0.004&&0.753&&0.000&&0.001&\\
&&&&(0.007&)&(0.010&)&&&(0.016&)&(0.022&)\\

					Reading material sim.&&0.814&&0.000&&0.000&&0.761&&0.010&&0.014&\\
&&&&(0.009&)&(0.013&)&&&(0.014&)&(0.020&)\\

					\midrule
					\(N\) (Students)&&104&&210&&210&&102&&204&&204&\\

					\midrule
					\multicolumn{13}{l}{\hspace{-1em}\textbf{Panel C. Overall quality}} \\ \addlinespace
					Overall quality (average)&&5.503&&0.453&\sym{**}&0.657&\sym{**}&4.466&&0.327&\sym{*}&0.470&\sym{*}\\
&&&&(0.200&)&(0.290&)&&&(0.188&)&(0.274&)\\

					Quality index (average)&&5.758&&0.344&\sym{*}&0.498&\sym{*}&4.645&&0.301&\sym{*}&0.431&\sym{*}\\
&&&&(0.180&)&(0.262&)&&&(0.173&)&(0.253&)\\

					\midrule
					\(N\) (Students)&&103&&207&&207&&97&&198&&198&\\

					\midrule
				\end{tabular}
			\end{center}
			
			\begin{singlespace}\vspace{-0.5cm}
				\footnotesize\noindent\justify
				
				\textit{Notes:} This table reports treatment effects of AI access on essay outcomes: AI-detection and plagiarism measures, textual similarity, and overall quality. Each row reports the effect of being assigned to the AI-allowed group on the indicated outcome. Panel A reports AI-detection and plagiarism measures: AI-detected (fraction) is the fraction of text classified as AI-generated by the Pangram 4 AI content detector; Plagiarism (fraction) is the fraction of text flagged as plagiarized by Pangram's plagiarism checker. Panel B reports homogeneity and similarity measures: within-group similarity is the average pairwise cosine similarity between a student's essay embedding and all other essays in the same treatment $\times$ topic $\times$ prompt cell; reading material similarity is the cosine similarity between a student's essay embedding and the embedding of the provided reading material. Both are computed using a sentence-embedding model \citep{reimers.gurevych2019} fine-tuned from the MPNet architecture \citep{song.etal2020}. Panel C reports overall quality (the grader's holistic assessment) and the quality index (the average of the five dimension scores: accuracy, evidence, relevance, organization, and writing style), each scored on a 0--10 scale and averaged across human and AI graders; human grades are averaged across each essay's graders before estimation, and Appendix Table~\ref{tab:essay_quality_dims} reports the five dimensions separately. The number of observations is reported below each panel. Columns 1 and 4 report the control-group mean. Columns 2 and 5 report intent-to-treat (ITT) estimates. Columns 3 and 6 report treatment-on-the-treated (TOT) estimates from two-stage least squares, instrumenting actual ChatGPT use with random treatment assignment. All specifications include controls selected through a double-lasso procedure \citep{belloni.etal2014}. Heteroskedasticity-robust standard errors in parentheses. $^{***}$ $p < 0.01$; $^{**}$ $p < 0.05$; $^{*}$ $p < 0.10$.  \par
				
		\end{singlespace}}
	\end{table}
	
	\begin{table}[H]{\footnotesize
			\begin{center}
				\caption{Time Use and Experience}
				\label{tab:mech_long}
				\newcommand\w{1.5}
				\begin{tabular}{l@{}lR{\w cm}@{}L{0cm}R{\w cm}@{}L{0.45cm}R{\w cm}@{}L{0.45cm}}
					\midrule
					&& Control &&  &&  \\
					Outcome && mean && ITT && TOT \\
					&& (1) && (2) && (3) \\\midrule
					\multicolumn{7}{l}{\hspace{-1em} \textbf{Panel A. Time spent learning}} \\ \addlinespace
					Time spent learning (Qualtrics, minutes)&&32.611&&$-$0.205&&$-$0.292&\\
&&&&(0.678&)&(0.968&)\\

					Time spent learning (self-reported, minutes)&&34.312&&$-$0.532&&$-$0.769&\\
&&&&(0.669&)&(0.966&)\\

					\midrule
					\multicolumn{7}{l}{\hspace{-1em} \textbf{Panel B. Time allocation}} \\ \addlinespace
					Share on research activities (percent)&&44.462&&4.435&\sym{**}&6.403&\sym{**}\\
&&&&(2.245&)&(3.244&)\\

					Share on writing activities (percent)&&53.381&&$-$5.347&\sym{**}&$-$7.719&\sym{**}\\
&&&&(2.407&)&(3.450&)\\

					Share on other activities (percent)&&2.157&&0.911&&1.316&\\
&&&&(1.310&)&(1.883&)\\

					\midrule
					\multicolumn{7}{l}{\hspace{-1em} \textbf{Panel C. Learning experience}} \\ \addlinespace
					Enjoyed learning (0--10)&&5.221&&0.579&\sym{*}&0.847&\sym{*}\\
&&&&(0.310&)&(0.461&)\\

					Found effective (0--10)&&5.000&&0.059&&0.084&\\
&&&&(0.283&)&(0.400&)\\

					\midrule
					\(N\) (Students)&&104&&210&&210&\\

					\midrule
				\end{tabular}
			\end{center}
			
			\begin{singlespace}\vspace{-0.5cm}
				\footnotesize\noindent\justify
				\textit{Notes:} This table reports treatment effects of AI access on total time spent learning (Panel A), how that time is allocated across activities (Panel B), and the learning experience (Panel C). Each row reports the effect of being assigned to the AI-allowed group on the indicated outcome. The two time spent learning rows give the duration of the learning phase in minutes, recorded by Qualtrics and in the post-learning survey, respectively. The share on research, writing, and other activities rows express each category's self-reported minutes as a percentage of the student's total self-reported learning time; these rows report effects in percentage points. Enjoyed learning and Found effective are 0--10 self-reports. All specifications include controls selected through a double-lasso procedure \citep{belloni.etal2014}. Heteroskedasticity-robust standard errors in parentheses. $^{***}$ $p < 0.01$; $^{**}$ $p < 0.05$; $^{*}$ $p < 0.10$. \par
		\end{singlespace}}
	\end{table}
	
	\clearpage
	\begin{table}[H]{\footnotesize
			\begin{center}
				\caption{Outcomes by Type of AI Use: Automation versus Augmentation}
				\label{tab:heterogeneity_ai}
				\newcommand\ww{1.18}
				\begin{tabular}{l@{}lR{\ww cm}@{}L{0.45cm}R{\ww cm}@{}L{0.45cm}R{\ww cm}@{}L{0.45cm}R{\ww cm}@{}L{0.45cm}R{\ww cm}@{}L{0.45cm}R{\ww cm}@{}L{0.45cm}}
					\midrule
					&& \multicolumn{5}{c}{Session One} && \multicolumn{5}{c}{Session Two} \\ \cmidrule{3-7} \cmidrule{9-13}
					&& Test && Overall && Quality && Test && Overall && Quality \\
					&& score && quality && index && score && quality && index \\
					&& (1) && (2) && (3) && (4) && (5) && (6) \\\midrule
					\multicolumn{13}{l}{\hspace{-1em} \textbf{Panel A. All students}} \\ \addlinespace
					Overall (TOT)&&0.408&\sym{**}&0.486&\sym{**}&0.405&\sym{*}&0.300&\sym{*}&0.348&\sym{*}&0.341&\sym{*}\\
&&(0.176&)&(0.215&)&(0.213&)&(0.166&)&(0.203&)&(0.200&)\\

					\(N\) (Students)&&210&&207&&207&&205&&198&&198&\\

					\midrule
					\multicolumn{13}{l}{\hspace{-1em} \textbf{Panel B. By type of AI use}} \\ \addlinespace
					Automation users&&0.293&&1.498&\sym{***}&1.279&\sym{**}&$-$0.244&&$-$0.144&&$-$0.091&\\
&&(0.292&)&(0.535&)&(0.514&)&(0.395&)&(0.451&)&(0.484&)\\

					\(N\) (Students)&&111&&110&&110&&109&&104&&104&\\

					\midrule
					Augmentation users&&0.345&\sym{**}&0.092&&0.062&&0.243&\sym{*}&0.231&&0.142&\\
&&(0.158&)&(0.152&)&(0.160&)&(0.145&)&(0.186&)&(0.179&)\\

					\(N\) (Students)&&150&&147&&147&&147&&140&&140&\\

					\midrule
					Mixed users&&0.581&\sym{***}&0.890&\sym{***}&0.780&\sym{***}&0.282&&0.246&&0.282&\\
&&(0.218&)&(0.312&)&(0.290&)&(0.228&)&(0.247&)&(0.251&)\\

					\(N\) (Students)&&124&&122&&122&&121&&116&&116&\\

					\midrule
				\end{tabular}
			\end{center}
			
			\begin{singlespace}\vspace{-0.5cm}
				\footnotesize\noindent\justify
				\textit{Notes:} This table reports treatment effects on test scores and essay quality, separately for the full sample (Panel A) and by how students used AI (Panel B). Overall (TOT) is the treatment-on-the-treated estimate, instrumenting actual ChatGPT use with random assignment to the AI-allowed group. A research assistant classified treated students' conversation logs into four mutually exclusive categories (Section~\ref{sub:aug_auto}): \textit{Automation}, if the AI did the work \textit{for} the student; \textit{Augmentation}, if the AI worked \textit{with} the student (e.g., explaining concepts or giving feedback on the student's own writing); \textit{Mixed}, if both occurred within the same conversation; and \textit{Other}, if the conversation was off-topic. Students with more than one conversation pool the labels: a student is a mixed user if any conversation is labeled Mixed or if the conversations include both an Automation and an Augmentation label, and otherwise takes the single substantive label; off-topic conversations do not affect the classification. Panel B reports differences between each of three mutually exclusive groups of treated students and the full control group. Treated students whose conversations were classified as Other, or who did not use ChatGPT at all, appear only in Panel A. All six outcomes are standardized by the control-group mean and standard deviation of the corresponding session, so the coefficients are in control-group SD units. All specifications use controls selected by double-lasso on the full sample \citep{belloni.etal2014}, with strata fixed effects. Heteroskedasticity-robust standard errors in parentheses. $^{***}$ $p < 0.01$; $^{**}$ $p < 0.05$; $^{*}$ $p < 0.10$. \par
		\end{singlespace}}
	\end{table}
	
	\clearpage 

\appendix
\begin{center}
	\noindent {\LARGE \textbf{Appendix}}
\end{center}
\label{app:figs}

\setcounter{table}{0}
\setcounter{figure}{0}
\setcounter{equation}{0}	
\renewcommand{\thetable}{A\arabic{table}}
\renewcommand{\thefigure}{A\arabic{figure}}
\renewcommand{\theequation}{A\arabic{equation}}

\definecolor{blockchainblue}{RGB}{37,99,235}
\definecolor{carbongreen}{RGB}{22,163,74}
\definecolor{crisprpurple}{RGB}{147,51,234}

\newcommand{\singleprompt}[3]{%
\begin{tikzpicture}
	\node[
		fill=white,
		rounded corners=6pt,
		draw=#1!30,
		line width=0.4pt,
		inner sep=10pt,
		text width=\linewidth-26pt,
		font=\footnotesize,
	] (prompt) {#3};
	\node[
		circle,
		fill=#1,
		text=white,
		font=\footnotesize\sffamily\bfseries,
		minimum size=18pt,
		inner sep=0pt,
		anchor=south east,
	] at ([xshift=-4pt]prompt.north west) {#2};
\end{tikzpicture}%
}

\newcommand{\topicprompts}[4]{%
\par\medskip\noindent\textbf{#2}\par\nopagebreak
\begin{quote}
\textit{Prompt 1.} #3\\[6pt]
\textit{Prompt 2.} #4
\end{quote}
}

\section{Appendix Figures and Tables}

\begin{figure}[H]
	\caption{Computer Lab Setup with Privacy Dividers}\label{fig:lab}
	\centering
	\includegraphics[width=.75\textwidth]{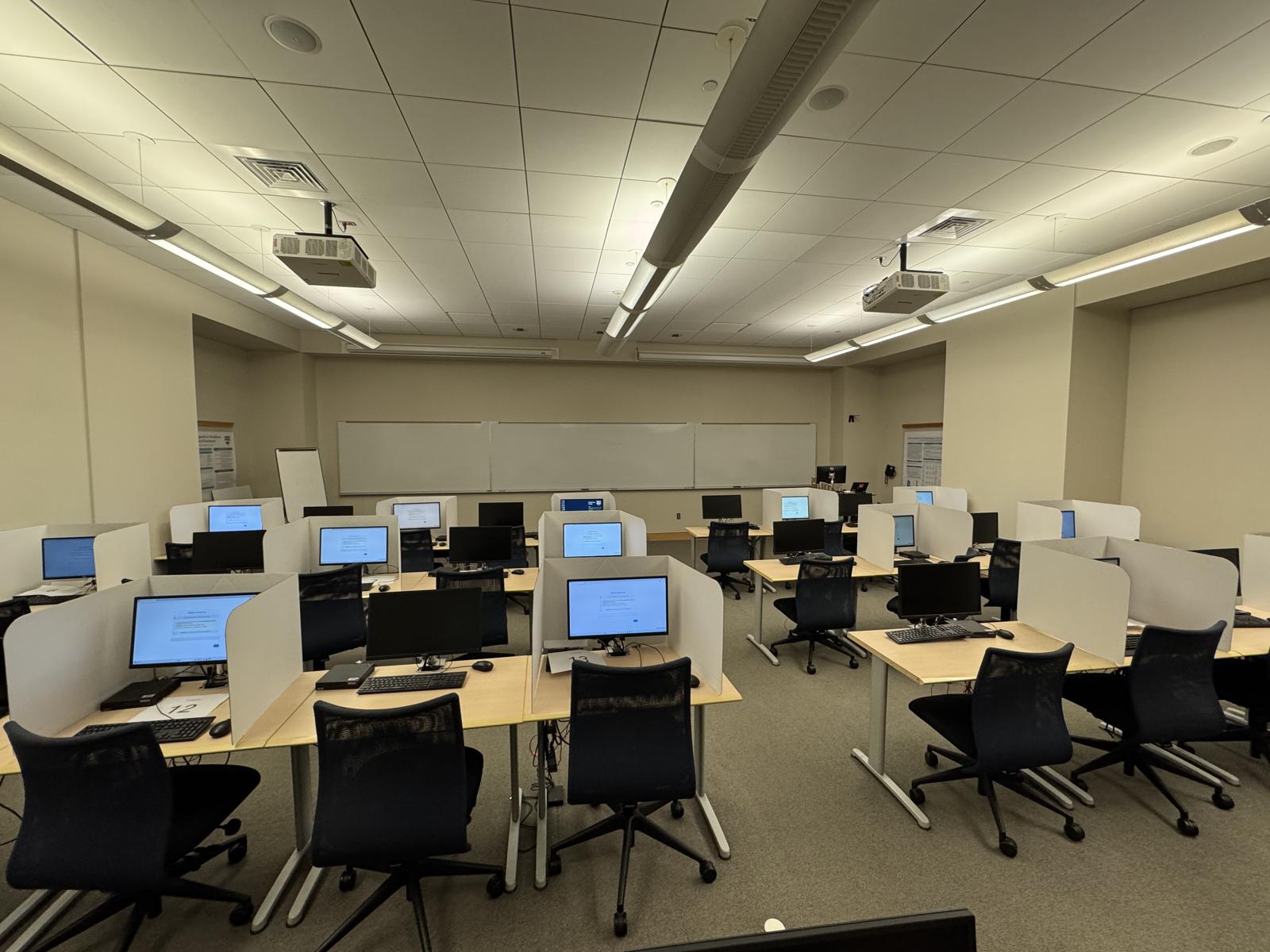}
	{\footnotesize \singlespacing \justify \textit{Notes:} This figure shows the computer lab setup used during experimental sessions. Each workstation was equipped with privacy dividers to minimize distractions and prevent participants from viewing other screens. \par

	}
	
\end{figure}

\begin{figure}[htpb]
	\caption{Treatment Instructions Displayed to Participants}
	\label{fig:treatment_instructions}
	\begin{center}
		\textbf{Panel A. AI-allowed condition}\\[0.3cm]
		\fbox{\includegraphics[width=0.52\linewidth]{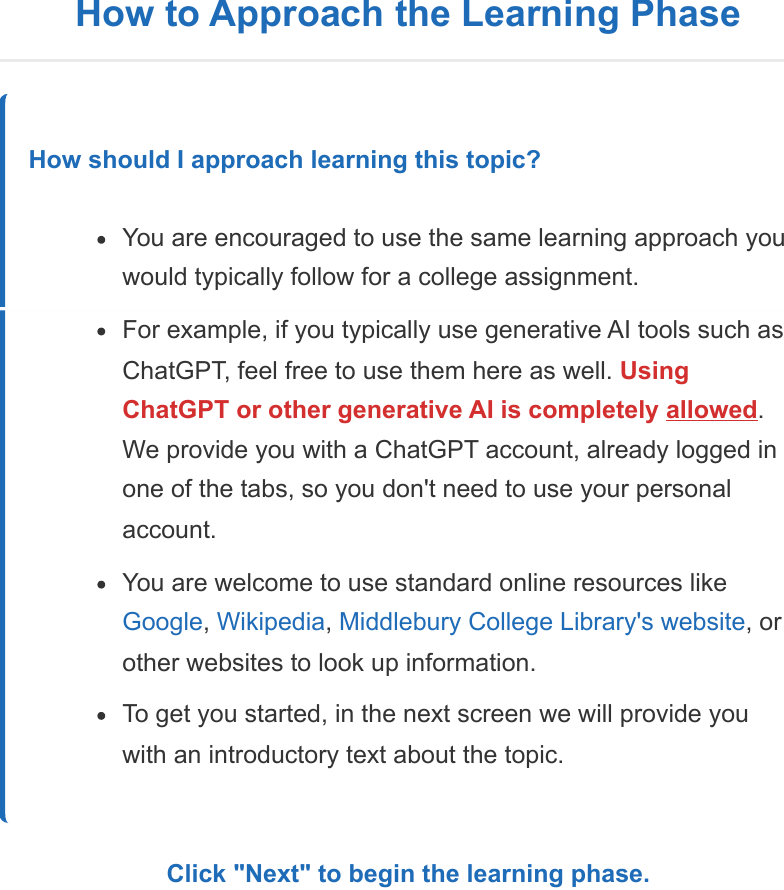}}
		\\[0.5cm]
		\textbf{Panel B. AI-forbidden condition}\\[0.3cm]
		\fbox{\includegraphics[width=0.52\linewidth]{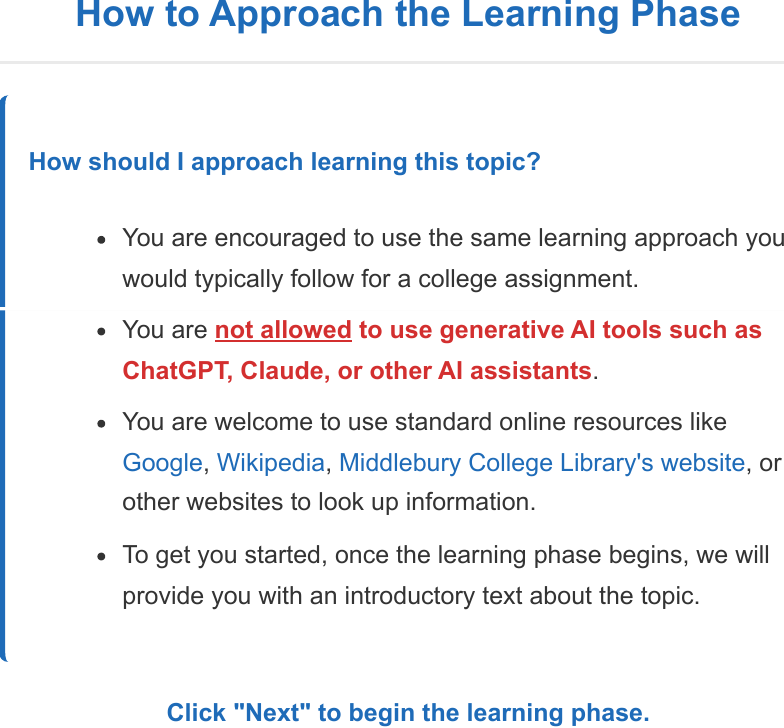}}
	\end{center}
	{\footnotesize \textit{Notes:} This figure shows the treatment-specific instructions displayed to participants in the survey interface. \par}
\end{figure}

\begin{figure}[H]
	\caption{Laboratory Seating Charts}\label{fig:map_labs}
	\centering
	\vspace{.5cm}
	\begin{subfigure}[t]{1\textwidth}
		\caption*{Panel A. Library computer lab (LIB 140)}
		\centering
		\includegraphics[width=.7\textwidth]{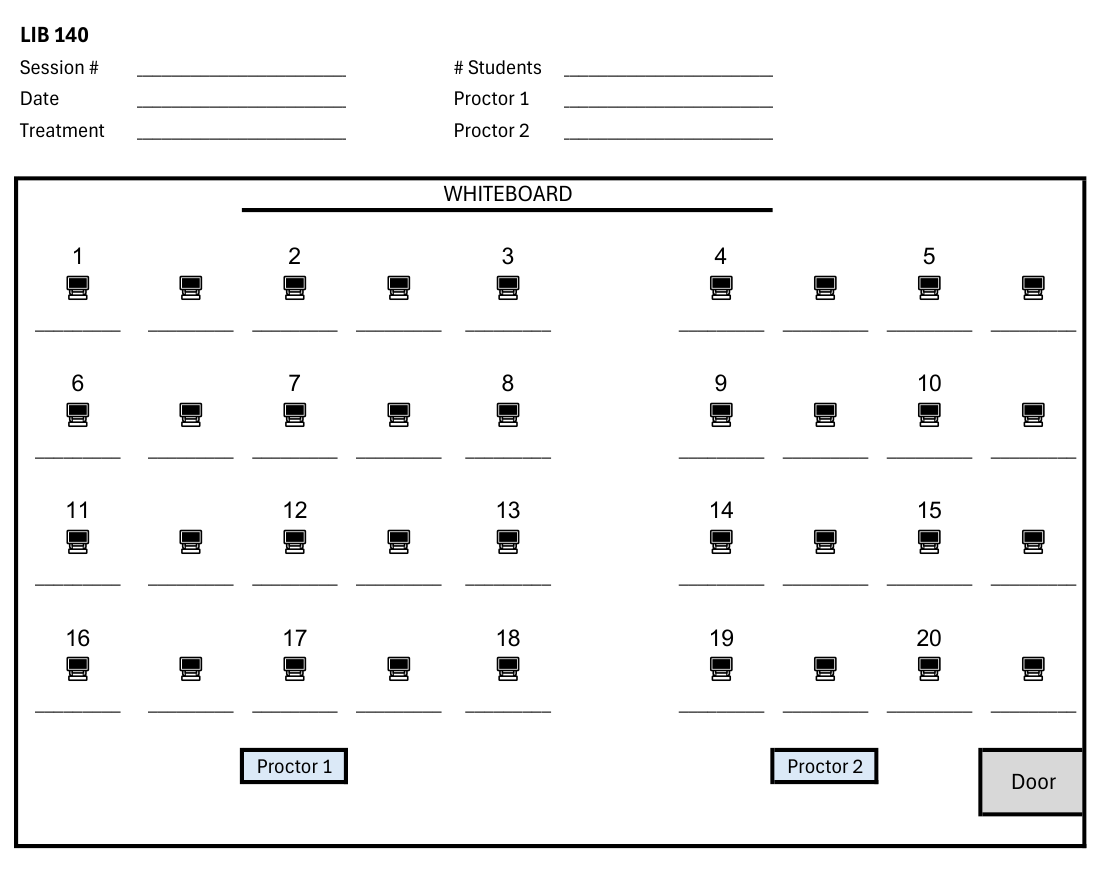}
	\end{subfigure} \vspace{.5cm}
	
	\begin{subfigure}[t]{0.7\textwidth}
		\caption*{Panel B. Language Building computer lab (SDL 122)}
		\centering
		\includegraphics[width=\textwidth]{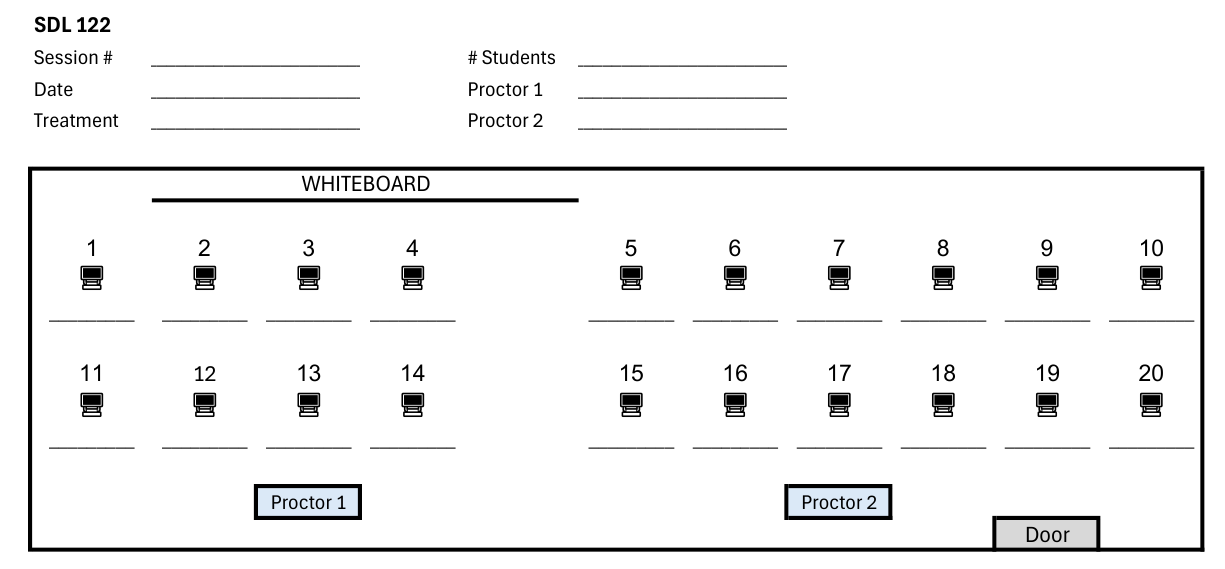} 
	\end{subfigure}
	{\footnotesize \singlespacing \justify \textit{Notes:} This figure shows the seating charts used by laboratory staff to monitor compliance during experimental sessions. Each numbered square represents a computer workstation. For each session time slot, one lab was randomly assigned to the AI-allowed condition and the other to the AI-forbidden condition. Proctors used these charts to record any instances of unauthorized resource use, with two proctors assigned to each laboratory. The physical separation of treatment conditions across different buildings helped prevent cross-contamination between experimental groups. \par

	}
\end{figure}

\clearpage
\begin{figure}[H]
	\caption{Test-Score Effects Excluding One Time Slot at a Time}\label{fig:loo_slots}
	\centering
	\begin{subfigure}[t]{.48\textwidth}
		\caption*{Panel A. Session One}
		\centering
		\includegraphics[width=\linewidth]{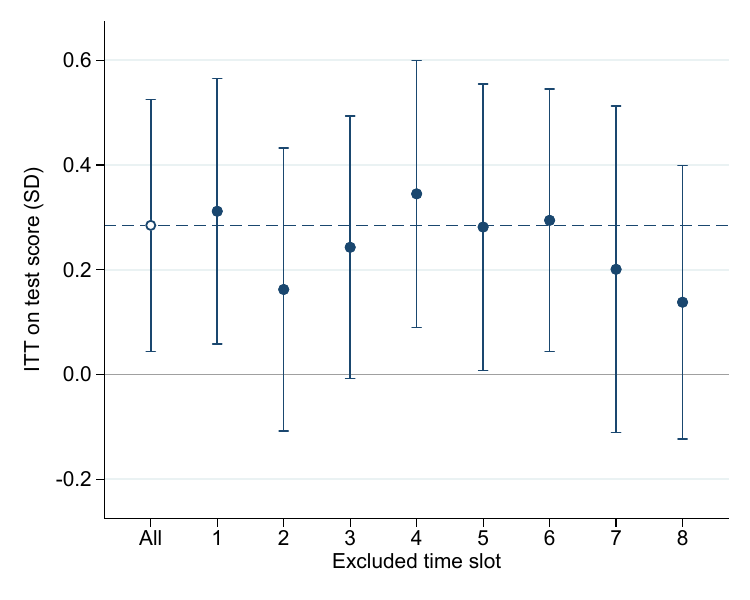}
	\end{subfigure}
	\hfill
	\begin{subfigure}[t]{0.48\textwidth}
		\caption*{Panel B. Session Two}
		\centering
		\includegraphics[width=\linewidth]{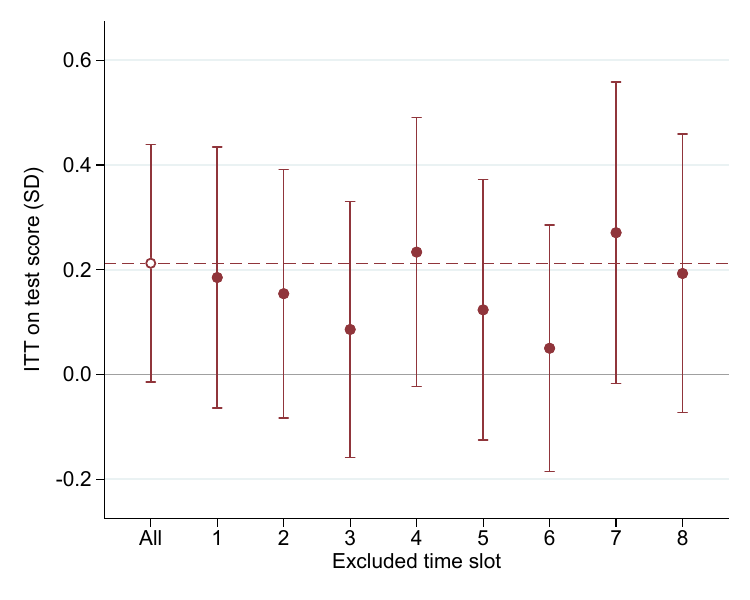}
	\end{subfigure}
	{\footnotesize
		\singlespacing \justify

		\textit{Notes:} This figure re-estimates the test-score ITT for Session One (Panel A) and Session Two (Panel B) excluding one of the eight time slots (the student's Session One time slot, the randomization stratum) at a time. The hollow marker keeps all slots and reproduces the estimates in Table~\ref{tab:test_score_long} (the dashed line marks that benchmark). Each estimate re-selects double-lasso controls on its subsample \citep{belloni.etal2014} and includes strata fixed effects. Vertical bars represent 95 percent confidence intervals from heteroskedasticity-robust standard errors. Excluding a slot leaves \getval{loo_s1_n_min}--\getval{loo_s1_n_max} students in Session One and \getval{loo_s2_n_min}--\getval{loo_s2_n_max} in Session Two.

	}
\end{figure}

\clearpage
\begin{figure}[H]
	\caption{Test-Score Effects by Item and Excluding One Item at a Time}\label{fig:item_level}
	\centering
	\medskip
	\textbf{Session One} \\[0.5em]
	\begin{subfigure}[t]{.48\textwidth}
		\caption*{Panel A. Effect by item}
		\centering
		\includegraphics[width=\linewidth]{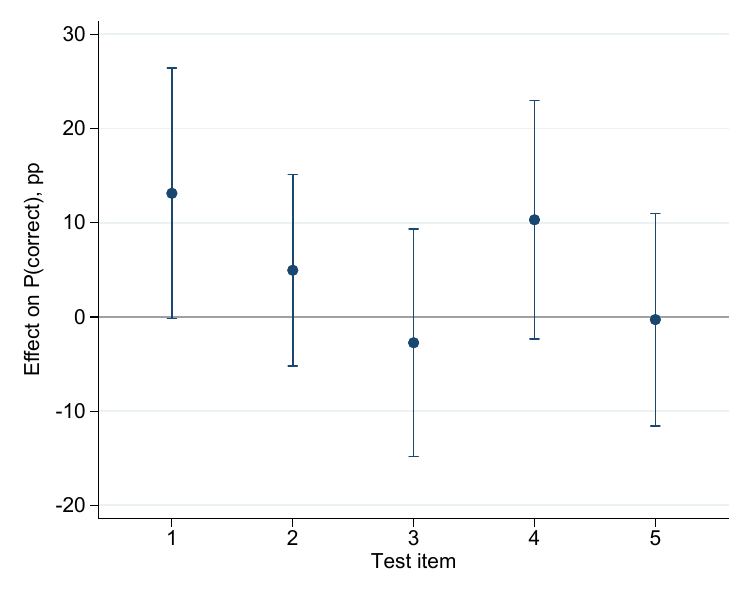}
	\end{subfigure}
	\hfill
	\begin{subfigure}[t]{0.48\textwidth}
		\caption*{Panel B. Leave one item out}
		\centering
		\includegraphics[width=\linewidth]{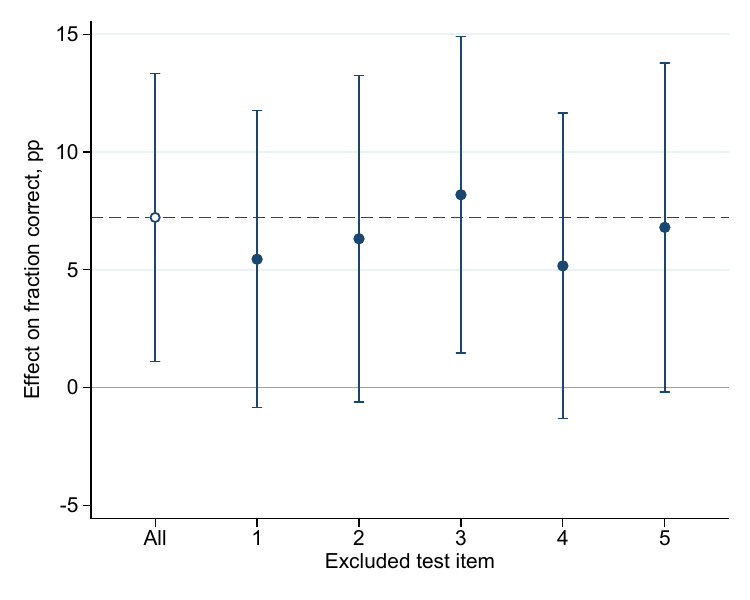}
	\end{subfigure}
	
	\bigskip
	\textbf{Session Two} \\[0.5em]
	\begin{subfigure}[t]{.48\textwidth}
		\caption*{Panel C. Effect by item}
		\centering
		\includegraphics[width=\linewidth]{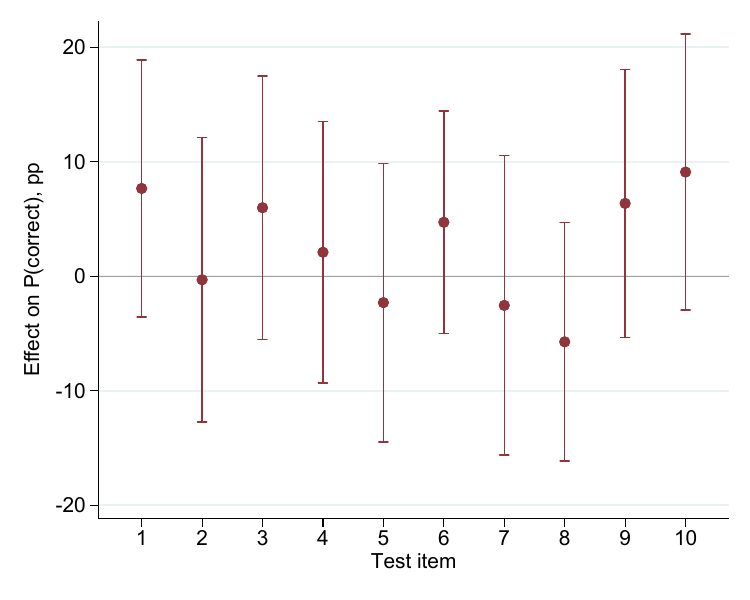}
	\end{subfigure}                                                                                                                                               \hfill
	\begin{subfigure}[t]{0.48\textwidth}
		\caption*{Panel D. Leave one item out}
		\centering
		\includegraphics[width=\linewidth]{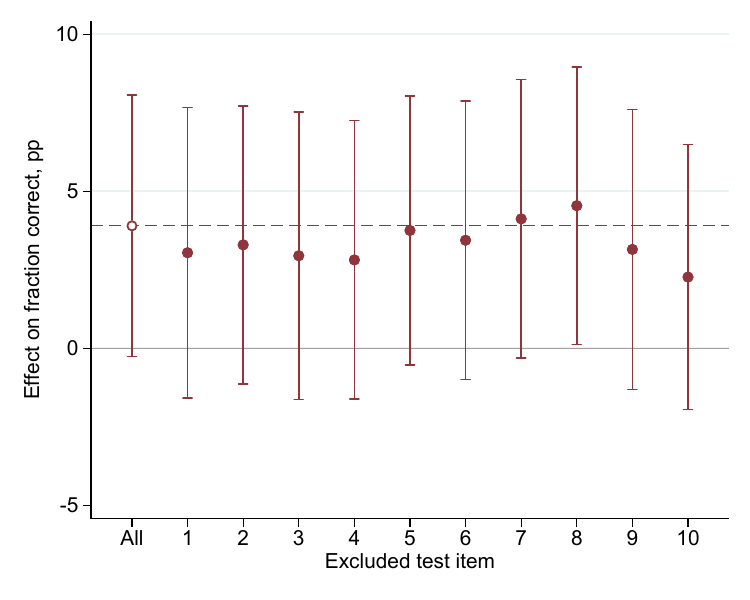}
	\end{subfigure}
	{\footnotesize
		\singlespacing \justify
		\textit{Notes:} This figure decomposes the test-score effects of Table~\ref{tab:test_score_long} by item. The Session One post-learning test has five items and the Session Two test has ten. Panels~A and~C report the ITT on the probability of answering each item correctly, in percentage points. Panels~B and~D report the ITT on the fraction of items answered correctly when one item at a time is excluded. The hollow marker keeps all items and reproduces the fraction-correct estimates in Table~\ref{tab:test_score_long} (in percentage points), and the dashed line marks that benchmark. Each estimate selects double-lasso controls for its own outcome \citep{belloni.etal2014} and includes strata fixed effects. Vertical bars are 95 percent confidence intervals from heteroskedasticity-robust standard errors. \par
	}
\end{figure}

\begin{figure}[H]
	\caption{Effects of AI Access on Essay Characteristics, by Dimension}\label{fig:writing_char_detailed}
	\centering
	\includegraphics[width=.85\linewidth]{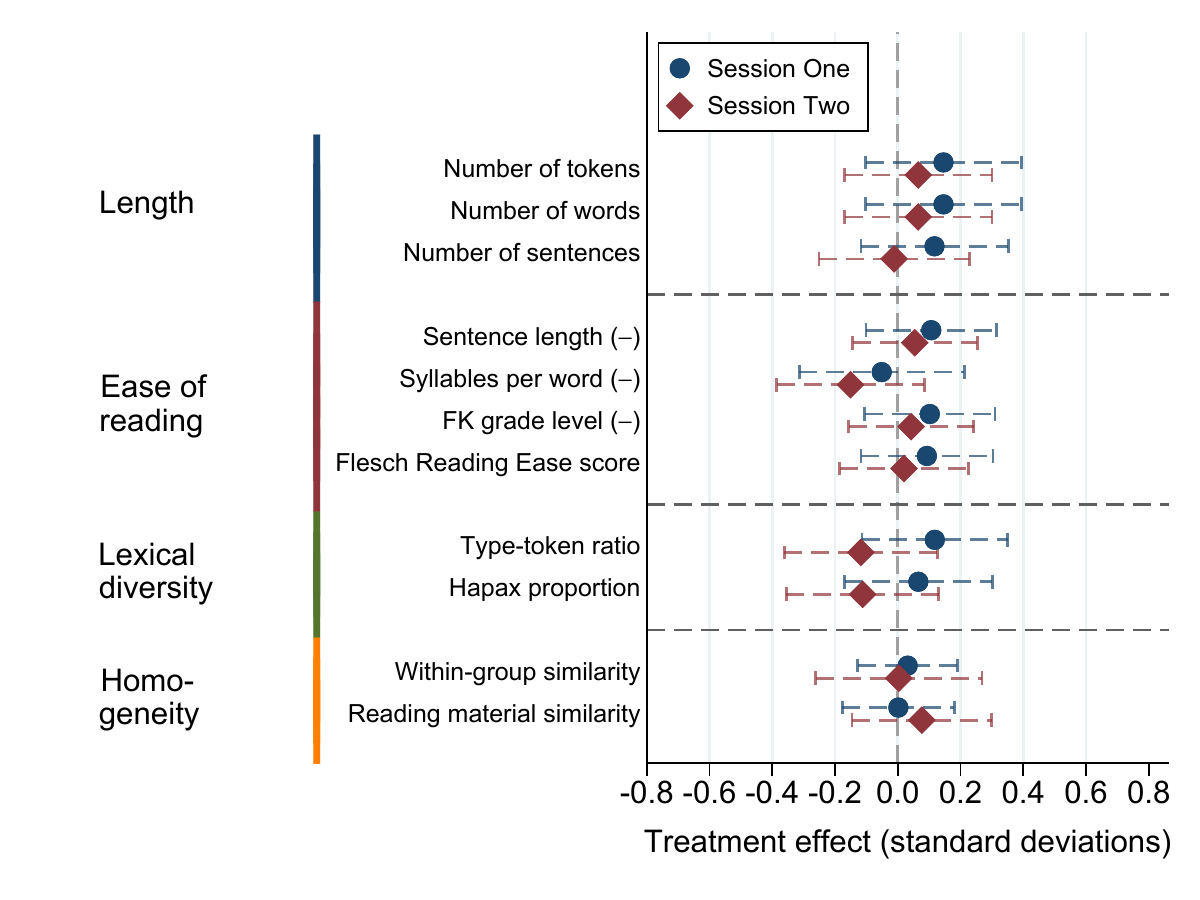}

	\hfill
	{\footnotesize
		\singlespacing \justify

		\textit{Notes:} This figure presents treatment effects of AI access on individual essay characteristics, measured in standard deviations. Each variable is standardized to have mean zero and standard deviation one in the control group. Variables are grouped into four categories, indicated by colored vertical bars: length (number of tokens, words, and sentences), ease of reading (sentence length, syllables per word, Flesch--Kincaid grade level, and Flesch Reading Ease score, with difficulty measures reversed so that higher values indicate easier readability), lexical diversity (type--token ratio and hapax proportion), and homogeneity and similarity (within-group cosine similarity and reading material cosine similarity). Circles represent Session One effects; diamonds represent Session Two effects. Horizontal lines represent 95 percent confidence intervals based on heteroskedasticity-robust standard errors. \par

	}
\end{figure}

\begin{figure}[H]
	\caption{Effects of AI Access on Essay Quality, by Dimension and Grader}\label{fig:essay_quality_detailed}
	\centering
	\includegraphics[width=.85\linewidth]{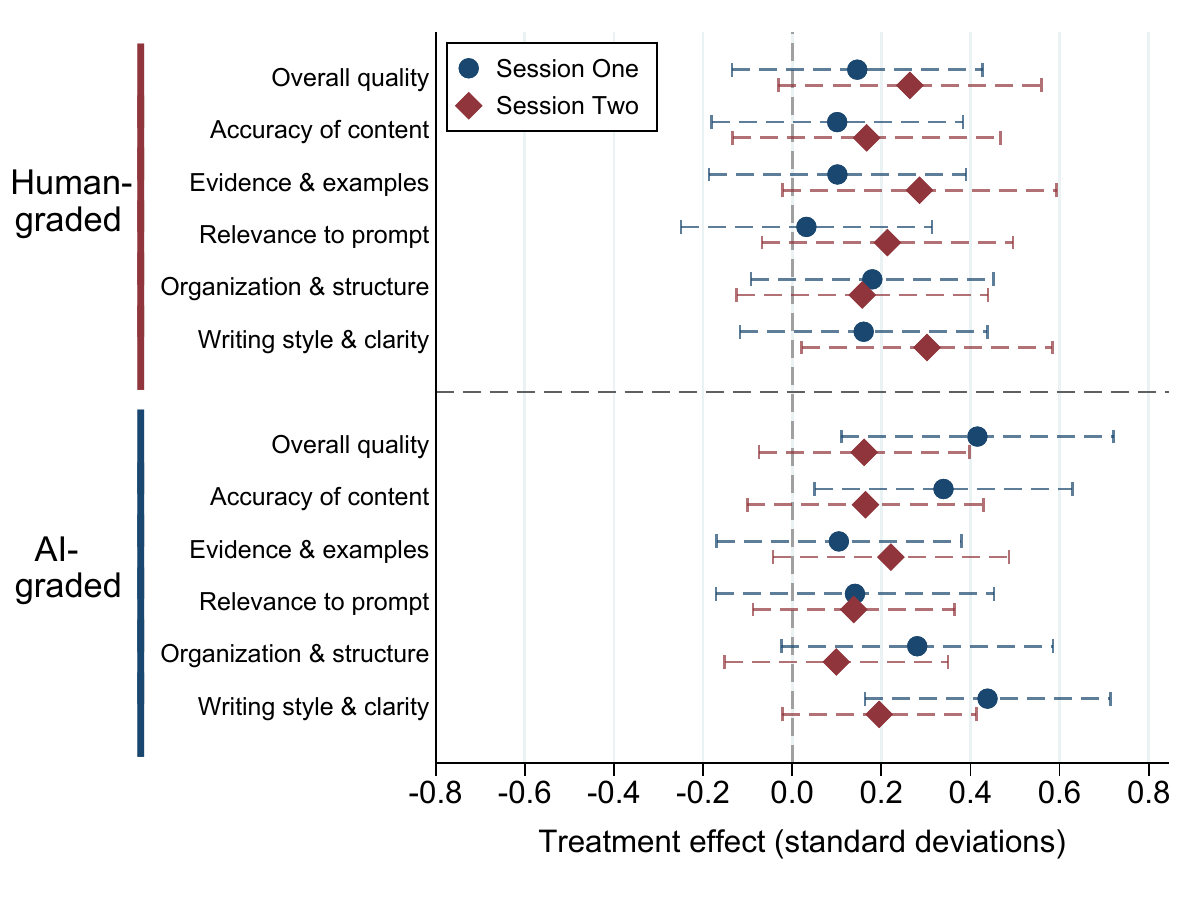}

	\hfill
	{\footnotesize
		\singlespacing \justify

		\textit{Notes:} This figure presents treatment effects of AI access on individual essay quality dimensions, measured in standard deviations. The top panel shows human-graded dimensions: essays were evaluated by independent graders on five rubric dimensions plus an overall quality rating (accuracy of content, use of evidence and examples, relevance to the prompt, organization and structure, writing style and clarity, and overall quality). Human-graded regressions are at the student level, averaging each essay's scores across its graders. The bottom panel shows the same dimensions as scored by an AI grader (see Appendix~\ref{app:ai_grading}). Each dimension is scored on a scale from 0 to 10 and standardized to have mean zero and standard deviation one in the control group. Circles represent Session One effects (essays written with or without AI access); diamonds represent Session Two effects (essays written one week later without AI access). Horizontal lines represent 95 percent confidence intervals based on heteroskedasticity-robust standard errors. \par

	}
\end{figure}

\clearpage
\begin{figure}[H]
	\caption{Distribution of Learning-Phase Duration by Treatment Arm}\label{fig:time_learning_dist}
	\centering
	\includegraphics[width=.75\linewidth]{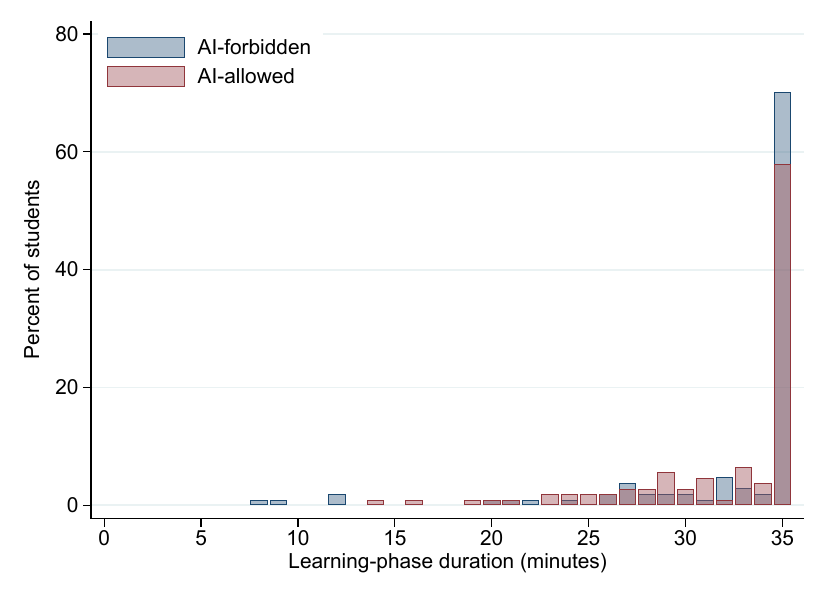}
	\hfill
	{\footnotesize
		\singlespacing \justify

		\textit{Notes:} This figure plots the distribution of the Qualtrics-recorded duration of the Session One learning phase, by whole minute, separately for the AI-forbidden and AI-allowed conditions.

	}
\end{figure}

\clearpage
\begin{figure}[H]
	\caption{Mechanisms: Time Allocation Across Learning Activities}\label{fig:mech_inputs}
	\centering
	\includegraphics[width=.75\linewidth]{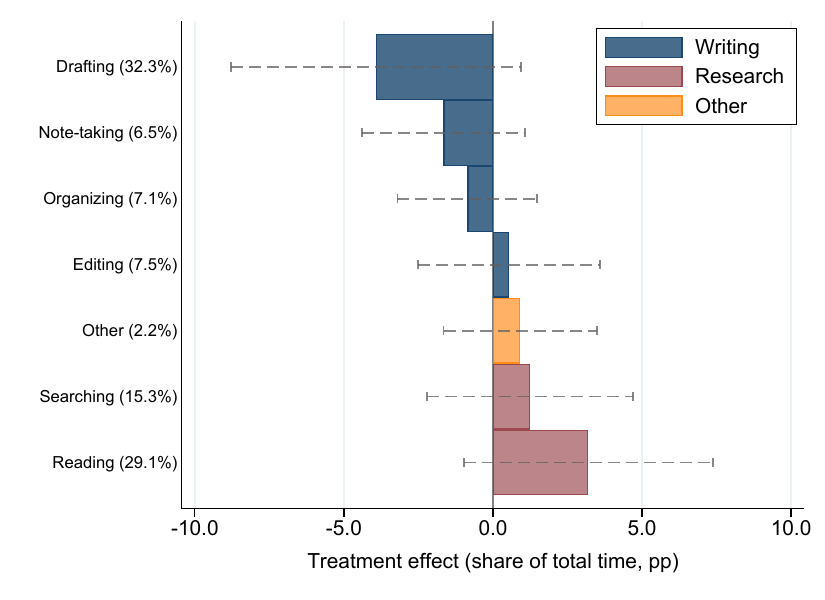}
	\hfill
	{\footnotesize
		\singlespacing \justify

		\textit{Notes:} This figure presents treatment effects of AI access on self-reported time allocation across different learning activities, measured as each activity's share of the student's total self-reported learning time (in percentage points). Students reported how many minutes they spent on each activity during the 35-minute learning phase; each activity's share is its minutes divided by the sum of minutes across all activities. The survey labels were: ``Writing a first draft of your write-up'' (Drafting), ``Revising and editing your draft'' (Editing), ``Taking notes on key concepts and facts'' (Note-taking), ``Planning your writing task structure and arguments'' (Organizing), ``Finding information about the topic, searching for relevant sources'' (Searching), ``Reading and comprehending information about the topic'' (Reading), and ``Casually browsing the web, unrelated to the topic'' and ``Other activities'' combined (Other). Navy bars represent writing activities, maroon bars represent research activities, and orange bars represent other activities. Bars represent the estimated treatment effect for each activity, with horizontal lines showing 95 percent confidence intervals. Numbers in parentheses indicate the control-group mean share. \par

	}
\end{figure}

\begin{figure}[H]
	\caption{Types of Generative AI Use by Augmentation, Automation, and Mixed Users}\label{fig:ai_usage_by_type}
	\centering
	\medskip
	\textbf{Augmentation users} \\[0.5em]
	\begin{subfigure}[t]{.40\textwidth}
		\caption*{Panel A. Self-reported usage}
		\centering
		\includegraphics[width=\linewidth]{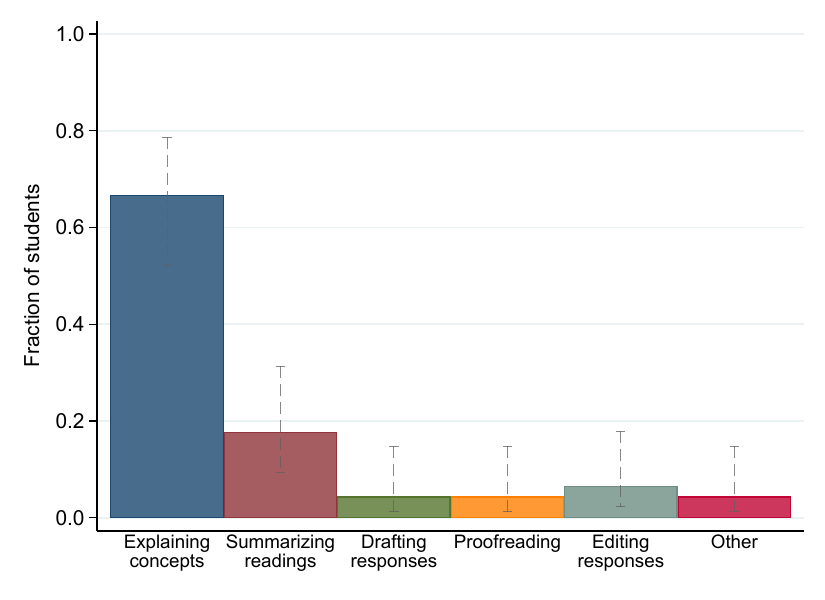}
	\end{subfigure}
	\hfill
	\begin{subfigure}[t]{.40\textwidth}
		\caption*{Panel B. ChatGPT logs}
		\centering
		\includegraphics[width=\linewidth]{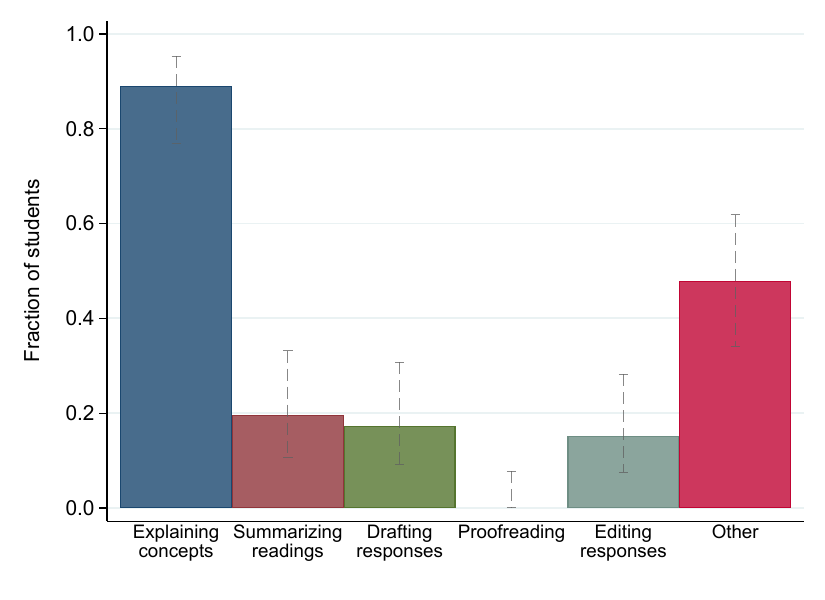}
	\end{subfigure}

	\textbf{Automation users} \\[0.5em]
	\begin{subfigure}[t]{.40\textwidth}
		\caption*{Panel C. Self-reported usage}
		\centering
		\includegraphics[width=\linewidth]{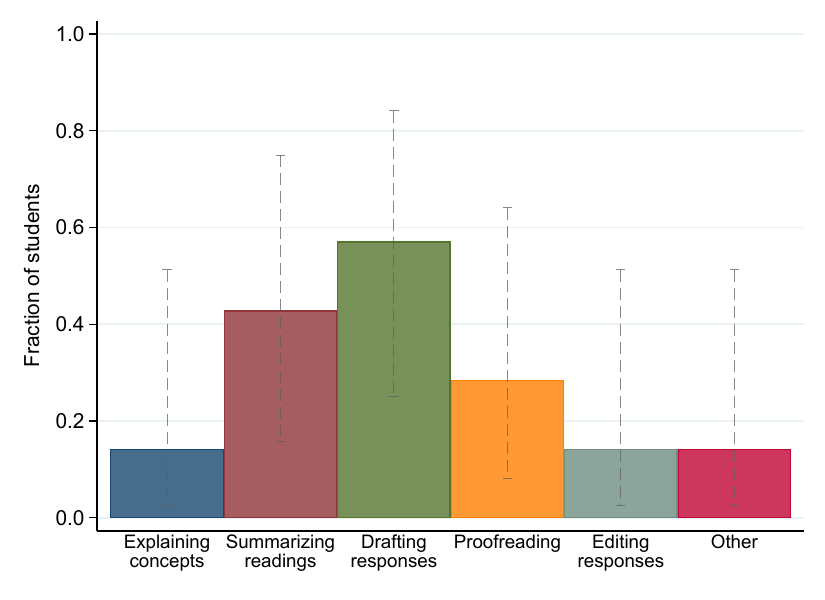}
	\end{subfigure}
	\hfill
	\begin{subfigure}[t]{.40\textwidth}
		\caption*{Panel D. ChatGPT logs}
		\centering
		\includegraphics[width=\linewidth]{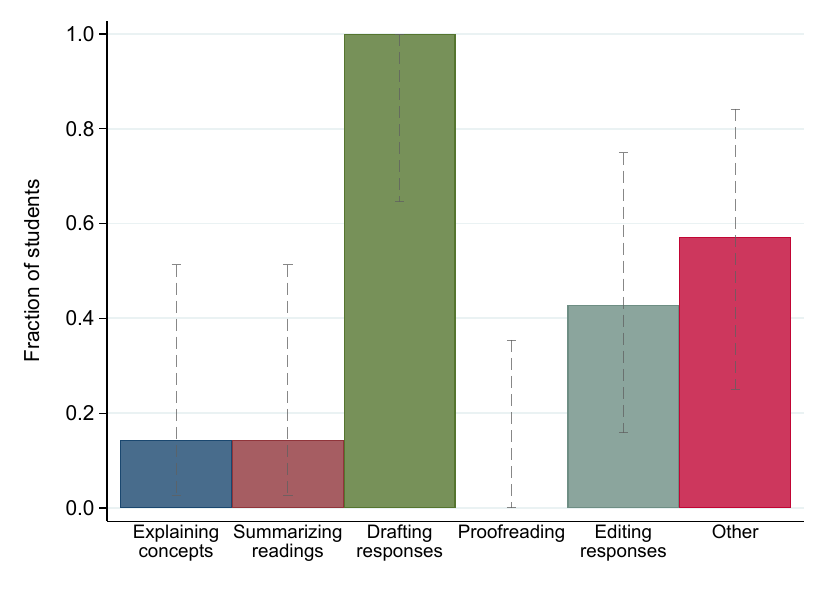}
	\end{subfigure}

	\textbf{Mixed users} \\[0.5em]
	\begin{subfigure}[t]{.40\textwidth}
		\caption*{Panel E. Self-reported usage}
		\centering
		\includegraphics[width=\linewidth]{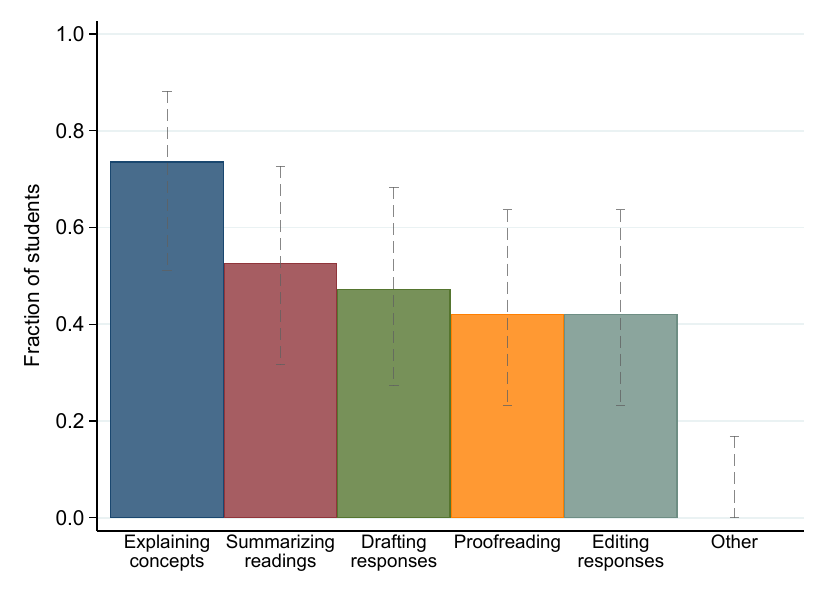}
	\end{subfigure}
	\hfill
	\begin{subfigure}[t]{.40\textwidth}
		\caption*{Panel F. ChatGPT logs}
		\centering
		\includegraphics[width=\linewidth]{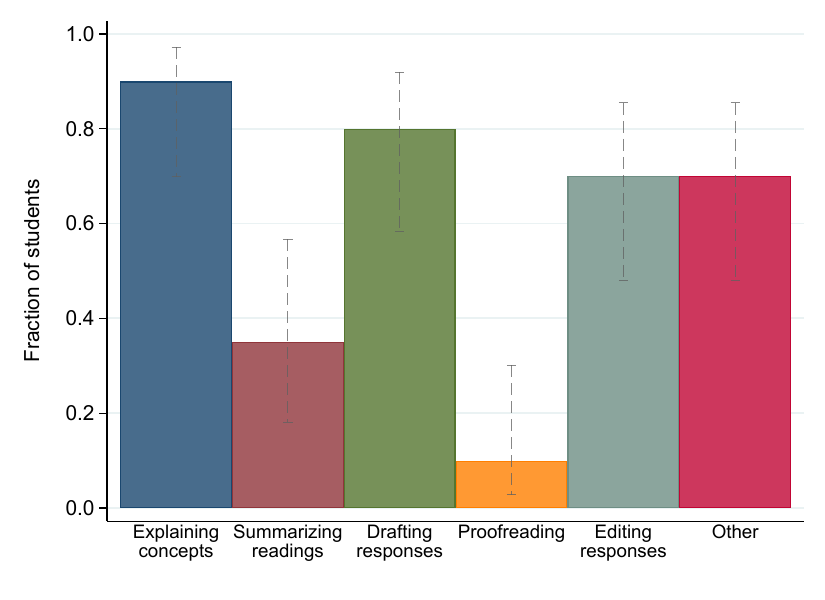}
	\end{subfigure}
	{\footnotesize
		\singlespacing \justify

		\textit{Notes:} This figure replicates Figure~\ref{fig:ai_usage_type} separately for augmentation, automation, and mixed users, as classified in Section~\ref{sub:aug_auto}. See notes to that figure for details. \par

	}
\end{figure}

\begin{figure}[H]
	\caption{Distribution of Perceived Treatment Effects on Own Test Score}\label{fig:hist_perceived_te}
	\centering
	\includegraphics[width=.75\linewidth]{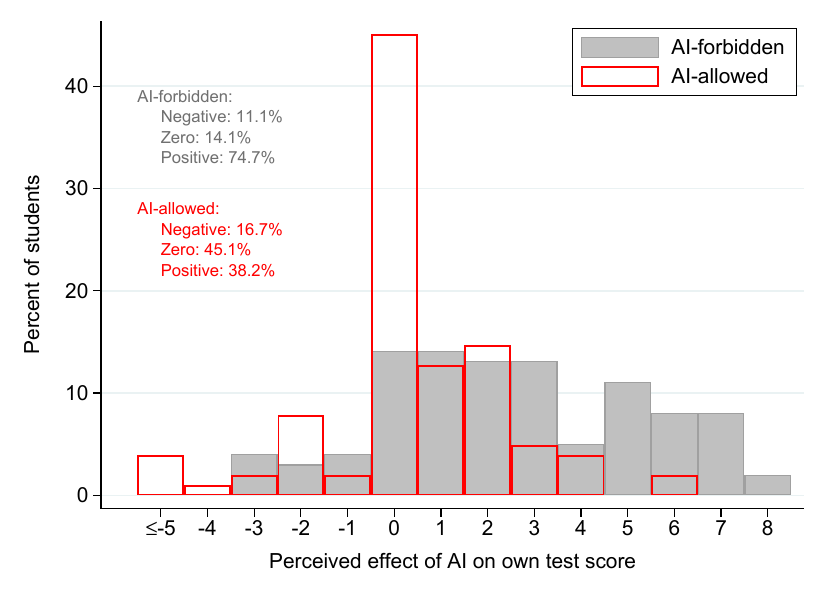}
	{\footnotesize
		\singlespacing \justify

		\textit{Notes:} This figure plots the distribution of students' perceived treatment effect of AI access on their own Session~Two test score, separately for AI-forbidden (control) and AI-allowed (treated) students. The perceived effect is each student's believed test score with AI access minus their believed score without it, expressed in number of questions answered correctly (for treated students, the believed own score minus the believed counterfactual; for control students, the believed counterfactual minus the believed own score). Values below $-5$ are bottom-coded at $-5$. Bin width is 1 and bins are centered on integers. In-figure annotations report the share of each group reporting a negative, zero, or positive perceived effect. \par
	}
\end{figure}

\begin{table}[H]{\footnotesize
		\begin{center}
			\caption{Experimental Sample Versus the Middlebury Student Body}
			\label{tab:sample_vs_body}
			\newcommand\w{2.2}
			\begin{tabular}{l@{}lR{\w cm}@{}L{0.45cm}R{\w cm}@{}L{0.45cm}}
				\midrule
				&& Experimental && Student \\
				&& sample && body \\
				&& (1) && (2) \\\midrule
				Female && 0.635 && 0.533 &\\
Male && 0.351 && 0.463 &\\
White && 0.500 && 0.538 &\\
Black && 0.086 && 0.052 &\\
Hispanic/Latino && 0.243 && 0.124 &\\
Asian && 0.286 && 0.073 &\\
International student && 0.269 && -- &\\

				\midrule
				Freshman && 0.374 && 0.255 &\\
Sophomore && 0.137 && 0.257 &\\
Junior && 0.223 && 0.201 &\\
Senior && 0.265 && 0.287 &\\

				\midrule
				College GPA && 3.683 && 3.670 &\\
Has not declared major && 0.220 && 0.417 &\\

				\midrule
				Natural-science major && 0.308 && 0.244 &\\
Social-science major && 0.365 && 0.243 &\\
Humanities/Arts/Languages major && 0.100 && 0.145 &\\

				\midrule
				\(N\) (Students) && 211 && 2,760 &\\

				\midrule
			\end{tabular}
		\end{center}

		\begin{singlespace}\vspace{-0.5cm}
			\footnotesize\noindent\justify

			\textit{Notes:} Column 1 reports means among the Session One attendees; characteristics are self-reported at intake. Column 2 reports the Middlebury student-body figures compiled in \citet{contractor2026generative}, Table~1, from administrative enrollment records (Fall 2024) and the Middlebury Common Data Set 2024--25. The race rows are not strictly comparable: the sample indicators are self-reported and not mutually exclusive, while the administrative categories are exclusive and exclude international students, so the Asian and Hispanic/Latino rows overstate the difference. College GPA is self-reported in column 1 and the registrar's Spring 2024 grade point average in column 2. Fields of study in column 2 are classified by declared degree; the humanities row sums the compilation's arts, humanities, languages, and literature categories. \par

	\end{singlespace}}
\end{table}

\clearpage
\begin{table}[H]{\footnotesize\renewcommand{\arraystretch}{0.9}
		\begin{center}
			\caption{Determinants of AI Use Among AI-Allowed Students} \label{tab:compliance}
			\begin{tabular}{l@{}lR{1.2cm}@{}L{0.5cm}R{1.2cm}@{}L{0.5cm}R{1.2cm}@{}L{0.5cm}R{1.2cm}@{}L{0.5cm}}
				\midrule
				&& \multicolumn{4}{c}{Bivariate regression} & \multicolumn{4}{c}{Multivariate regression} \\ \cmidrule{3-5} \cmidrule{7-9}
				&& Used account && Self-reported && Used account && Self-reported \\
				&& (1) && (2) && (3) && (4) \\ 
				\midrule
				\multicolumn{9}{l}{\hspace{-1em}\textbf{Panel A. Demographic characteristics}} \\ \addlinespace
				Male&&0.102&&0.113&&0.064&&0.014&\\
&&(0.102&)&(0.110&)&(0.114&)&(0.114&)\\

				White&&$-$0.202&\sym{*}&$-$0.065&&$-$0.170&&0.059&\\
&&(0.106&)&(0.112&)&(0.119&)&(0.141&)\\

				International student&&0.141&&0.088&&0.106&&0.171&\\
&&(0.114&)&(0.127&)&(0.151&)&(0.148&)\\

				Private high school&&0.010&&$-$0.077&&$-$0.126&&$-$0.244&\sym{**}\\
&&(0.096&)&(0.106&)&(0.113&)&(0.107&)\\

				\midrule
				\multicolumn{9}{l}{\hspace{-1em}\textbf{Panel B. Academic background}} \\ \addlinespace
				Natural-science major&&$-$0.063&&$-$0.167&&$-$0.109&&$-$0.123&\\
&&(0.100&)&(0.106&)&(0.133&)&(0.133&)\\

				Social-science major&&0.118&&0.258&\sym{**}&0.125&&0.281&\sym{**}\\
&&(0.096&)&(0.101&)&(0.124&)&(0.130&)\\

				High GPA (at or above median)&&$-$0.144&&$-$0.114&&0.007&&$-$0.005&\\
&&(0.090&)&(0.100&)&(0.113&)&(0.127&)\\

				Freshman&&$-$0.143&&$-$0.071&&$-$0.214&&0.015&\\
&&(0.099&)&(0.109&)&(0.140&)&(0.150&)\\

				Sophomore&&0.059&&0.119&&$-$0.104&&0.043&\\
&&(0.105&)&(0.125&)&(0.131&)&(0.178&)\\

				Junior&&$-$0.034&&$-$0.016&&$-$0.125&&$-$0.029&\\
&&(0.124&)&(0.133&)&(0.146&)&(0.161&)\\

				\midrule
				\multicolumn{9}{l}{\hspace{-1em}\textbf{Panel C. AI experience and beliefs}} \\ \addlinespace
				Frequent AI user (at or above median)&&0.357&\sym{***}&0.359&\sym{***}&0.283&\sym{*}&0.325&\sym{**}\\
&&(0.113&)&(0.116&)&(0.148&)&(0.162&)\\

				Early AI adopter (before Fall 2024)&&0.117&&0.161&&$-$0.050&&$-$0.069&\\
&&(0.096&)&(0.103&)&(0.113&)&(0.131&)\\

				High AI proficiency (at or above median)&&0.208&\sym{**}&0.264&\sym{**}&0.096&&0.165&\\
&&(0.096&)&(0.103&)&(0.119&)&(0.148&)\\

				High perceived AI effect (at or above median)&&$-$0.025&&$-$0.171&&0.066&&$-$0.089&\\
&&(0.121&)&(0.139&)&(0.116&)&(0.133&)\\

				\midrule
				\(N\) (Students)&&107&&102&&103&&98&\\
 \midrule
				
			\end{tabular}
		\end{center}
		\begin{singlespace}  \vspace{-.5cm}
			\noindent \justify \textit{Notes:} This table reports the association between student characteristics and AI use among treated students. Columns 1--2 are bivariate; columns 3--4 include all characteristics simultaneously. ``Used account'' (columns 1, 3) is based on ChatGPT server logs; ``Self-reported'' (columns 2, 4) is collected in Session Two. All variables are indicators. High GPA is at or above the sample median. Frequent AI user reports AI use during the semester at or above the median frequency (occasional or more frequent). Early AI adopter first used AI for academics before Fall 2024. High AI proficiency is self-assessed proficiency at or above the median. High perceived AI effect is a perceived treatment effect on test performance (believed score with AI minus believed counterfactual) at or above the median of \getval{compliance_perceived_te_median} questions, so the reference group is the students who expect AI to lower their score. Missing characteristics are set to zero. Every specification includes an indicator for missingness. All specifications include session (strata) fixed effects. Heteroskedasticity-robust standard errors in parentheses. $^{***}$ $p < 0.01$; $^{**}$ $p < 0.05$; $^{*}$ $p < 0.10$. \par
			
		\end{singlespace}
	}
\end{table}

\clearpage
\begin{table}[H]{\footnotesize
		\begin{center}
			\caption{Effects of Generative AI Access on Academic Integrity}
			\label{tab:integrity}
			\newcommand\w{1.5}
			\begin{tabular}{l@{}lR{\w cm}@{}L{0cm}R{\w cm}@{}L{0.45cm}R{\w cm}@{}L{0.45cm}}
				\midrule
				&& Control &&  &&  \\
				Outcome && mean && ITT && TOT \\
				&& (1) && (2) && (3) \\\midrule
				Caught by proctor&&0.029&&0.056&\sym{*}&0.080&\sym{*}\\
&&&&(0.031&)&(0.045&)\\

				Used unauthorized resources, Session One&&0.010&&0.096&\sym{***}&0.135&\sym{***}\\
&&&&(0.032&)&(0.045&)\\

				Any integrity violation&&0.039&&0.136&\sym{***}&0.191&\sym{***}\\
&&&&(0.043&)&(0.060&)\\

				\midrule
				\(N\) (Students)&&102&&205&&205&\\

				\midrule
			\end{tabular}
		\end{center}

		\begin{singlespace}\vspace{-0.5cm}
			\footnotesize\noindent\justify
			\textit{Notes:} This table reports treatment effects of AI access on rule violations during the Session One knowledge test. The knowledge tests prohibited all students from using external resources. Each row reports the effect of being assigned to the AI-allowed group on the indicated outcome. Caught by proctor is an indicator for being flagged by a proctor during the Session One knowledge test. Used unauthorized resources, Session One is an indicator for answering ``Yes, during Session 1 only'' to the exit-survey question, asked at the end of Session Two, ``Did you use any unauthorized resources or assistance (such as looking up answers online or using AI tools) while completing the multiple choice questions in either session?''; no student reported a violation in Session Two. Any integrity violation is an indicator for either. Column 1 reports the control-group mean. Column 2 reports intent-to-treat (ITT) estimates. Column 3 reports treatment-on-the-treated (TOT) estimates from two-stage least squares, instrumenting actual ChatGPT use with random treatment assignment. All specifications include controls selected through a double-lasso procedure \citep{belloni.etal2014}. Heteroskedasticity-robust standard errors in parentheses. $^{***}$ $p < 0.01$; $^{**}$ $p < 0.05$; $^{*}$ $p < 0.10$. \par

	\end{singlespace}}
\end{table}

\clearpage
\begin{table}[H]{\footnotesize
		\begin{center}
			\caption{Effects of AI Access on Essay Writing Style}
			\label{tab:writ_style_long}
			\newcommand\w{1.4}
			\begin{tabular}{l@{}lR{\w cm}@{}L{0cm}R{\w cm}@{}L{0.45cm}R{\w cm}@{}L{0.45cm}R{\w cm}@{}L{0cm}R{\w cm}@{}L{0.45cm}R{\w cm}@{}L{0.45cm}}
				\midrule
				&& \multicolumn{5}{c}{Session One (with AI)} && \multicolumn{5}{c}{Session Two (without AI)} \\ \cmidrule{3-7} \cmidrule{9-13}
				&& Control &&  &&  && Control &&  &&  \\
				Outcome && mean && ITT && TOT && mean && ITT && TOT \\
				&& (1) && (2) && (3) && (4) && (5) && (6) \\\midrule
				Length index&&0.000&&0.132&&0.190&&0.000&&0.042&&0.059&\\
&&&&(0.120&)&(0.173&)&&&(0.115&)&(0.163&)\\

				Readability index&&0.000&&0.061&&0.087&&0.000&&0.004&&0.006&\\
&&&&(0.088&)&(0.127&)&&&(0.082&)&(0.116&)\\

				Lexical diversity index&&0.000&&0.106&&0.154&&0.000&&$-$0.115&&$-$0.162&\\
&&&&(0.115&)&(0.165&)&&&(0.123&)&(0.175&)\\

				\midrule
				\(N\) (Students)&&104&&210&&210&&102&&204&&204&\\

				\midrule
			\end{tabular}
		\end{center}

		\begin{singlespace}\vspace{-0.5cm}
			\footnotesize\noindent\justify

			\textit{Notes:} This table reports treatment effects of AI access on three z-scored writing style indices, standardized relative to the control group. Each row reports the effect of being assigned to the AI-allowed group on the indicated outcome. Length index averages the z-scores of tokens, words, and sentences. Readability index averages four z-scored measures (sentence length, syllables per word, Flesch--Kincaid grade level, and Flesch Reading Ease), with signs oriented so that higher values indicate easier readability. Lexical diversity index averages the z-scored type--token ratio and hapax proportion. Appendix Figure~\ref{fig:writing_char_detailed} reports results for the individual characteristics. Columns 1 and 4 report the control-group mean. Columns 2 and 5 report intent-to-treat (ITT) estimates. Columns 3 and 6 report treatment-on-the-treated (TOT) estimates from two-stage least squares, instrumenting actual ChatGPT use with random treatment assignment. All specifications include controls selected through a double-lasso procedure \citep{belloni.etal2014}. Heteroskedasticity-robust standard errors in parentheses. $^{***}$ $p < 0.01$; $^{**}$ $p < 0.05$; $^{*}$ $p < 0.10$. \par

	\end{singlespace}}
\end{table}

\clearpage
\begin{table}[H]{\footnotesize
		\begin{center}
			\caption{Effects of AI Access on Essay Quality, by Dimension}
			\label{tab:essay_quality_dims}
			\newcommand\w{1.3}
			\begin{tabular}{l@{}lR{\w cm}@{}L{0cm}R{\w cm}@{}L{0.45cm}R{\w cm}@{}L{0.45cm}R{\w cm}@{}L{0cm}R{\w cm}@{}L{0.45cm}R{\w cm}@{}L{0.45cm}}
				\midrule
				&& \multicolumn{5}{c}{Session One} && \multicolumn{5}{c}{Session Two} \\ \cmidrule{3-7} \cmidrule{9-13}
				&& Control &&  &&  && Control &&  &&  \\
				Outcome && mean && ITT && TOT && mean && ITT && TOT \\
				&& (1) && (2) && (3) && (4) && (5) && (6) \\\midrule
				Accuracy&&6.049&&0.340&\sym{*}&0.489&\sym{*}&4.626&&0.261&&0.375&\\
&&&&(0.174&)&(0.249&)&&&(0.169&)&(0.246&)\\

				Evidence&&5.568&&0.231&&0.335&&3.654&&0.431&\sym{**}&0.623&\sym{**}\\
&&&&(0.209&)&(0.304&)&&&(0.180&)&(0.269&)\\

				Relevance&&6.229&&0.167&&0.240&&5.313&&0.291&&0.405&\\
&&&&(0.219&)&(0.315&)&&&(0.193&)&(0.271&)\\

				Organization&&5.235&&0.446&\sym{**}&0.651&\sym{**}&4.611&&0.236&&0.333&\\
&&&&(0.211&)&(0.308&)&&&(0.210&)&(0.297&)\\

				Writing style&&5.707&&0.398&\sym{**}&0.592&\sym{**}&5.020&&0.378&\sym{**}&0.546&\sym{**}\\
&&&&(0.194&)&(0.292&)&&&(0.171&)&(0.255&)\\

				\midrule
				\(N\) (Students)&&103&&207&&207&&97&&198&&198&\\
 \midrule
			\end{tabular}
		\end{center}

		\begin{singlespace}\vspace{-0.5cm}
			\footnotesize\noindent\justify

			\textit{Notes:} This table reports treatment effects of AI access on the five essay quality dimensions, averaging human and AI grades. Each row reports the effect of being assigned to the AI-allowed group on the indicated dimension. Each dimension is scored on a 0--10 scale. Each row is one student-level regression on the average of the human graders' mean score and the AI grader's score for that dimension. Columns 1--3 report Session One results (when AI-allowed students had access to ChatGPT); columns~4--6 report Session Two results (when all students wrote without AI access). Intent-to-treat (ITT) estimates report the effect of being assigned to the AI-allowed group; treatment-on-the-treated (TOT) estimates instrument ChatGPT use with random assignment. All specifications include controls selected through a double-lasso procedure \citep{belloni.etal2014}. Heteroskedasticity-robust standard errors in parentheses. $^{***}$ $p < 0.01$; $^{**}$ $p < 0.05$; $^{*}$ $p < 0.10$. \par

	\end{singlespace}}
\end{table}

\clearpage
\begin{landscape}
\begin{table}[H]{\footnotesize\setlength{\tabcolsep}{2.5pt}
		\begin{center}
			\caption{Baseline Characteristics by Type of AI Use}
			\label{tab:usage_baseline}
			\newcommand\w{2.0}
			\begin{tabular}{l@{}lR{\w cm}@{}L{0.3cm}R{\w cm}@{}L{0.3cm}R{\w cm}@{}L{0.3cm}R{\w cm}@{}L{0.3cm}R{\w cm}@{}L{0.3cm}}
				\midrule
				&& \multicolumn{1}{r}{Augmentation} && \multicolumn{1}{r}{Automation} && \multicolumn{1}{r}{Mixed} && \multicolumn{1}{r}{Non-user} && \multicolumn{1}{r}{AI-forbidden} \\
				&& (1) && (2) && (3) && (4) && (5) \\\midrule
				\multicolumn{12}{l}{\hspace{-1em} \textbf{Panel A. Research assistant's classification}} \\ \addlinespace
				Age && 20.3 && 20.0 && 20.6 && 19.9 && 19.9 &\\
Male (percent) && 26.1 && 71.4 && 30.0 && 21.9 && 41.3 &\\
White (percent) && 45.7 && 42.9 && 30.0 && 62.5 && 52.4 &\\
International student (percent) && 23.3 && 14.3 && 40.0 && 12.5 && 31.7 &\\
College GPA && 3.63 && 3.75 && 3.68 && 3.67 && 3.71 &\\
SAT score && 1348 && 1355 && 1377 && 1434 && 1391 &\\
Study hours per week && 16.2 && 15.6 && 16.8 && 16.7 && 16.7 &\\
Baseline self-assessed knowledge (0--10) && 1.4 && 1.3 && 1.4 && 1.7 && 1.6 &\\
Baseline percent correct && 26.5 && 42.9 && 37.0 && 28.8 && 30.6 &\\

				\midrule
				\(N\) (Students) && 46 && 7 && 20 && 32 && 104 &\\

				\midrule
				\multicolumn{12}{l}{\hspace{-1em} \textbf{Panel B. LLM classification}} \\ \addlinespace
				Age && 20.4 && 20.1 && 20.5 && 19.9 && 19.9 &\\
Male (percent) && 26.3 && 55.6 && 17.6 && 21.9 && 41.3 &\\
White (percent) && 52.6 && 33.3 && 23.5 && 62.5 && 52.4 &\\
International student (percent) && 22.2 && 33.3 && 31.2 && 12.5 && 31.7 &\\
College GPA && 3.68 && 3.68 && 3.58 && 3.67 && 3.71 &\\
SAT score && 1337 && 1390 && 1360 && 1434 && 1391 &\\
Study hours per week && 16.4 && 14.3 && 18.1 && 16.7 && 16.7 &\\
Baseline self-assessed knowledge (0--10) && 1.3 && 1.6 && 1.2 && 1.7 && 1.6 &\\
Baseline percent correct && 25.3 && 42.2 && 31.8 && 28.8 && 30.6 &\\

				\midrule
				\(N\) (Students) && 38 && 18 && 17 && 32 && 104 &\\

				\midrule
			\end{tabular}
		\end{center}

		\begin{singlespace}\vspace{-0.6cm} %
			\footnotesize\noindent\justify

			\textit{Notes:} This table reports unadjusted means by how students in the AI-allowed condition used the provided ChatGPT account, with the AI-forbidden group in column 5 as the reference. In Panel A, a research assistant labeled each conversation (Section~\ref{sub:aug_auto}). In Panel B, an LLM (Anthropic's Claude Fable~5.1) labeled the same conversations (Appendix~\ref{app:chatgpt_classification}). In both panels, we group students by the same rule. A student is an automation user if at least one conversation is labeled Automation and none Augmentation or Mixed, an augmentation user if at least one is labeled Augmentation and none Automation or Mixed, a mixed user if the conversations include a Mixed conversation or both an Automation and an Augmentation conversation, and a non-user if the account has no conversation. Students whose conversations were all classified as Other are omitted. Columns 4 and 5 do not depend on the labels and are the same in both panels. The rows are characteristics measured at intake and the baseline test. \par

	\end{singlespace}}
\end{table}
\end{landscape}

\clearpage

\setcounter{table}{0}
\setcounter{figure}{0}
\setcounter{equation}{0}
\renewcommand{\thetable}{B\arabic{table}}
\renewcommand{\thefigure}{B\arabic{figure}}
\renewcommand{\theequation}{B\arabic{equation}}

\section{Empirical Appendix} \label{app:texts}

\subsection{Reading Materials}  

We wrote original texts on each of the three topics, designing them so that participants faced similar cognitive demands across topics. Each text follows a parallel structure, beginning with an overview of the technology, followed by its historical development, technical underpinnings, core applications, and limitations. The texts are information-rich but self-contained, requiring no prior knowledge of the topic. The full texts are available in the \href{https://www.dropbox.com/scl/fi/pzzmqwikxweqzrofrnzt9/ai-learning-supplementary.pdf?rlkey=zjg1lf7cmzno304na3bgq58bn&st=gtj3nlqa&dl=0}{Supplementary Materials}.

The reading materials are 1,248--1,382 words long, or about 5.0--5.5 minutes of reading at an average silent reading speed of 250 words per minute for college students \citep{carver1992reading, brysbaert2019many}. They are closely matched across several readability metrics (Appendix Table~\ref{tab:readability}). Flesch--Kincaid grade levels of 15.1--15.9 place them at a late-undergraduate reading level. Measures of lexical and syntactic complexity, including sentence and word length, also indicate a consistent level of difficulty across topics.

\begin{table}[H]
	\begin{center}
		{\footnotesize
			
			\caption{Readability Metrics for Topic Texts} \label{tab:readability}
			\begin{tabular}{lccc}
				\toprule
				& 	\multicolumn{3}{c}{Learning topic:} \\ \cmidrule{2-4}
				Metric & Blockchain & Carbon Capture & CRISPR \\
				\midrule
				Flesch--Kincaid grade level & 15.90 & 15.76 & 15.11 \\
				SMOG index & 16.75 & 17.28 & 16.50 \\
				Automated readability index & 17.75 & 16.26 & 15.86 \\
				Type--token ratio & 0.51 & 0.49 & 0.51 \\
				Average sentence length & 21.59 & 24.15 & 20.80 \\
				Average word length & 6.43 & 6.09 & 6.07 \\
				Word count & 1,248 & 1,315 & 1,382 \\
				Estimated reading time (250 wpm) & 5.0 min & 5.3 min & 5.5 min \\
				\bottomrule
			\end{tabular}
		}
	\end{center}

	\begin{singlespace}\vspace{-0.2cm}
		\footnotesize\noindent\justify
		\textit{Notes:} This table presents readability metrics for the three reading materials. Estimated reading times assume an average silent reading speed of 250 words per minute for college students \citep{carver1992reading, brysbaert2019many}. Word counts are those of the texts as delivered to participants. \par
	\end{singlespace}
\end{table}

\subsection{Writing Assessment Prompts} \label{app:prompts}

This appendix presents the essay prompts we developed for each of the three topics. For each topic, we designed two complementary prompts of comparable difficulty. Each prompt requires participants to analyze relationships between concepts, evaluate the relative importance of different factors, and support their analysis with specific examples. Each participant is randomly assigned to receive one prompt in Session One and the other in Session Two.\footnote{Three participants received the same prompt in both sessions: two who completed Session Two through the emergency link and one whose Session Two login was recovered from a wrong account.}

\topicprompts{blockchainblue}{Blockchain Technology}{%
``Examine how blockchain has grown beyond cryptocurrency. Analyze which two application areas have the strongest potential for transformative impact and explain the specific factors that support your analysis. Draw on examples from the reading to develop your response.''%
}{%
``Examine key differences between blockchain technology and traditional databases. Analyze which specific features make blockchain unique, which types of applications benefit most from these differences, and why. Support your analysis with concrete evidence or examples.''%
}

\topicprompts{carbongreen}{Carbon Capture}{%
``Analyze the three main barriers to scaling carbon capture technologies (technical limitations, economic costs, and policy challenges). Identify which barrier you believe is most critical to overcome first, how it impacts the other challenges, and what approaches might address it. Support your analysis with specific examples.''%
}{%
``Compare direct air capture with point-source carbon capture approaches. Analyze the specific advantages and limitations of each method, identify which contexts each approach is better suited for, and explain what factors most influence their effectiveness. Draw on evidence from the reading to support your analysis.''%
}

\topicprompts{crisprpurple}{CRISPR Gene Editing}{%
``Analyze how CRISPR differs from previous gene editing technologies in terms of precision, accessibility, and versatility. Identify which two differences have been most consequential for scientific advancement, explain why these particular differences matter, and provide specific examples that demonstrate their impact.''%
}{%
``Examine the three main technical challenges that limit CRISPR today (delivery methods, off-target effects, and ethical considerations). Analyze which challenge is most important to address first, how it relates to the others, and what approaches show promise for overcoming it. Support your analysis with specific examples.''%
}

\subsection{Human Essay Grading} \label{app:prolific_graders}

We recruit independent graders through Prolific to evaluate participants' analytical essays, restricting eligibility to individuals holding a master's or PhD degree. Graders are blind to treatment condition, participant identity, and all other experimental variables. Each grader evaluates five essays and is compensated with a \$15 fixed payment plus eligibility for a \$10 lottery bonus. To incentivize consistency, we tell graders that their chances of winning the bonus rise if their scores resemble those of other graders evaluating the same essays.

Before grading, each grader completes a comprehension check verifying their understanding of the task requirements and reviews two example essays with suggested scores and detailed rationales. The examples are chosen to illustrate different quality levels---one strong response (receiving scores of 6--9 across dimensions) and one weaker response (receiving scores of 4--8)---so that graders can calibrate their standards. Graders are instructed to read each essay carefully, consult external resources to verify factual claims if needed, and maintain consistent scoring standards across all five essays.

Graders evaluate each essay on five rubric dimensions plus an overall quality rating, each on a 0--10 scale. \textit{Accuracy of Content} measures the factual correctness of the information presented in the essay. \textit{Use of Evidence and Examples} assesses the integration of specific evidence, examples, and data from the reading to support arguments. \textit{Relevance to the Prompt} evaluates how well the essay addresses the specific prompt and stays on topic throughout. \textit{Organization and Structure} measures the logical structure, coherence, flow, and effectiveness of transitions. \textit{Writing Style and Clarity} assesses clarity, readability, grammar, and precision of language. \textit{Overall Essay Quality} captures the grader's holistic assessment of the essay. Graders also report their prior familiarity with each of the three topics (blockchain, carbon capture, and CRISPR) on a 0--10 scale before beginning the grading task.

Each essay is evaluated by three to four independent graders.\footnote{Two Session Two essays that were assigned to two grading batches received seven and eight ratings.} We use the overall quality rating as our benchmark measure of essay quality and report results for each dimension separately.\footnote{Each grading packet listed the prompt of the essay. For three Session One essays the packet showed the topic's other prompt rather than the prompt the student had received, because those students saw the same prompt in both sessions and the packets assumed the prompts alternated. We set to missing the human ratings of those three essays and three further ratings whose packet recorded no prompt. Six Session Two essays restored after the grading round have AI grades but no human ratings. These scores enter only the AI-graded outcomes, since the combined human-and-AI measures require both graders.}

\subsection{AI Essay Grading} \label{app:ai_grading}

We complement our human grading with AI-generated scores to provide an independent assessment of essay quality. We use Anthropic's Claude Fable~5.1 to grade each essay on the same five rubric dimensions plus the overall quality rating, using the same 0--10 scale as the Prolific graders.

For each essay, we make six separate API calls---one per grading dimension---to avoid cross-contamination between scores. Each call includes a system prompt establishing the grading context, a description of the dimension, two calibration essays with their suggested scores and rationales for that dimension, and the student essay to be graded. The calibration essays, scores, and rationales are the ones shown to human graders during training, ensuring that human and AI graders share a common scoring anchor. We ask the model for a brief justification followed by an integer score on the 0--10 scale.

The system prompt is as follows: \\

\begin{centering}
	\fbox{\begin{minipage}{\textwidth}
			\ttfamily \footnotesize
			You are grading essays for a university research study on the determinants of learning. The essays were written by college students, who were given 35 minutes to learn about a topic and respond to a prompt. Each essay is approximately 300--500 words long. Grade each essay on a 0--10 scale. Use the two example essays, their suggested scores, and their rationales to calibrate your standards, and keep those standards consistent across essays. You may use your own knowledge to verify facts.
	\end{minipage}}
\end{centering}

For each dimension, the user prompt includes the dimension name and a description of what the dimension measures. The descriptions are as follows:

\begin{centering}
	\fbox{\begin{minipage}{\textwidth}
			\ttfamily \footnotesize
			\textbf{Accuracy of Content.} Evaluate the factual correctness and depth of understanding demonstrated in the essay. Consider whether key concepts are accurately described, whether technical details are correct, and whether the student shows genuine comprehension of the topic. \\

			\textbf{Use of Evidence \& Examples.} Assess whether the essay draws on evidence from the reading the students were given. Consider how specific and well-integrated the references are. \\

			\textbf{Relevance to the Prompt.} Measure whether the essay directly addresses the prompt requirements. Consider whether all parts of the prompt are addressed and whether the essay stays focused throughout. \\

			\textbf{Organization and Structure.} Evaluate the essay's organization, clarity of argument flow, and coherence. Consider whether there is a clear introduction, body, and conclusion, and whether transitions between ideas are smooth. \\

			\textbf{Writing Style and Clarity.} Assess the clarity, readability, grammar, and language precision of the essay. Consider vocabulary, sentence variety, and overall polish of the writing. \\

			\textbf{Overall Essay Quality.} Provide a holistic assessment of the essay's overall quality. Consider all previous dimensions together: accuracy, evidence, relevance, structure, and writing style.
	\end{minipage}}
\end{centering}

\medskip

The prompt also includes the two calibration essays, the essay prompt they answer, and their suggested scores and rationales for the relevant dimension. After presenting the calibration material, the prompt provides the student's assigned topic, the exact prompt the student was shown in that session, and the essay text. Empty essays, or those with fewer than 20 characters, are skipped and recorded as missing without querying the model.

\subsection{SAT Score Imputation} \label{app:sat_imputation}

We lack admissions test scores (SAT or ACT) for approximately \getval{pct_sat_missing} percent of the \getval{n_experiment_base} students who signed up (Middlebury College has test-optional admissions). To include these students in our regressions, we impute SAT scores for students with missing scores using baseline covariates and assign them to decile bins alongside students with observed scores. 

We estimate an OLS regression of SAT scores on baseline characteristics among the \getval{n_sat_observed} students with observed scores. The predictors are college GPA, high school GPA, age, gender, race and ethnicity indicators (Black, Latino, Asian, with white as the omitted category), international student status, public and private high school indicators, academic field indicators (natural sciences, social sciences, humanities and arts), weekly study hours, whether the student has declared a major, cohort fixed effects, baseline self-assessment of topic knowledge, and baseline number of test questions answered correctly (out of five).\footnote{Predictors with missing values enter as zero together with an indicator for missingness, so all \getval{n_sat_imputation} students with an observed score contribute to the regression and every student without a score receives a prediction.}

We include a missing indicator (equal to one for imputed observations) in the pool of potential controls, so the double-lasso procedure can account for any residual differences between observed and imputed groups. The imputed SAT scores are used only for bin assignment; they do not enter the regression directly as a continuous covariate.

\subsection{Classification of ChatGPT Conversations} \label{app:chatgpt_classification}

During Session One, participants in the AI-allowed condition interacted with ChatGPT through monitored accounts, generating conversation logs that we classified using an LLM (Anthropic's Claude Fable~5.1). The use-type groups in the paper come from a research assistant's hand coding of the same logs (Section~\ref{sub:aug_auto}). We use the LLM labels described here as a check on that coding.

\subsubsection{Prompt-Level Classification.}

We classify each student prompt into one of six categories that mirror the self-reported usage types from Figure~\ref{fig:ai_usage_type}: (1)~\textit{Explaining concepts}---asking ChatGPT to explain, clarify, or teach a concept from the reading; (2)~\textit{Summarizing readings}---asking ChatGPT to summarize or condense the reading material; (3)~\textit{Drafting responses}---asking ChatGPT to write, draft, or generate essay text; (4)~\textit{Proofreading}---asking ChatGPT to check grammar, spelling, or typos; (5)~\textit{Editing responses}---asking ChatGPT to revise, improve, rephrase, or restructure existing text; and (6)~\textit{Other}---anything that does not fit the above categories.

For each prompt, we provide the LLM with the full system prompt establishing the experimental context---that students were given a reading passage on a science topic and asked to write an essay---along with descriptions of each category. The classifier returns a single category label. Across all \getval{n_prompts_classified} student prompts, the distribution is: Explaining concepts (\getval{prompt_pct_explaining_concepts} percent), Other (\getval{prompt_pct_other} percent), Editing responses (\getval{prompt_pct_editing_responses} percent), Drafting responses (\getval{prompt_pct_drafting_responses} percent), Summarizing readings (\getval{prompt_pct_summarizing_readings} percent), and Proofreading (\getval{prompt_pct_proofreading} percent).

The exact system prompt provided to the LLM is: \\

\begin{centering}
	\fbox{\begin{minipage}{\textwidth}
			\ttfamily \footnotesize
			You are classifying student prompts sent to ChatGPT during a learning experiment. Students were given a reading passage on a science topic (Blockchain, CRISPR, or Carbon Capture) and asked to write an essay. Some students had access to ChatGPT during this process.\\
			\\
			Classify each student prompt into exactly ONE of these categories:\\
			\\
			1. Explaining concepts \textemdash{} Asking ChatGPT to explain, clarify, or teach a concept from the reading\\
			2. Summarizing readings \textemdash{} Asking ChatGPT to summarize or condense the reading material\\
			3. Drafting responses \textemdash{} Asking ChatGPT to write, draft, or generate essay text\\
			4. Proofreading \textemdash{} Asking ChatGPT to check grammar, spelling, or typos\\
			5. Editing responses \textemdash{} Asking ChatGPT to revise, improve, rephrase, or restructure existing text\\
			6. Other \textemdash{} Anything that does not fit the above (e.g., off-topic chat, testing the tool)\\
			\\
			Respond with ONLY the category name, exactly as written above. Nothing else.
	\end{minipage}}
\end{centering}

\medskip

To illustrate, Appendix Table~\ref{tab:prompt_examples} provides representative examples of student prompts classified under each category.

\begin{table}[H]{\footnotesize
	\begin{center}
		\caption{Examples of Student Prompts by Classification Category}
		\label{tab:prompt_examples}
		\begin{tabular}{p{3.5cm}p{10cm}}
			\toprule
			Category & Example prompt \\ \midrule
			Explaining concepts & ``easy definition and understanding of crispr'' \\[0.3em]
			& ``how might the cas9 attach to the wrong spot in crispr'' \\[0.3em]
			& ``Give a brief overview on CRISPR technology and then explain why CRISPR differs from other gene editing technology\ldots'' \\ \addlinespace
			Summarizing readings & ``summary with bullet points and main takeaways from this document\ldots'' \\[0.3em]
			& ``Can you give me a summary of this article?'' \\ \addlinespace
			Drafting responses & ``wirte approximately 500 word response to the following prompt: analyze the three main barriers to scaling carbon capture technologies\ldots'' \\[0.3em]
			& ``Write a response paper of at least 500 words\ldots'' \\ \addlinespace
			Editing responses & ``Make this more approachable and simpler for a regular reader'' \\[0.3em]
			& ``rewrite the whole thing now'' \\ \addlinespace
			Proofreading & ``use appropriate grammer and capitalization'' \\ \addlinespace
			Other & ``keyboard shortcuts to copy and paste'' \\[0.3em]
			& ``i think that policy challenges is most difficult what with trump in office and everything'' \\
			\bottomrule
		\end{tabular}
	\end{center}
	\begin{singlespace}\vspace{-0.5cm}
		\footnotesize\noindent\justify
		\textit{Notes:} This table presents example ChatGPT prompts illustrating the classification categories. Each row shows a verbatim student prompt from the ChatGPT conversation logs. Prompts are reproduced exactly as typed, including spelling and grammatical errors. \par
	\end{singlespace}}
\end{table}

\subsubsection{Conversation-Level Classification.}

We also classify each full conversation---the complete sequence of student and ChatGPT messages---into one of four categories: (1)~\emph{Automation}, where the AI does the work for the student (generating essay text, producing draft paragraphs, or creating content that could be pasted directly into the essay); (2)~\emph{Augmentation}, where the AI works with the student (explaining concepts, answering clarifying questions, or providing feedback on the student's own writing); (3)~\emph{Mixed}, where the conversation contains substantial elements of both; and (4)~\emph{Other}, where the conversation is off-topic, unrelated to the academic task, or consists of testing or experimenting with ChatGPT without engaging with the reading or essay.

For each conversation, we provide the LLM with the full exchange between the student and ChatGPT, along with descriptions of each category. Across all ChatGPT conversations, the distribution is: Augmentation (\getval{conv_pct_augmentation} percent), Automation (\getval{conv_pct_automation} percent), Mixed (\getval{conv_pct_mixed} percent), and Other (\getval{conv_pct_other} percent).

To illustrate the distinction, we reproduce excerpts from conversations classified under each category. Appendix Figures~\ref{fig:chat_auto1} and~\ref{fig:chat_aug1} show selected exchanges.

The exact system prompt provided to the LLM is: \\

\begin{centering}
	\fbox{\begin{minipage}{\textwidth}
			\ttfamily \footnotesize
			You are classifying a full ChatGPT conversation from a learning experiment. Students were given a reading passage on a science topic (Blockchain, CRISPR, or Carbon Capture) and asked to write an essay. Some students had access to ChatGPT during this process.\\
			\\
			Classify the overall conversation into exactly ONE of these categories:\\
			\\
			1. Automation \textemdash{} The AI is doing the work FOR the student. The student asks ChatGPT to generate essay text, produce draft paragraphs, write responses, or create content that the student could paste directly into their essay. The student outsources the writing/thinking to the AI.\\
			\\
			2. Augmentation \textemdash{} The AI is working WITH the student. The student uses ChatGPT to understand concepts, get explanations, ask clarifying questions, get feedback on their own writing, or check their work. The student retains ownership of the thinking and writing process.\\
			\\
			3. Mixed \textemdash{} The conversation contains substantial elements of both automation and augmentation, making it hard to assign a single label.\\
			\\
			4. Other \textemdash{} The conversation is off-topic, unrelated to the academic task, or consists of testing/experimenting with ChatGPT without engaging with the essay or reading material.\\
			\\
			Respond with ONLY the category name (Automation, Augmentation, Mixed, or Other). Nothing else.
	\end{minipage}}
\end{centering}

\definecolor{studentblue}{RGB}{219,234,254}   %
\definecolor{studentborder}{RGB}{147,197,253}  %
\definecolor{gptfill}{RGB}{244,244,245}        %
\definecolor{gptborder}{RGB}{209,213,219}      %
\definecolor{gptgreen}{RGB}{16,163,127}        %

\tikzset{
	chat base/.style={
		rounded corners=10pt,
		text width=0.88\linewidth,
		inner sep=10pt,
		outer sep=0pt,
		font=\footnotesize,
		align=left,
	},
	student bubble/.style={
		chat base,
		fill=studentblue,
		draw=studentborder,
		line width=0.4pt,
	},
	gpt bubble/.style={
		chat base,
		fill=gptfill,
		draw=gptborder,
		line width=0.4pt,
	},
	role student/.style={
		font=\footnotesize\sffamily\bfseries,
		text=blue!60!black,
		anchor=south west,
		inner sep=0pt,
	},
	role gpt/.style={
		font=\footnotesize\sffamily\bfseries,
		text=gptgreen!90!black,
		anchor=south west,
		inner sep=0pt,
	},
	avatar/.style={
		circle,
		minimum size=14pt,
		inner sep=0pt,
		font=\tiny\sffamily\bfseries,
		text=white,
	},
	student avatar/.style={avatar, fill=blue!50!black},
	gpt avatar/.style={avatar, fill=gptgreen},
}

\newcommand{\chatexchange}[1]{%
	{\footnotesize
	\begin{tikzpicture}[node distance=0pt]
		#1
	\end{tikzpicture}}%
}

\newcommand{\studentmsg}[2]{%
	\node[student bubble] (#1) {#2};
	\node[student avatar, anchor=south east] at ([xshift=-2pt]#1.north west) {S};
	\node[role student, anchor=south west] at ([xshift=2pt]#1.north west) {Student};
}

\newcommand{\studentmsgbelow}[3]{%
	\node[student bubble, below=12pt of #2] (#1) {#3};
	\node[student avatar, anchor=south east] at ([xshift=-2pt]#1.north west) {S};
	\node[role student, anchor=south west] at ([xshift=2pt]#1.north west) {Student};
}

\newcommand{\gptmsg}[3]{%
	\node[gpt bubble, below=12pt of #2] (#1) {#3};
	\node[gpt avatar, anchor=south east] at ([xshift=-2pt]#1.north west) {G};
	\node[role gpt, anchor=south west] at ([xshift=2pt]#1.north west) {ChatGPT};
}

\begin{figure}[H]
	\caption{Automation Example: Student Delegates Writing to ChatGPT}
	\label{fig:chat_auto1}
	\centering\vspace{0.3cm}
	\chatexchange{
		\studentmsg{s1}{Write a response paper of at least 500 words\ldots{} Analyze which two application areas have the strongest potential for transformative impact and explain the specific factors that support your analysis.\ \ldots{}}
		\gptmsg{g1}{s1}{\textit{[Produces a complete 500-word essay on supply chain management and healthcare.]}}
		\studentmsgbelow{s2}{g1}{\textit{[Pastes a lightly edited version of the essay and asks for its word count.]}}
		\gptmsg{g2}{s2}{\textit{[Replies that the essay has 567 words.]}}
	}
	\begin{singlespace}\vspace{-0.1cm}
		\footnotesize\noindent\justify
		\textit{Notes:} This figure shows an excerpt of a ChatGPT conversation classified as automation. The student's first prompt is verbatim. The bracketed text summarizes the other turns. %
	\end{singlespace}
\end{figure}

\begin{figure}[H]
	\caption{Augmentation Example: Student Uses ChatGPT as an Explainer}
	\label{fig:chat_aug1}
	\centering\vspace{0.3cm}
	\chatexchange{
		\studentmsg{s1}{what are main technical challenges of crispr}
		\gptmsg{g1}{s1}{\textit{[Lists four challenges: off-target effects, delivery mechanisms, control over DNA repair outcomes, and mosaicism, with a short explanation of each.]}}
		\studentmsgbelow{s2}{g1}{which challenge is most important to address first}
		\gptmsg{g2}{s2}{\textit{[Argues that off-target effects come first because safety and regulatory approval depend on precision, and notes that delivery or mosaicism can matter more for in vivo or germline editing.]}}
		\studentmsgbelow{s3}{g2}{what are promising approaches to overcome off target effects}
		\gptmsg{g3}{s3}{\textit{[Describes high-fidelity Cas variants, optimized guide RNAs, base and prime editing, and genome-wide off-target detection methods.]}}
	}
	\begin{singlespace}\vspace{-0.1cm}
		\footnotesize\noindent\justify
		\textit{Notes:} This figure shows an excerpt of a ChatGPT conversation classified as augmentation. The student's three prompts are verbatim. The bracketed text summarizes ChatGPT's replies. \par
	\end{singlespace}
\end{figure}

\clearpage
\subsection{Literature Comparison} \label{app:literature_comparison}

We surveyed experimental papers on generative AI and learning outcomes published or circulated between 2023 and 2026. We required that each study satisfy seven criteria: (1) a randomized experimental design; (2) an effect size reported in standard deviations or convertible from reported statistics; (3) a standard error or confidence interval (reported or back-calculable); (4) an outcome measuring learning on an unassisted assessment, excluding tasks where AI was available during evaluation; (5) a clean no-AI control, rather than an active comparison such as human tutoring or structured hints; (6) a total sample size of at least 50 participants; and (7) a sample composed of students, rather than working professionals or general online participants. 

Appendix Table~\ref{tab:literature_summary} summarizes the 14 included studies; we describe each below.

\begin{table}[H]
	\centering
	\caption{Included Studies in Literature Comparison} \label{tab:literature_summary}
	\footnotesize
	\begin{tabular}{llrll}
		\toprule
		Study & Domain & $N$ & Design & Outcome \\
		\midrule
		\citet{ba.etal2024}           & Medicine          & 77        & RCT                  & Unassisted exam \\
		\citet{bassner.etal2026}      & Coding            & 275       & RCT                  & Unassisted post-test \\
		\citet{bastani.etal2025}      & Math              & 839       & Cluster RCT          & Unassisted exam \\
		\citet{dai.etal2025}          & Physics           & 349       & RCT ($\times 2$)     & Unassisted exam \\
		\citet{desimone.etal2025}     & Language          & 759       & RCT                  & Unassisted post-test \\
		\citet{fischer.etal2025}      & Economics         & 334       & Lab RCT              & Unassisted exam \\
		\citet{franco.etal2026} & Language & 572 & Lab RCT & Unassisted exam \\
		\citet{gan.etal2024}          & Medicine          & 110       & RCT                  & Unassisted exam \\
		\citet{hou.etal2026}          & Engineering       & 95        & RCT                  & Unassisted post-test \\
		\citet{huang.etal2025}        & Medicine          & 187       & RCT                  & Unassisted skill test \\
		\citet{kavadella.etal2024}    & Medicine          & 70        & RCT                  & Unassisted exam \\
		\citet{kazemitabaar.etal2023} & Coding            & 69        & RCT                  & Post-test + 1-wk retention \\
		\citet{learnlm.team2026}      & Math              & 1{,}423   & Cluster RCT          & Unassisted exam \\
		\citet{lehmann.etal2025}      & Coding            & 176       & Lab RCT ($\times 2$) & Unassisted post-test \\
		\bottomrule
	\end{tabular}
	\begin{singlespace}\vspace{-0.1cm}
		\footnotesize\noindent\justify
		\textit{Notes:} This table lists the studies included in our literature comparison. $N$ is the total sample across all arms included in our comparison. All studies compare AI access to a no-AI control on unassisted assessments. \par
	\end{singlespace}
\end{table}

\citet{ba.etal2024} randomized 77 medical interns rotating through pediatric cardiology at Sun Yat-sen University to two weeks of ChatGPT-assisted instruction or standard bedside teaching with identical cases and instructors. Scores on a closed-book knowledge exam were statistically indistinguishable ($-$0.07~SD), with both groups near the test ceiling. We computed the effect size and standard error from reported means and standard deviations.

\citet{bassner.etal2026} randomized 452 introductory-programming students at TU Munich to a 90-minute Java exercise with a scaffolded LLM tutor (Iris), unrestricted ChatGPT, or no AI (275 analyzed after exclusions). Both AI arms scored higher on the exercise itself, but on the supervised, unassisted knowledge test neither arm learned more than the control: baseline-adjusted gains are $-$0.07~SD (Iris) and $-$0.01~SD (ChatGPT)---the paper's titular dissociation between performance and learning. We standardized the difference in pre--post gains by the pooled pre-test standard deviation.

\citet{bastani.etal2025} conducted a cluster-randomized trial with 839 Turkish high school students learning math on an online platform. Students were assigned to base GPT-4 access, GPT-4 with a tutoring prompt that withheld direct answers, or a no-AI control, and took an unassisted math exam after the practice period. Although base GPT-4 students solved more practice problems during the AI-assisted phase, this advantage reversed on the unassisted exam: the base group scored $-$0.19~SD relative to the control, while the tutor-prompted group scored $-$0.01~SD. We converted the authors' raw-scale standard errors to SD units by dividing by the control-arm standard deviation (0.277).

\citet{dai.etal2025} ran two randomized experiments with grade-10 physics students at a Chinese high school (Experiment~1: $N = 83$ in the two arms we use; Experiment~2: $N = 266$), providing LLM-generated feedback on regular homework over five weeks. In Experiment~1, treated students received recommended problems with AI heuristic hints, and we exclude the study's third arm, which received recommended problems without AI hints; in Experiment~2, students could request AI help on demand while studying, either choosing the feedback type themselves or receiving the system's choice. Effects on the unassisted end-of-term physics exam are small and insignificant in all three AI arms relative to no-intervention controls (0.03--0.21~SD). Exam scores are standardized, so regression coefficients are effect sizes directly. %

\citet{desimone.etal2025} studied 759 secondary school students in Nigeria randomly assigned to practice English language skills with an AI chatbot or a no-AI control. Students showed gains on unassisted post-tests in English skills (0.24~SD) and on the school's regular third-term curricular exam at the end of the intervention (0.21~SD). The authors also report a ``total'' estimate bundling AI and digital literacy skills directly taught by the intervention; we exclude it because it captures treatment-specific content rather than general learning.

\citet{fischer.etal2025} randomized 334 university students in Berlin to study two microeconomics textbook excerpts with a GPT-4-based tutor grounded in the readings---available either throughout the session or only after ten minutes of independent reading---or with the textbook alone. On an incentivized, unassisted 25-question exam, unrestricted AI tutoring raised performance by 0.34~SD, while restricted access produced a statistically insignificant 0.13~SD. Test scores are standardized, so the reported coefficients are effect sizes directly.

\citet{franco.etal2026} randomized 572 students at the University of Nottingham to study Esperanto in a proctored lab session---a 15-minute lesson followed by about 20 minutes of practice questions---with a browsing-only control, unrestricted ChatGPT access, or ChatGPT access plus brief written guidance on learning-oriented use. On an incentivized, unassisted 15-question exam, neither AI arm differs from the control (0.04~SD for unrestricted access, $-$0.12~SD for guided access), and students with AI attempt fewer practice questions.\footnote{The paper reports coefficients on the 0--15 exam scale but not the score's standard deviation; we standardize by a standard deviation of 1.9, backed out from the residual standard deviation implied by the reported regression standard errors and consistent with the 1.7--2.0 range implied by the control group's mean and tail shares under normality.}

\citet{gan.etal2024} randomized 125 third-year medical students at Jinan University to a one-week orthopedics review using ChatGPT-4 or conventional internet resources, 110 of whom were analyzed after 15 withdrawals. The ChatGPT group scored 0.40~SD higher on an unassisted 214-item orthopedics exam. We computed the effect size and standard error from reported means and standard deviations.

\citet{hou.etal2026} randomized 95 undergraduates with no prior construction-related coursework to review construction-engineering material with a generative AI assistant---with or without a structured prompting framework---or with lecture slides only. On an unassisted post-test combining multiple-choice and open-ended items, guided AI use raised total scores by 0.86~SD, while unguided use produced a near-zero effect (0.09~SD). We computed effect sizes and standard errors from reported means and standard deviations.

\citet{huang.etal2025} randomized 187 dental students at Wuhan University to one week of operative-skills training with instructional videos plus ChatGPT-3.5 or the same videos alone. The ChatGPT group scored 0.67~SD higher on an unassisted virtual-reality skill assessment. We computed the effect size and standard error from reported means and standard deviations.

\citet{kavadella.etal2024} randomized 77 second-year dental students at European University Cyprus to complete a radiation-biology assignment using ChatGPT or conventional literature search. On an unannounced, unassisted ten-question exam, which 70 of the 77 sat, the ChatGPT group scored 0.52~SD higher. We computed the effect size and standard error from reported means and standard deviations.

\citet{kazemitabaar.etal2023} ran a ten-session randomized trial with 69 K--12 learners (ages 10--17, novice programmers) using OpenAI Codex to support Python authoring tasks. AI access produced large performance gains during training (1.67~SD) but no detectable difference on an immediate unassisted post-test ($-$0.05~SD; the baseline group scored slightly higher, 62.9 versus 61.3 percent) and only a marginal difference on a one-week retention test of code modification (0.41~SD). Standard errors were back-calculated from the reported $N$ and standardized effect size.

\citet{learnlm.team2026} conducted a preregistered trial randomizing 48 mathematics classrooms (1,763 grade 7--8 students enrolled, 1,423 in the analyzed balanced panel) in Sierra Leone to integrate Gemini's Guided Learning feature into half of weekly lessons for eight weeks or to continue standard instruction. The intent-to-treat effect on an unassisted, IRT-scaled endline mathematics assessment is 0.26~SD, with standard errors clustered at the classroom level.

\citet{lehmann.etal2025} ran two laboratory experiments with German university students learning Python programming (Study~2: $N = 107$; Study~3: $N = 69$). In both, treated students used ChatGPT during practice and then completed a 20-question unassisted coding post-test. Effects are 0.25~SD (Study~2) and 0.42~SD (Study~3), neither individually significant at conventional levels. Study~2 included an unintended copy-paste restriction that may have affected the treatment.

The random-effects grand mean in Figure~\ref{fig:literature_comparison} uses \citet{dersimonian.laird1986} weights, accounting for between-study heterogeneity through a variance component estimated from study-level effects. The pool is the 24 literature estimates, aggregated within each study by inverse-variance weighting so that each of the 14 studies enters once (our own estimates are excluded). 

\clearpage
\section{Student Beliefs About AI and Learning} \label{app:open_ended}

\setcounter{table}{0}
\setcounter{figure}{0}
\setcounter{equation}{0}
\renewcommand{\thetable}{C\arabic{table}}
\renewcommand{\thefigure}{C\arabic{figure}}
\renewcommand{\theequation}{C\arabic{equation}}

This appendix analyzes student responses to an open-ended question included in the Session Two survey: ``In your opinion, how does generative AI (e.g., ChatGPT) affect student learning in college? Please explain your reasoning.'' All \getval{n_both_sessions} students who attended both sessions answered the question. We analyze these responses in two ways: sentiment analysis to show the measure carries signal rather than noise, and causal-graph coding, following \citet{andre.etal2026}, to characterize how students reason about AI and learning.

\subsection{Validating the Open-Ended Response Measure} \label{app:open_ended_validation}

\begin{figure}[H]
	\caption{Distribution of AI Sentiment Scores}\label{fig:vader_hist}
	\centering
	\includegraphics[width=.62\linewidth]{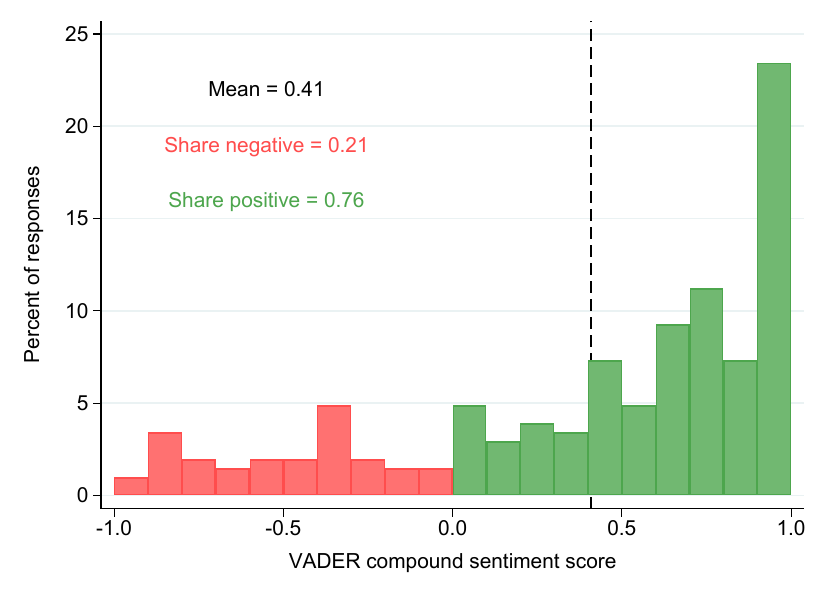}
	{\footnotesize
		\singlespacing \justify

		\textit{Notes:} This figure plots the distribution of compound sentiment scores across students who responded to the open-ended question about how generative AI affects student learning. Scores are computed using the VADER algorithm of \citet{hutto2014vader} and range from $-1$ (most negative) to $+1$ (most positive). The dashed vertical line marks the mean. \par
	}
\end{figure}

We validate our open-ended response measure using VADER (Valence Aware Dictionary and sEntiment Reasoner), a lexicon-based sentiment analysis tool \citep{hutto2014vader}. VADER assigns each response a compound sentiment score ranging from $-1$ (most negative) to $+1$ (most positive). Appendix Figure~\ref{fig:vader_hist} plots the distribution of these scores. Responses skew positive, with most students expressing net-positive sentiment about AI's effect on learning and a minority expressing negative sentiment.

\begin{figure}[H]
	\caption{Relationship Between AI Sentiment and AI Usage}\label{fig:vader_experiment}
	\centering
	\begin{subfigure}[t]{.48\textwidth}
		\caption*{Panel A. Used provided ChatGPT account}
		\centering
		\includegraphics[width=\linewidth]{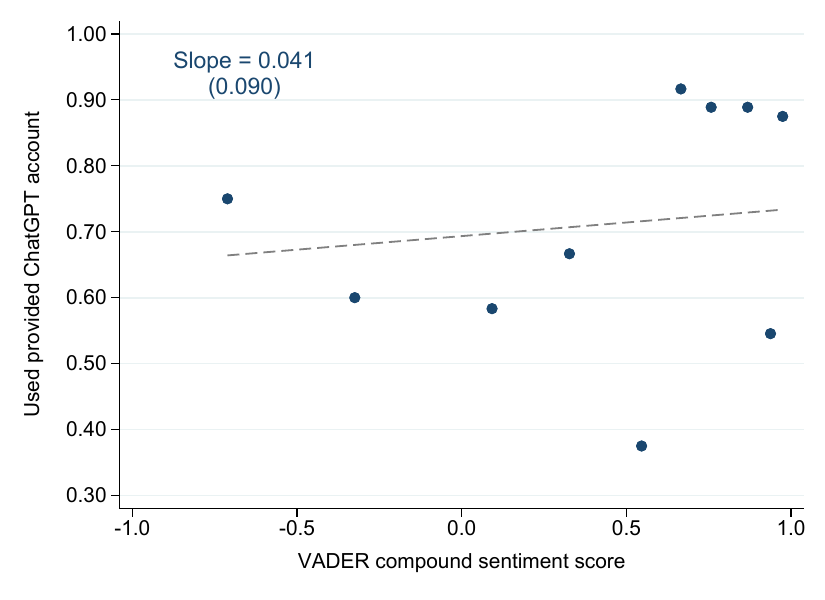}
	\end{subfigure}
	\hfill
	\begin{subfigure}[t]{0.48\textwidth}
		\caption*{Panel B. Frequently uses AI for academics}
		\centering
		\includegraphics[width=\linewidth]{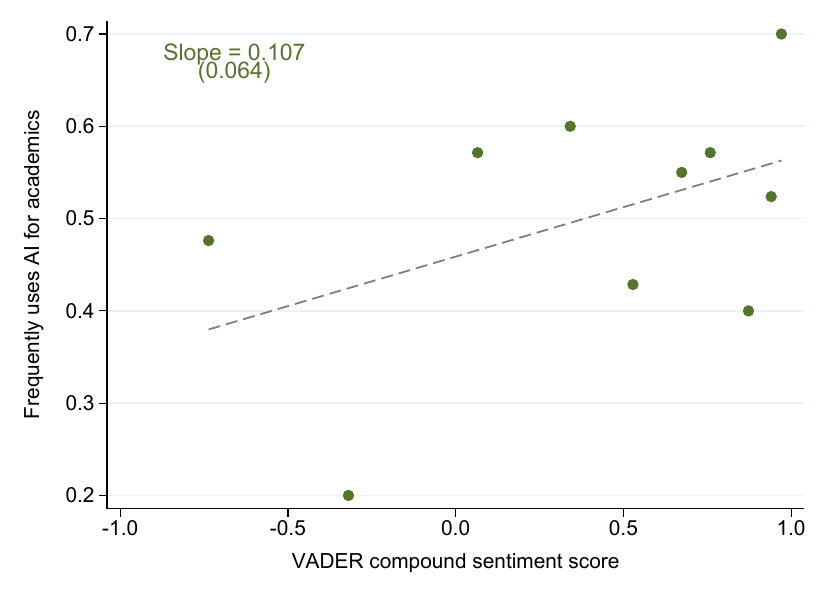}
	\end{subfigure}
	{\footnotesize
		\singlespacing \justify

		\textit{Notes:} This figure presents the relationship between AI sentiment and AI usage. Panel~A restricts to students in the AI-allowed condition and shows the proportion who used AI during Session One (from usage logs). Panel~B shows an indicator for using AI for academic purposes frequently or very frequently during the semester, for all respondents. Each point represents the mean outcome for respondents within sentiment score bins. Sentiment scores are compound scores computed using the VADER algorithm of \citet{hutto2014vader} applied to responses to the open-ended question about AI and learning. Positive values indicate positive sentiment and negative values indicate negative sentiment. The dashed lines show OLS best-fit lines estimated on the microdata. \par
	}
\end{figure}

To assess whether the open-ended responses carry a relevant signal, we test whether their sentiment tracks students' actual AI use. Appendix Figure~\ref{fig:vader_experiment} presents binned scatterplots relating compound sentiment to two measures of AI use. Sentiment is weakly related to both measures. Panel~A relates sentiment to whether an AI-allowed student used AI during Session One. A regression of AI use on the sentiment score yields a positive but statistically insignificant coefficient ($\hat{\beta} = \getval{vader_takeup_b}$, $p = \getval{vader_takeup_p}$). Panel~B shows that students who express more positive views about AI's effect on learning use AI somewhat more for academic work during the semester ($\hat{\beta} = \getval{vader_semester_b}$, $p = \getval{vader_semester_p}$). This pattern is consistent with the open-ended responses capturing a relevant signal.

\subsection{Narratives as Causal Graphs} \label{app:open_ended_method}

We characterize the content of the open-ended responses by coding each one as a causal graph, following \citet{andre.etal2026}, who analyze how households and experts explain macroeconomic events. In their setting respondents explain why inflation rose, and the explanation is coded as a chain of cause-and-effect links running from a driving factor to inflation. We adapt their method in one substantive way: their links are unsigned, because in the episode they study inflation only rose, whereas AI can raise or lower learning. Each link in our graphs therefore carries a sign. We represent each response as a signed causal graph from a source node, AI use, to a sink node, learning, passing through the mechanisms the student names. A link is positive if the student describes the mechanism as raising learning and negative if as lowering it.

We proceed in two steps. First, we develop the codebook. Using the full set of student responses, we catalog the distinct mechanisms students raise, and consolidate near-synonyms into fifteen mechanisms. Second, we code every response against this fixed codebook with an LLM. We group the mechanisms by their effect on learning. Seven promote it: explanation and tutoring, efficiency, information access, brainstorming, personalized feedback, access and equity, and engagement. Eight harm it: shortcutting and cheating, cognitive offloading, skill atrophy, dependency, inaccuracy, reduced productive struggle, retention loss, and reduced human contact. A final node, \emph{mode of use}, appears when the student's channels branch according to how AI is used (``it depends on how you use it''). This node is a mediator between AI use and the channels rather than a channel itself: AI use feeds mode of use, which branches into channels that raise learning and channels that lower it, so conditionality is encoded by the node carrying both a promoting and a harming arm. 

\begin{figure}[H]
	\caption{Three Example Student Narratives, Coded as Signed Causal Graphs}\label{fig:dag_examples}
	\centering
	\begin{subfigure}{\textwidth}
		\caption*{Panel A. A narrative in which AI helps learning}
		\centering
		\includegraphics[width=.9\linewidth]{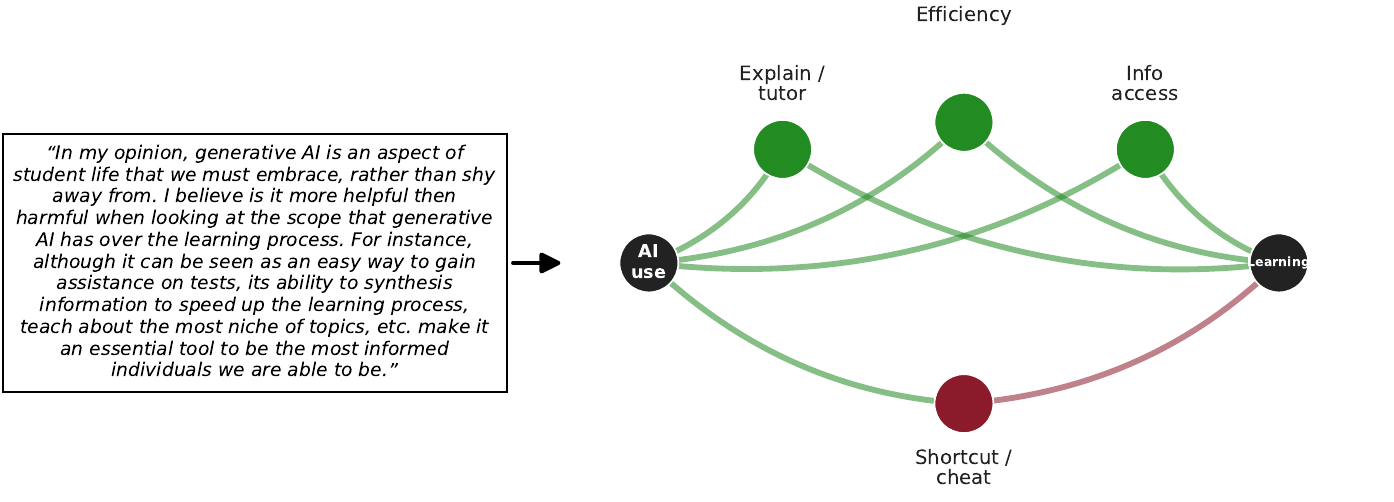}
	\end{subfigure}

	\vspace{0.2cm}
	\begin{subfigure}{\textwidth}
		\caption*{Panel B. A narrative in which AI harms learning}
		\centering
		\includegraphics[width=.9\linewidth]{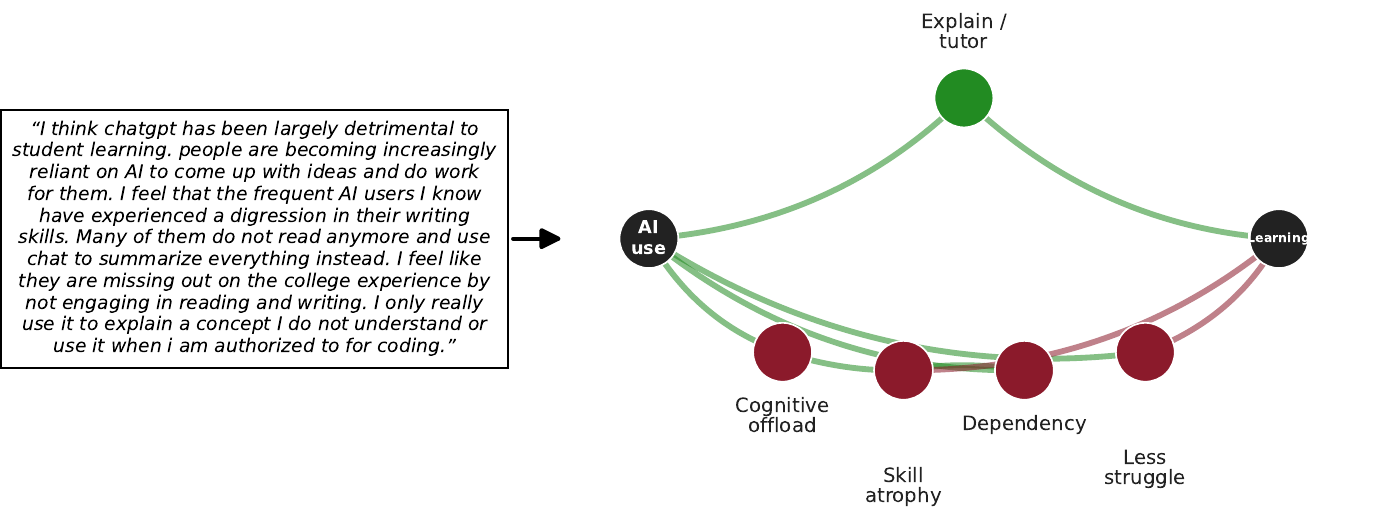}
	\end{subfigure}

	\vspace{0.2cm}
	\begin{subfigure}{\textwidth}
		\caption*{Panel C. A conditional narrative}
		\centering
		\includegraphics[width=.9\linewidth]{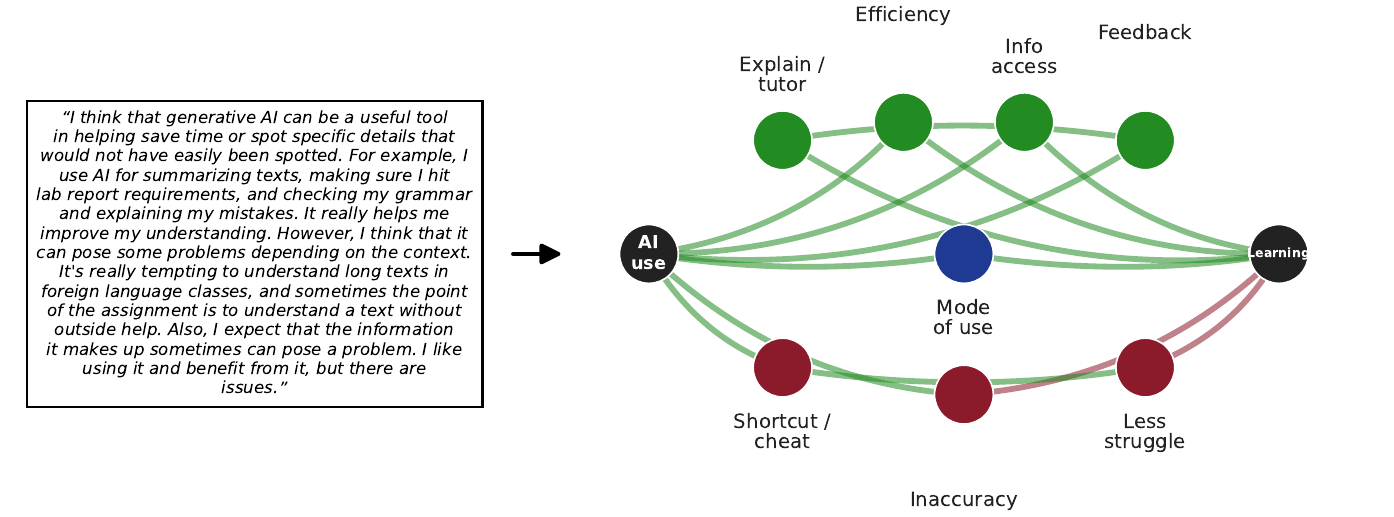}
	\end{subfigure}
	{\footnotesize
		\singlespacing \justify

		\textit{Notes:} Each panel pairs a student's verbatim open-ended response (left) with the response coded as a signed causal graph from AI use to learning (right), following \citet{andre.etal2026}, so the mapping from text to graph can be assessed directly. Green links denote mechanisms the student describes as promoting learning, red links mechanisms that harm it, and navy the mode-of-use node. \par
	}
\end{figure}

To illustrate how the coding works, Appendix Figure~\ref{fig:dag_examples} shows the directed acyclic graphs of three specific students. The coder assigns each response an overall valence: the student's bottom-line judgment of whether AI helps learning (Panel~A), harms it (Panel~B), or depends on how it is used (Panel~C). The Panel~A student concludes that AI helps learning yet flags the temptation to shortcut; the Panel~B student concludes that AI harms learning yet credits AI with explaining difficult concepts. The pattern extends beyond these two students: \getval{narr_helps_concede_pct} percent of net-positive narratives name at least one harming channel, and \getval{narr_harms_concede_pct} percent of net-negative narratives name at least one promoting channel.

The taxonomy also maps onto the use classification in the main text. Explanation, feedback, and brainstorming are channels in which AI works \textit{with} the student; shortcutting and offloading are channels in which AI works \textit{for} the student. The coder accordingly classifies how each response frames AI use: \emph{augmentation} when the student describes working with AI (asking it to explain, give feedback, or check work), \emph{automation} when the student describes AI doing the work in their place, both, or unspecified.

We code each response with an LLM (Claude Opus 4.8). The majority of responses ($98$ percent) can be represented as a causal graph; the rest are non-substantive (e.g., ``I don't know''). As a validation, we code every response a second time with a model from a different developer (OpenAI's GPT-5.5) and compute their inter-rater agreement statistic, the pooled probability that a code assigned by one rater is also assigned by the other. Agreement matches that of the trained human coders in \citet{andre.etal2026}: \getval{narr_irr_factor} for the individual mechanisms (0.88 in their data), \getval{narr_irr_group} and \getval{narr_irr_group_harm} for the coarser grouping into any promoting and any harming channel (0.94 and 0.93), and \getval{narr_irr_link} for the signed links (0.77), with the mechanism and link statistics far above their random-coding benchmarks of \getval{narr_irr_factor_rand} and \getval{narr_irr_link_rand}, respectively. The two models agree on a response's overall valence \getval{narr_irr_valence_pct} percent of the time and on its augmentation-versus-automation framing \getval{narr_irr_framing_pct} percent of the time.\footnote{Two checks confirm that the coding is not model-specific. Coding every response a third time with a different model from the primary coder's developer (Anthropic's Claude Sonnet 4.6) gives nearly identical agreement with the primary coder (0.85 for mechanisms, 0.66 for links, and 92 percent for valence). Agreement rises with model capability: within the OpenAI line, GPT-5.5 reaches mechanism and link agreement of 0.88 and 0.73, whereas two smaller models reach 0.77 and 0.53, and 0.70 and 0.47. All remain far above the random-coding benchmarks, but we use frontier models for both coders.}

\subsection{The Narratives of College Students} \label{app:narratives_results}

\subsubsection{The Average Narrative.} \label{app:open_ended_average}

 The average narrative contains \getval{narr_mean_factors} mechanisms and \getval{narr_mean_links} links, and \getval{narr_multi_pct} percent of coded responses mention more than one mechanism. Among students naming any mechanism, \getval{narr_both_pct} percent name both a promoting and a harming channel, and \getval{narr_cond_pct} percent of coded responses say that the effect depends on how AI is used. The most common single node is mode of use (\getval{narr_usemode_pct} percent), and the most common promoting and harming mechanisms are explanation and tutoring (\getval{narr_explain_pct} percent) and shortcutting and cheating (\getval{narr_shortcut_pct} percent).

\subsubsection{Variation Across Students.} \label{app:open_ended_variation}

We ask how narratives vary with student characteristics. Appendix Table~\ref{tab:narr_covariates} regresses each of six narrative features---three \textit{sophistication} measures (the number of mechanisms, the number of signed links, and an indicator for naming more than one mechanism) and three \textit{framing} measures (augmentation, automation, and conditional-on-use)---on the randomized treatment and the full set of student characteristics from Appendix Table~\ref{tab:compliance}.

Among the AI-experience measures, only self-assessed proficiency predicts how students frame AI. Students with proficiency at or above the median are \getval{narr_mv_aug_prof_pp} percentage points (pp) more likely than less proficient students to frame AI as augmentation ($p = \getval{narr_mv_aug_prof_pval}$) and \getval{narr_mv_auto_prof_pp}~pp less likely to frame it as automation ($p = \getval{narr_mv_auto_prof_pval}$). Early adopters (those who first used AI for academics before Fall 2024) do not differ from later adopters on any narrative feature. Randomized AI access lowers the share framing AI as automation by \getval{narr_mv_auto_treat_pp}~pp ($p = \getval{narr_mv_auto_treat_pval}$) and the share naming more than one mechanism by \getval{narr_mv_multi_treat_pp}~pp ($p = \getval{narr_mv_multi_treat_pval}$). Framing thus tracks self-assessed proficiency rather than frequency of use or early adoption.

Demographics, academic background, and beliefs, by contrast, do predict narratives. Men name \getval{narr_mv_male_factors} fewer mechanisms and \getval{narr_mv_male_links} fewer links than women ($p = \getval{narr_mv_male_factors_pval}$ and $p = \getval{narr_mv_male_links_pval}$), the same gender difference \citet{andre.etal2026} document in narratives about inflation. Students with GPAs at or above the median name \getval{narr_mv_gpa_factors} more mechanisms ($p = \getval{narr_mv_gpa_factors_pval}$). Framing varies instead with field of study and beliefs about AI. Natural-science majors are \getval{narr_mv_natsci_cond_pp}~pp more likely to condition AI's effect on how it is used ($p = \getval{narr_mv_natsci_cond_pval}$) and \getval{narr_mv_natsci_auto_pp}~pp less likely to frame it as automation ($p = \getval{narr_mv_natsci_auto_pval}$). Students who expect large AI effects on their own test scores frame AI the same way: they are \getval{narr_mv_perc_auto_pp}~pp less likely to frame it as automation ($p = \getval{narr_mv_perc_auto_pval}$) and \getval{narr_mv_perc_cond_pp}~pp more likely to condition its effect on use ($p = \getval{narr_mv_perc_cond_pval}$).

\begin{table}[H]{\footnotesize\renewcommand{\arraystretch}{0.9}
		\begin{center}
			\caption{Correlates of Students' Narratives}\label{tab:narr_covariates}
			\newcommand\w{1.25}
			\begin{tabular}{l@{}lR{\w cm}@{}L{0.45cm}R{\w cm}@{}L{0.45cm}R{\w cm}@{}L{0.45cm}R{\w cm}@{}L{0.45cm}R{\w cm}@{}L{0.45cm}R{\w cm}@{}L{0.45cm}}
				\midrule
				&& \multicolumn{6}{c}{Narrative sophistication} & \multicolumn{6}{c}{Framing of AI use} \\ \cmidrule(lr){3-8} \cmidrule(lr){9-14}
				&& \# mech. && \# links && $>$1 mech. && Augment. && Automat. && Condit. \\
				&& (1) && (2) && (3) && (4) && (5) && (6) \\ \midrule
				AI access (randomized)&&0.131&&0.425&&$-$0.073&\sym{*}&$-$0.018&&$-$0.098&\sym{*}&0.066&\\
&&(0.208&)&(0.439&)&(0.043&)&(0.061&)&(0.056&)&(0.078&)\\

				\midrule
				\multicolumn{13}{l}{\hspace{-1em}\textbf{Panel A. Demographic characteristics}} \\ \addlinespace
				Male&&$-$0.407&\sym{**}&$-$0.782&\sym{*}&$-$0.039&&$-$0.060&&$-$0.008&&$-$0.035&\\
&&(0.199&)&(0.404&)&(0.040&)&(0.057&)&(0.056&)&(0.074&)\\

				White&&0.301&&0.590&&0.008&&0.084&&$-$0.005&&$-$0.106&\\
&&(0.200&)&(0.405&)&(0.043&)&(0.071&)&(0.062&)&(0.076&)\\

				International student&&0.026&&$-$0.089&&0.016&&0.102&&0.022&&$-$0.120&\\
&&(0.266&)&(0.558&)&(0.047&)&(0.094&)&(0.075&)&(0.107&)\\

				Private high school&&0.122&&0.213&&0.004&&$-$0.007&&$-$0.058&&0.035&\\
&&(0.195&)&(0.421&)&(0.043&)&(0.069&)&(0.070&)&(0.088&)\\

				\midrule
				\multicolumn{13}{l}{\hspace{-1em}\textbf{Panel B. Academic background}} \\ \addlinespace
				Natural-science major&&$-$0.324&&0.027&&$-$0.018&&$-$0.106&&$-$0.214&\sym{***}&0.306&\sym{***}\\
&&(0.238&)&(0.497&)&(0.064&)&(0.072&)&(0.068&)&(0.094&)\\

				Social-science major&&$-$0.026&&0.141&&0.015&&0.053&&$-$0.084&&0.070&\\
&&(0.231&)&(0.461&)&(0.045&)&(0.070&)&(0.074&)&(0.091&)\\

				High GPA&&0.451&\sym{**}&0.605&&0.069&&$-$0.004&&$-$0.025&&0.005&\\
&&(0.200&)&(0.423&)&(0.047&)&(0.070&)&(0.060&)&(0.079&)\\

				Freshman&&0.468&\sym{*}&1.069&\sym{**}&$-$0.005&&$-$0.115&&$-$0.084&&0.156&\sym{*}\\
&&(0.249&)&(0.507&)&(0.058&)&(0.075&)&(0.077&)&(0.092&)\\

				Sophomore&&0.268&&0.522&&0.036&&0.062&&$-$0.040&&0.044&\\
&&(0.312&)&(0.675&)&(0.068&)&(0.116&)&(0.086&)&(0.115&)\\

				Junior&&0.506&\sym{*}&0.773&&0.055&&$-$0.106&&0.022&&0.030&\\
&&(0.257&)&(0.541&)&(0.050&)&(0.077&)&(0.077&)&(0.097&)\\

				\midrule
				\multicolumn{13}{l}{\hspace{-1em}\textbf{Panel C. AI experience and beliefs}} \\ \addlinespace
				Frequent AI user&&0.293&&0.602&&$-$0.034&&0.018&&$-$0.082&&0.068&\\
&&(0.252&)&(0.505&)&(0.042&)&(0.074&)&(0.059&)&(0.079&)\\

				Early AI adopter&&$-$0.261&&$-$0.275&&$-$0.037&&0.033&&$-$0.011&&0.004&\\
&&(0.213&)&(0.431&)&(0.041&)&(0.059&)&(0.053&)&(0.069&)\\

				High AI proficiency&&0.110&&0.275&&0.039&&0.163&\sym{**}&$-$0.124&\sym{*}&$-$0.013&\\
&&(0.267&)&(0.528&)&(0.050&)&(0.072&)&(0.064&)&(0.084&)\\

				High perceived AI effect&&0.351&&0.964&\sym{**}&$-$0.022&&$-$0.063&&$-$0.164&\sym{***}&0.172&\sym{**}\\
&&(0.217&)&(0.454&)&(0.051&)&(0.069&)&(0.059&)&(0.085&)\\

				\midrule
				\(N\) (Students)&&196&&196&&196&&196&&196&&196&\\
 \midrule
			\end{tabular}
		\end{center}

		\begin{singlespace}\vspace{-0.5cm}
			\footnotesize\noindent\justify
			\textit{Notes:} Each column reports an OLS regression of the indicated narrative feature on all rows jointly, estimated on the codable open-ended responses. Outcomes come from each student's coded causal graph: the number of mechanisms, the number of signed links, an indicator for naming more than one mechanism, and indicators for the augmentation, automation, and conditional-on-use framings. All characteristics are indicators, defined as in Appendix Table~\ref{tab:compliance} with the medians recomputed in this sample: frequent AI user indicates use during the semester at or above this sample's median (frequently or more often), and high perceived AI effect is a perceived gain of at least one question; missing values are set to zero, with missing indicators included. Heteroskedasticity-robust standard errors in parentheses. $^{***}$ $p < 0.01$; $^{**}$ $p < 0.05$; $^{*}$ $p < 0.10$. \par
	\end{singlespace}}
\end{table}

\subsubsection{Narrative Clusters.} \label{app:open_ended_clusters}

We next ask whether students' narratives come in recurring types or ``clusters.'' Following \citet{andre.etal2026}, we measure the distance between two narratives as the fraction of their combined signed links that the two do not have in common (the Jaccard distance), group narratives with agglomerative hierarchical clustering with average linkage, and choose the number of clusters to maximize the average silhouette, a standard measure of how well each narrative fits its own cluster relative to the nearest alternative.

\begin{figure}[H]
	\caption{Narrative Clusters}\label{fig:narr_clusters}
	\centering
	\begin{subfigure}[t]{.48\textwidth}
		\caption*{Panel A. Double-edged (\getval{narr_cl_double_pct} percent)}
		\centering
		\includegraphics[width=\linewidth]{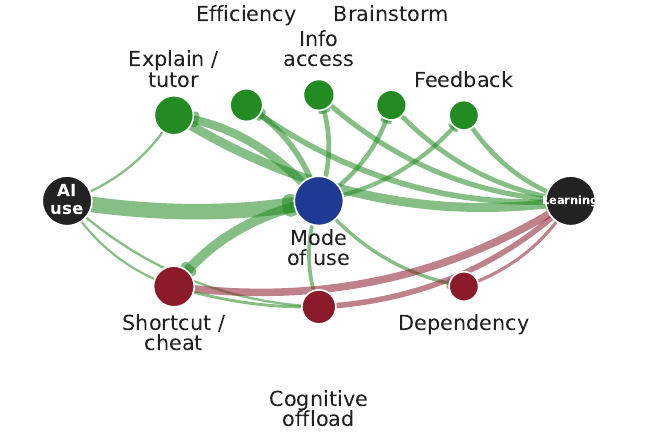}
	\end{subfigure}
	\hfill
	\begin{subfigure}[t]{.48\textwidth}
		\caption*{Panel B. Augmentation (\getval{narr_cl_aug_pct} percent)}
		\centering
		\includegraphics[width=\linewidth]{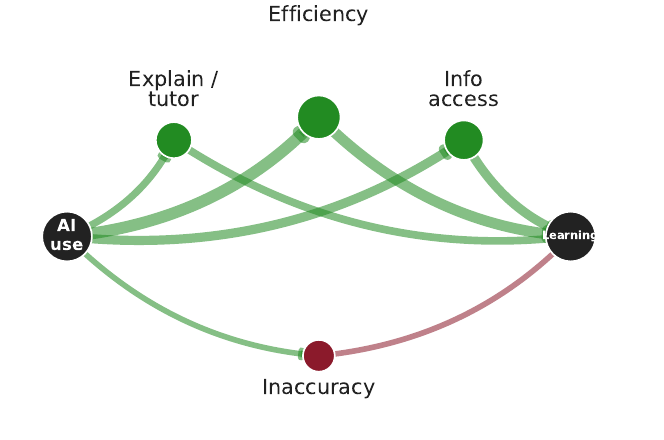}
	\end{subfigure}

	\vspace{0.2cm}
	\begin{subfigure}[t]{.48\textwidth}
		\caption*{Panel C. Automation (\getval{narr_cl_auto_pct} percent)}
		\centering
		\includegraphics[width=\linewidth]{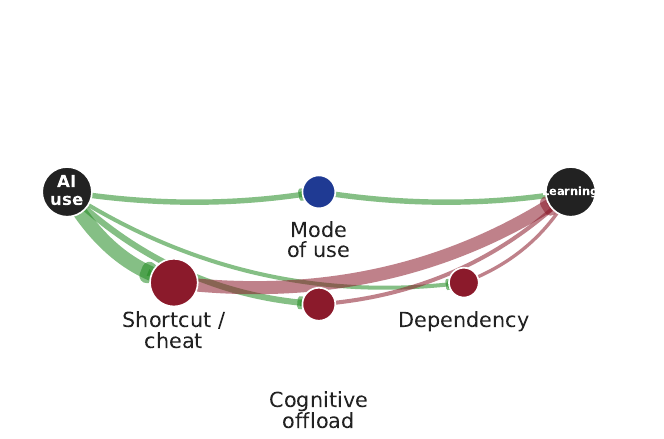}
	\end{subfigure}
	\hfill
	\begin{subfigure}[t]{.48\textwidth}
		\caption*{Panel D. Deskilling (\getval{narr_cl_deskill_pct} percent)}
		\centering
		\includegraphics[width=\linewidth]{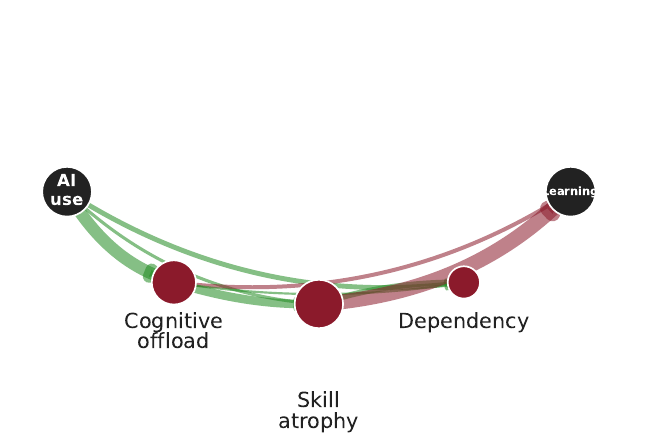}
	\end{subfigure}
	{\footnotesize
		\singlespacing \justify

		\textit{Notes:} This figure shows the average causal graph within each narrative cluster; panel headings report each cluster's share of coded narratives. We form the clusters with agglomerative hierarchical clustering, using average linkage on the Jaccard distance between signed-link sets and choosing the number of clusters to maximize the average silhouette. We show the four clusters with at least eight responses and, within each panel, omit mechanisms below 20 percent and links below 6 percent. \par
	}
\end{figure}

Four narrative types cover \getval{narr_cl_top4_pct} percent of coded responses, while the rest scatter across small clusters (Appendix Figure~\ref{fig:narr_clusters}). The modal type (\getval{narr_cl_double_pct} percent) is the double-edged narrative, in which mode of use feeds both an explanation channel and a shortcutting channel. An augmentation type (\getval{narr_cl_aug_pct} percent) chains efficiency, information access, and explanation to higher learning, though many of these narratives also flag accuracy concerns, and an automation type (\getval{narr_cl_auto_pct} percent) chains shortcutting and offloading to lower learning. The remaining type is smaller and one-sided: a deskilling narrative (\getval{narr_cl_deskill_pct} percent), in which skill atrophy and offloading erode critical thinking.

\clearpage
\section{Conceptual Framework} \label{app:framework}

\setcounter{table}{0}
\setcounter{figure}{0}
\setcounter{equation}{0}
\renewcommand{\thetable}{D\arabic{table}}
\renewcommand{\thefigure}{D\arabic{figure}}
\renewcommand{\theequation}{D\arabic{equation}}

This appendix presents the complete model of Section~\ref{sub:framework}. Students do research and write to produce an essay and acquire knowledge. We distinguish AI that helps students understand the material from AI that writes the essay in their place, and we show how the two capabilities affect effort, knowledge, and the incentive to delegate. We first present a baseline model without AI and then introduce the two AI capabilities and derive how each changes effort and knowledge relative to the baseline.

\subsection{Setup}

A student chooses cognitive effort in research, $r \geq 0$ (e.g., reading and searching), and writing, $w \geq 0$ (e.g., drafting, editing, note-taking, and organizing). Both activities produce knowledge (denoted by $K$), that is, lead to learning. Writing also produces the essay the student submits, and the quality of that essay reflects the knowledge the student has acquired. Without AI, knowledge and Session One essay quality are
\begin{align}
	K &= \theta_r r + \theta_w w, \notag \\
	Q_1 &= w + \phi K, \label{eq:app_baseline_outputs}
\end{align}
where $\theta_r, \theta_w > 0$ are the returns to effort in each activity and $\phi \geq 0$ measures how much of the student's knowledge shows in the essay. We focus on the case $\phi > 0$. Writing raises essay quality directly, and both activities raise it through the knowledge they produce, so the reward for essay quality enters both effort decisions. 

The student values knowledge at $\omega_K > 0$ per unit and essay quality at $\omega_Q > 0$ per unit. These valuations combine intrinsic interest with external rewards. In our experiment, the lottery tickets paid for correct test answers and for essay quality contribute to them. In ordinary coursework, grades and the future returns to skill are external incentives, while interest in the topic is an intrinsic incentive. Effort $x$ in each activity costs $x^2/(2\mu)$, with $\mu > 0$. The cost is convex, so each additional unit of effort costs more than the last, and a higher $\mu$ lowers the marginal cost of effort.%

\subsection{The Student's Problem Without AI}

We begin by studying the student's problem without access to AI. The student solves
\begin{align} \label{eq:app_baseline_problem}
	\max_{r, w \geq 0} \; \underbrace{\omega_K \big(\theta_r r + \theta_w w\big)}_{\substack{\text{Value of} \\ \text{knowledge}}} + \underbrace{\omega_Q \big[w + \phi \big(\theta_r r + \theta_w w\big)\big]}_{\substack{\text{Reward for} \\ \text{essay quality}}} - \underbrace{\frac{r^2 + w^2}{2\mu}}_{\substack{\text{Cost of} \\ \text{effort}}}.
\end{align}

The objective is strictly concave in $(r, w)$, so the first-order conditions give the unique solution
\begin{align}
	r_0 = \mu \big(\omega_K + \phi \omega_Q\big) \theta_r \quad \text{ and } \quad w_0 = \mu \big[\big(\omega_K + \phi \omega_Q\big) \theta_w + \omega_Q\big]. \label{eq:app_baseline_effort}
\end{align}

Substituting these choices into equation~\eqref{eq:app_baseline_outputs} gives an expression for learning without AI:

\begin{align} \label{eq:app_baseline_knowledge}
	K_0 = \underbrace{\mu \omega_K \big(\theta_r^2 + \theta_w^2\big)}_{\substack{\text{Value of} \\ \text{knowledge}}} + \underbrace{\mu \omega_Q \big[\theta_w + \phi \big(\theta_r^2 + \theta_w^2\big)\big]}_{\substack{\text{Reward for} \\ \text{essay quality}}}.
\end{align}

Equation~\eqref{eq:app_baseline_knowledge} makes explicit the two sources of learning. The first term is the knowledge the student acquires through research and writing because they value knowledge. The second term is the knowledge that the reward for essay quality generates: producing the essay requires writing, and writing produces knowledge; and because the essay shows what the student knows, the reward also induces research and writing that add to knowledge. 

\subsection{Introducing AI}

We extend the baseline model by giving the student access to AI, which enters through two non-mutually exclusive capabilities. First, used for augmentation, AI raises the return to research effort by $\tau \geq 0$. The parameter $\tau$ is a reduced form for the many channels through which AI makes each unit of research effort more productive. AI can act as a tutor that explains confusing material, as a research assistant that finds relevant and more rigorous sources faster, as a summarizer that condenses reading to what matters most, as an illustrator that produces visualizations on demand, and as an examiner that quizzes the student on what they have read. For concreteness, we refer to this capability as tutoring. Second, the student may automate the process of writing the essay by delegating this task to AI. Let $m \in \{0, 1\}$ indicate delegation.\footnote{We model delegation as a binary decision for simplicity. In practice there are gradations: students can have AI write parts of the essay, or use it for narrower purposes such as proofreading or paraphrasing. The model also abstracts from the combination of an AI draft with the student's own revisions. It treats delegation as the extreme case in which the submitted essay is AI-written, whereas the empirical automation category of Section~\ref{sub:aug_auto} includes partial delegation. Allowing $m$ to vary continuously between 0 and 1 would not change the solution: the student's utility, maximized over effort, is convex in $m$, so the student would still choose one of the two corners. Interior delegation would require, for example, diminishing returns to delegating a larger share of the essay.} Let $\bar{Q} \geq 0$ denote the quality of the AI-written essay.\footnote{Both $\tau$ and $\bar{Q}$ depend on the capabilities of the AI model, which may explain why students made little use of language models before ChatGPT. They may also depend on the student's academic preparation, knowledge of the topic, and prompting skill, which would make them vary across students and add a second source of selection into use types. For simplicity, we take them as given parameters when solving for effort and delegation.} 

With AI, knowledge and Session One essay quality are
\begin{align}
	K &= (\theta_r + \tau)\, r + \theta_w w, \notag  \\
	Q_1 &= (1 - m)\, (w + \phi K) + m \bar{Q}. \notag
\end{align}

Delegation saves the student the time that writing the essay would have taken. Let $v \geq 0$ denote the student's valuation of this time. It is distinct from the cognitive effort that delegation saves, whose cost the cost-of-effort term already captures. Given AI access and the tutoring gain, the student solves
\begin{align} \label{eq:app_problem}
	\max_{r, w \geq 0,\; m \in \{0,1\}} \; \underbrace{\omega_K K}_{\substack{\text{Value of} \\ \text{knowledge}}} + \underbrace{\omega_Q \big[(1 - m)\, (w + \phi K) + m \bar{Q}\big]}_{\substack{\text{Reward for} \\ \text{essay quality}}} + \underbrace{m v}_{\substack{\text{Value of} \\ \text{recovered time}}} - \underbrace{\frac{r^2 + w^2}{2\mu}}_{\substack{\text{Cost of} \\ \text{effort}}},
\end{align}
where $K = (\theta_r + \tau)\, r + \theta_w w$.

For a given delegation choice, the first-order conditions yield:
\begin{align}
	r^*(\tau, m) = \mu \tilde{\omega}_m (\theta_r + \tau) \quad \text{ and } \quad  w^*(\tau, m) = \mu \big[\tilde{\omega}_m \theta_w + \omega_Q (1 - m)\big], \label{eq:app_effort}
\end{align}
where $\tilde{\omega}_m \equiv \omega_K + \phi \omega_Q (1 - m)$ is the value of a unit of knowledge given the delegation choice: $\omega_K + \phi \omega_Q$ for a student who writes the essay, as in the baseline, and $\omega_K$ for a student who delegates, because the student's knowledge no longer raises the quality of the submitted essay.

Equation~\eqref{eq:app_effort} nests the baseline effort levels of equation~\eqref{eq:app_baseline_effort} and shows how each capability changes effort, holding the other fixed. Tutoring raises research effort by $\mu \tilde{\omega}_m \tau$, because the return to research effort is higher, and leaves writing effort unchanged. Delegation removes the reward for essay quality from the research and writing decisions. Thus, both types of effort decrease. In particular, writing effort falls by $\mu \omega_Q (1 + \phi \theta_w)$ because the student's writing no longer determines the submitted essay, and research effort falls by $\mu \phi \omega_Q (\theta_r + \tau)$, because the knowledge research produces no longer improves the essay. In proportional terms, research effort falls by the fraction $\phi \omega_Q / (\omega_K + \phi \omega_Q)$ and writing effort by the larger fraction $\omega_Q (1 + \phi \theta_w) / [(\omega_K + \phi \omega_Q) \theta_w + \omega_Q]$, so delegation lowers the share of effort devoted to writing.\footnote{The difference between the two fractions is $\omega_K \omega_Q / \{(\omega_K + \phi \omega_Q) [(\omega_K + \phi \omega_Q) \theta_w + \omega_Q]\} > 0$. Delegation removes the direct reward for writing, which is a larger part of the motive for writing than the essay value of knowledge is of the motive for research.} Notably, writing effort remains positive even with delegation ($m = 1$) because students still get the learning benefit of writing, for example when they take notes or organize their ideas. 

Substituting the effort choices into the knowledge equation yields the level of learning with AI:
\begin{align} \label{eq:app_knowledge}
	K^*(\tau, m) = \mu \tilde{\omega}_m \big[(\theta_r + \tau)^2 + \theta_w^2\big] + \mu \omega_Q \theta_w (1 - m).
\end{align}

The effect of AI access on knowledge, the model's counterpart of the treatment effect on test scores, is
\begin{align}
	\Delta K &= \underbrace{\mu \big(\omega_K + \phi \omega_Q\big) \big[(\theta_r + \tau)^2 - \theta_r^2\big]}_{\substack{\text{Gain from} \\ \text{tutoring}}} - \underbrace{\mu \omega_Q \big[\theta_w + \phi \big(\theta_r^2 + \theta_w^2\big)\big] m}_{\substack{\text{Learning displaced} \\ \text{by delegation}}} - \underbrace{\mu \phi \omega_Q \big[(\theta_r + \tau)^2 - \theta_r^2\big] m}_{\substack{\text{Gain from tutoring} \\ \text{lost by delegating}}}. \label{eq:app_deltaK}
\end{align}

Equation~\eqref{eq:app_deltaK} shows that the impact of AI on learning is ex ante ambiguous and it depends on the relative size of three forces. The first one is the gain from tutoring for a student who writes the essay. It combines the increase in the knowledge that each unit of research effort produces with the increase in research effort that the higher return induces. The second force is the learning displaced by delegation: the knowledge that the reward for essay quality generates when the student writes the essay, through writing directly and through the research and writing that the essay value of knowledge motivates. The third force is that the knowledge that tutoring produces no longer improves the submitted essay, so research effort responds less to the higher return. The effect of AI access on knowledge therefore depends on whether the student delegates and, if so, on whether the gain from tutoring exceeds the two losses from delegation. Appendix Figure~\ref{fig:app_deltaK} plots the change in knowledge against the tutoring gain for a student who writes the essay and for the same student when delegating, and shows the two losses from delegation, whose sum widens with the tutoring gain. We now turn to the choice of whether to delegate.

\begin{figure}[H]
	\caption{The Effect of AI Access on Knowledge With and Without Delegation} \label{fig:app_deltaK}
	\centering
	\begin{tikzpicture}[x=4cm, y=0.36cm, >=Latex]
		\fill[gray!30] plot[domain=0:2, samples=60] (\x, {\x*\x + 2*\x}) -- plot[domain=0:2, samples=60] ({2-\x}, {(2-\x)*(2-\x) + 2*(2-\x) - 3}) -- cycle;
		\fill[gray!12] plot[domain=0:2, samples=60] (\x, {2*\x*\x + 4*\x}) -- plot[domain=0:2, samples=60] ({2-\x}, {(2-\x)*(2-\x) + 2*(2-\x)}) -- cycle;
		\draw[thin, gray!60!black] plot[domain=0:2, samples=60, smooth] (\x, {\x*\x + 2*\x});
		\draw[->] (0,-4) -- (2.15,-4);
		\draw[->] (0,-4) -- (0,17.5);
		\node[below=0.25cm, font=\small] at (1.075,-4) {Tutoring gain, $\tau$};
		\node[rotate=90, anchor=south, font=\small] at ([xshift=-0.35cm]0,6.5) {Knowledge relative to no AI, $\Delta K$};
		\draw[densely dotted] (0,0) -- (2.15,0) node[right, font=\footnotesize] {No AI};
		\draw[very thick, blue!60!black] plot[domain=0:2, samples=60, smooth] (\x, {2*\x*\x + 4*\x});
		\draw[very thick, red!60!black, dashed] plot[domain=0:2, samples=60, smooth] (\x, {\x*\x + 2*\x - 3});
		\fill (1,0) circle (1.6pt);
		\node[below right, font=\footnotesize, align=left] at (1.03,-0.3) {Gain equals\\ losses};
		\draw[decorate, decoration={brace, amplitude=4pt}] (2.06,16) -- (2.06,8) node[midway, right=5pt, font=\footnotesize, align=left] {Gain from tutoring\\ lost by delegating};
		\draw[decorate, decoration={brace, amplitude=4pt}] (2.06,8) -- (2.06,5) node[midway, right=5pt, font=\footnotesize, align=left] {Learning displaced\\ by delegation};
		\node[blue!60!black, above left, font=\small] at (1.3,8.58) {Writes the essay ($m = 0$)};
		\node[red!60!black, below right, font=\small] at (1.5,2.25) {Delegates ($m = 1$)};
	\end{tikzpicture}
	\begin{singlespace}
		\footnotesize \justifying \noindent
		\textit{Notes:} The figure plots the change in knowledge relative to no AI, $K^*(\tau, m) - K_0$, against the tutoring gain, for a student who writes the essay ($m = 0$) and for the same student when delegating ($m = 1$), with research and writing effort chosen optimally in each case. The curves are drawn for $\mu = \omega_K = \omega_Q = \theta_r = \theta_w = \phi = 1$. The values are illustrative, so the axes show no scale. The two curves are two counterfactual choices for the same student, holding everything else fixed, not average outcomes for two groups of students. The shaded area between the curves is the loss from delegation, the sum of the last two terms of equation~\eqref{eq:app_deltaK}. The darker band is the learning displaced by delegation, which does not depend on the tutoring gain. The lighter band is the gain from tutoring lost by delegating, which grows with the tutoring gain, so the delegation curve rises more slowly and the gap between the curves widens. The braces show the two parts at the largest tutoring gain drawn. At the marked point on the no-AI line, the gain from tutoring equals the two losses from delegation. \par
	\end{singlespace}
\end{figure}

\subsection{The Decision to Delegate Work to the AI} \label{app:framework_delegation}

The student compares the utility obtained with and without delegation, each evaluated at the effort levels of equation~\eqref{eq:app_effort}. Substituting those effort levels into the objective function gives the utility under each choice,
\begin{align}
	V_m = \underbrace{\frac{\mu}{2}\, \tilde{\omega}_m^2 (\theta_r + \tau)^2}_{\substack{\text{Net benefit of} \\ \text{research effort}}} + \underbrace{\frac{\mu}{2} \big[\tilde{\omega}_m \theta_w + \omega_Q (1 - m)\big]^2}_{\substack{\text{Net benefit of} \\ \text{writing effort}}} + \underbrace{m \big(\omega_Q \bar{Q} + v\big)}_{\substack{\text{AI-written essay} \\ \text{and recovered time}}}. \notag
\end{align}

The student delegates when
\begin{align}
	V_1 - V_0 &= \omega_Q \bar{Q} + v - \frac{\mu}{2} \Big[\big(\tilde{\omega}_0^2 - \omega_K^2\big) (\theta_r + \tau)^2 + \big(\tilde{\omega}_0 \theta_w + \omega_Q\big)^2 - \omega_K^2 \theta_w^2\Big] \geq 0 \notag \\
	&\iff \overbrace{\underbrace{\omega_Q \bar{Q}}_{\substack{\text{Reward for the} \\ \text{AI-written essay}}} + \underbrace{v}_{\substack{\text{Value of} \\ \text{recovered time}}}}^{\text{Benefit of delegation}} \notag \\[1.5ex]
	&\qquad \geq \overbrace{\underbrace{\mu \omega_K \omega_Q \big[\theta_w + \phi \big((\theta_r + \tau)^2 + \theta_w^2\big)\big]}_{\substack{\text{Value of the} \\ \text{forgone knowledge}}} + \underbrace{\frac{\mu \omega_Q^2}{2} \big[(1 + \phi \theta_w)^2 + \phi^2 (\theta_r + \tau)^2\big]}_{\substack{\text{Own-essay reward} \\ \text{net of extra effort}}}}^{\text{Cost of delegation}}. \label{eq:app_delegate}
\end{align}

Equation~\eqref{eq:app_delegate} clarifies the sources that determine the decision to delegate. On the benefit side, a higher-quality AI-written essay ($\bar{Q}$) and a higher value of recovered time ($v$) make delegation more attractive. The cost of delegation is the value of the knowledge that a student with AI gives up by delegating, $\omega_K [K^*(\tau, 0) - K^*(\tau, 1)]$, plus the reward the student's own essay would have earned, net of the extra effort it would have taken. A higher value of knowledge makes delegation less attractive, since $\partial (V_1 - V_0)/\partial \omega_K = -\mu \omega_Q [\theta_w + \phi ((\theta_r + \tau)^2 + \theta_w^2)] < 0$: students who value learning more are less willing to give up the knowledge that the essay reward induces. Less costly effort (a higher $\mu$), a higher return to writing effort ($\theta_w$), a larger role of knowledge in the essay ($\phi$), and a larger tutoring gain ($\tau$) also make delegation less attractive, the last because a student who writes the essay captures the essay value of the knowledge that tutoring produces: $\partial (V_1 - V_0)/\partial \tau = -\mu \phi \omega_Q (2 \omega_K + \phi \omega_Q)(\theta_r + \tau) < 0$.

Whether delegation hurts or helps essay quality is ambiguous. A student who delegates submits an essay of quality $\bar{Q}$ in place of the essay they would have written, whose quality is $w^*(\tau, 0) + \phi K^*(\tau, 0)$ by equations~\eqref{eq:app_effort} and \eqref{eq:app_knowledge}. Solving equation~\eqref{eq:app_delegate} at equality for $\bar{Q}$ gives the quality of the AI-written essay at which the student is indifferent between delegating and writing. This threshold is below the quality of the student's own essay: the AI-written essay can be worse than the student's own by $(\mu \omega_Q / 2) [(1 + \phi \theta_w)^2 + \phi^2 (\theta_r + \tau)^2] + v / \omega_Q$ and the student still delegates. The first term is the effort that delegation saves, net of the knowledge that effort would have produced, and the second is the value of recovered time, both in units of essay quality. A student whose AI-written essay falls in that range delegates to a worse essay, trading essay quality for the effort and time that delegation saves. A student whose AI-written essay is better than their own delegates to a better one. The model therefore allows a gain in essay quality without a gain in knowledge, and it does not require an essay-quality gain whenever the student delegates. For a student who writes the essay, tutoring raises its quality through the knowledge it produces, by $\phi$ times the gain from tutoring in equation~\eqref{eq:app_deltaK}, so a gain in essay quality can also come with a gain in knowledge.

Equation~\eqref{eq:app_delegate} also makes clear that delegation is endogenous to the incentives the student faces. The value of knowledge, $\omega_K$, combines the student's intrinsic motivation to learn with the external rewards for learning: the lottery tickets for correct test answers in our experiment, or grades and the returns to skill in a course. A student with a higher $\omega_K$ is less willing to delegate and also supplies more effort in both activities, so students who delegate and students who do not delegate already differ in motivation and incentives before AI enters. The reward for essay quality and the value of recovered time, which outside the laboratory come from grades and from competing demands on the student's time, shift the decision as well. A higher reward for essay quality discourages delegation: it raises the quality the AI-written essay must reach, since the student at the threshold delegates to an essay worse than their own.

\subsection{Relationship Between the Model and the Experiment} \label{app:framework_time}

The experiment measures the model's outputs. Both tests measure $K$. The Session One essay measures $Q_1$. The Session Two essay is written without AI, so its quality reflects the student's writing effort during the assessment and, through $\phi$, the knowledge the student retains. The time-use survey reports minutes, which relate to effort loosely, as discussed below, and the learning-experience measures relate to the cost of effort. We take each result in turn and end with three implications.

\subsubsection{Test Scores.}

AI access raised test scores in both sessions (Section~\ref{sub:tests}). In the model, treated students' scores measure $K^*(\tau, m)$ and control students' scores measure $K_0$. Random assignment equalizes the two groups' preferences and returns to effort in expectation, so the Session One effect is the average of $\Delta K$ in equation~\eqref{eq:app_deltaK} across treated students, each with their own tutoring gain and delegation choice, including students who did not use AI, for whom $\tau = 0$ and $m = 0$, so $\Delta K = 0$. Through the lens of the model, a positive average effect means that, averaged across treated students, the gain from tutoring exceeded the two losses from delegation. 

\subsubsection{Essays.}

The Session One essay measures $Q_1$. For a student who writes the essay, tutoring raises its quality through the knowledge it produces. For a student who delegates (and the rise in the share of text flagged as AI-generated shows that many treated students did), the submitted essay can be better or worse than the student's own. The model therefore does not sign the average Session One essay effect. The unaided Session Two essay reflects retained knowledge and writing effort during the assessment. The direct contribution of AI-written text to the first essay, $m \bar{Q}$, does not enter it. Knowledge gains can appear in both the follow-up test and the unaided essay, whereas the features of AI-written text cannot.

\subsubsection{Time Allocation.}

Treated students shifted time from writing toward research (Table~\ref{tab:mech_long}, Panel~B). In the model, both AI capabilities lower the share of effort devoted to writing: a higher $\tau$ raises research effort without changing writing effort, and delegation lowers effort in both activities, writing by proportionally more.\footnote{The time-use estimates measure minutes, whereas $r$ and $w$ in equation~\eqref{eq:app_effort} measure cognitive effort, which also depends on how intensively a student works during those minutes. The two move in the same direction here, but not necessarily one for one.}

\subsubsection{The Learning Experience.}

Treated students enjoyed the task more (Table~\ref{tab:mech_long}, Panel~C). One interpretation is that AI changes the cost of effort, a channel that the decomposition in equation~\eqref{eq:app_deltaK} holds fixed.\footnote{We note that other interpretations are possible: students may enjoy the task more because they are learning more or because less writing is required.} Holding constant the delegation choice, equation~\eqref{eq:app_knowledge} implies
\begin{align}
	\frac{\partial K^*}{\partial \mu} = \tilde{\omega}_m \big[(\theta_r + \tau)^2 + \theta_w^2\big] + \omega_Q \theta_w (1 - m) > 0. \notag
\end{align}

Less costly effort induces more research and writing effort and raises knowledge.

\subsubsection{Additional Implications.}

The model delivers three implications that the experiment can examine descriptively by use type. 

First, a delegating student's Session One essay advantage comes from $\bar{Q}$ and is absent from unaided work, whereas knowledge, whether from tutoring or from the student's own research and writing, persists into the unaided assessments, including the unaided essay, which expresses knowledge through $\phi$. The essay advantage of automation users should therefore be concentrated in Session One, any test-score advantage should appear in both sessions, and an essay advantage of augmentation users, which can arise only through knowledge, should appear in both sessions.

Second, the two capabilities move effort in different directions. Relative to a student without AI, tutoring raises research effort and leaves writing effort unchanged, whereas delegation lowers writing effort and frees time, which in the laboratory students could use only by stopping early and browsing. Augmentation users should therefore do more research than control students, and automation users should write less than control students and stop earlier. The model does not sign automation users' research effort relative to control students, since tutoring raises it and delegation lowers it. Holding the tutoring gain fixed, however, automation users differ from augmentation users only by the delegation effect in equation~\eqref{eq:app_effort}, which lowers effort in both activities. Section~\ref{sub:aug_auto} reports that automation users spend less time on research and less time overall than augmentation users. 

Third, the choice to delegate reveals the student's incentives and type. By the delegation condition, students who delegate place a lower value on knowledge or a higher value on recovered time, expect a better AI-written essay, or gain less from tutoring, so predetermined characteristics that correlate with these parameters should predict use type. Appendix Table~\ref{tab:usage_baseline} shows that they do: automation users enter with higher baseline test scores than augmentation users, which the model can rationalize through a higher $\bar{Q}$ for better-prepared students.

	\clearpage
	\begin{singlespace}
		\bibliographystyle{apa}
		\bibliography{add_gr.bib}
	\end{singlespace}
	
\end{document}